\documentclass[twoside,a4paper,12pt,titlepage]{article}
\usepackage[latin1]{inputenc} 
 \usepackage[T1]{fontenc}      % Police contenant les caractres franais
\usepackage{geometry}         % Dfinir les marges
\RequirePackage{textcomp}
\RequirePackage{amsthm}
\RequirePackage{amsmath}
\RequirePackage{bm}
\RequirePackage{amssymb}
\RequirePackage{upref}
\RequirePackage{amsthm}
\RequirePackage{enumerate}
\RequirePackage{pb-diagram}
\RequirePackage{amsfonts}
\RequirePackage[mathscr]{eucal}
\RequirePackage{verbatim}
\RequirePackage{xr}
\newcommand{\al}{\alpha}
\newcommand{\bet}{\beta}
\newcommand{\ga}{\gamma}
\newcommand{\de}{\delta }
\newcommand{\e}{\epsilon}

\newcommand{\h}{\eta}

\newcommand{\tht}{\theta}

\newcommand{\m}{\mu}

\newcommand{\p}{\pi}

\newcommand{\vr}{\varrho}
\newcommand{\s}{\sigma}
\newcommand{\x}{\xi}
\newcommand{\z}{\zeta}
\newcommand{\bo}{\mathbf}

\newcommand{\Hyp}{\mathcal{H}}
\newcommand{\Lb}{\underline{L}}

\newcommand{\deb}{\bar{\de}\hspace{.5mm}}

\newcommand{\ebaa}{\bar{\e}}
\newcommand{\hb}{\underline{\h}}
\newcommand{\mub}{\tilde{\mu}}
\newcommand{\demi}{\frac{1}{2}}
\newcommand{\ls}{\leqslant}
\newcommand{\rs}{\geqslant}
\newcommand{\hs}{\hspace{5mm}}
\newcommand{\dope}{\mathcal{D}}

\newcommand{\nd}[1]{\|#1\|_{L^2(S_t)}}
\newcommand{\ndleb}[2]{\|#1\|_{L^{#2}(S_t)}}
\newcommand{\chiu}{\underline{\chi}}
\newcommand{\nun}[1]{\mathcal{N}_1(#1)}
\newcommand{\norme}[1]{\|#1\|}
\newcommand{\normeb}[1]{\|#1\|_{\mathcal{B}^0}}
\newcommand{\normep}[1]{\|#1\|_{\mathcal{P}^0}}
\newcommand{\abs}[1]{\lvert#1\rvert}

\newcommand{\norm}[1]{\lVert#1\rVert}

\newcommand{\normf}[1]{\left\|#1\right\|}
\newcommand{\nl}[3]{\left\|#1\right\|_{L^{#2}_x L^{#3}_{[0,T]}}}
\newcommand{\nli}[2]{\left\|#1\right\|_{L^{#2}_x(S_0)}}

\newcommand{\ntsl}[3]{\left\|#1\right\|_{L^{#2}_{[0,T]}L^{#3}_x }}
\newcommand{\nlsl}[2]{\left\|#1\right\|_{L^{#2}_x(S_t)}}
\newcommand{\ns}[2]{\left\|#1\right\|_{L^{#2}(\Sigma_t)}}
\newcommand{\nm}[2]{\left\|#1\right\|_{L^{#2}(M_t)}}
\newcommand{\nnc}[2]{\left\|#1\right\|_{L^{#2}(\mathcal{N}^{-}(p,\delta))}}
\usepackage{xspace}

\newcommand{\sstrike}[1]{\ensuremath{\not\!\text{#1}}\xspace}
\newcommand{\nablab}{\sstrike{$\nabla$}}
\newcommand{\Deltab}{\sstrike{$\Delta$}}
\newtheorem{thm}{Theorem}[subsection]
\newtheorem{lem}{Lemma}[subsection]
\newtheorem{prop}{Proposition}[subsection]
\newtheorem{nota}{Notation}[subsection]
\newtheorem{defi}{Definition}[subsection]

\begin{document}

\title{An integral  breakdown criterion for Einstein vacuum equations in the case of  asymptotically flat spacetimes.}   

\author{David Parlongue, Nice University}

\maketitle              

\begin{abstract}
  We will give in this paper the proof of an integral breakdown criterion for Einstein vacuum equations. In \cite{krb} a new breakdown criterion was proved as a result of the sequence of articles \cite{krg}, \cite{krk}, \cite{krc} and \cite{kri}. This result is a large improvement of preexisting results as it will be explained later on. However, in this article, the authors mentioned that it was likely possible to prove a sharper result involving an integral condition instead of a pointwise one. This paper is concerned with giving the proof of this improvement. Moreover the proof of  this breakdown criterion was written in \cite{krb} for a foliation of constant mean curvature, we will present it here for a maximal foliation which leads to some difficulties due to the non-compacity of the leaves of such a foliation and the use of weighted Sobolev norms.
 \end{abstract}
\pagenumbering{arabic} \setcounter{page}{3} 

\tableofcontents         

% \listoffigures              % Table des figures

% \listoftables               % Liste des tableaux
\cleardoublepage

\section{Breakdown for Einstein Equations :}
\subsection{Introduction :}                  % Commencer une partie...
We will consider here a $3+1$ dimensional Lorentzian manifold $\mathbf{(M,g)}$ satisfying the Einstein  vacuum equations (EVE).

\begin{equation}
\begin{aligned}
\mathbf{R_{\al \bet}(g)=0}
\end{aligned}
\end{equation}

In order to describe our problem as a PDE evolution problem, that is to state our problem as a Cauchy problem, we need a $3+1$ splitting of our space-time. To do that, we will suppose that a part of the space-time denoted by  $\mathcal{M} \subset \mathbf{M}$ is globally hyperbolic with respect to a spatial hypersurface $\Sigma_0$ and foliated by the level hypersurfaces of a regular time function denoted by $t$. We suppose moreover that this time function is monotically increasing towards future, with lapse function $n$ and second fundamental form $k$. We will use the following sign convention  :
\begin{eqnarray}
&& k(X,Y)=-\bo{g}(\b(D_{X}T,Y) ,\\
&& n=(-\bo{g}(\bo{D}t,\bo{D}t))^{-1/2}
\end{eqnarray}

Throughout this paper $\bo{T}$, will denote the future unit normal to the leaf $\Sigma_{t}$ of the time foliation and $\bo{D}$ the covariant derivative associated with  $\bo{g}$. Let $\Sigma_{0}$ be a slice of the time foliation wich will be referred as the initial slice. The EVE can be seen as an initial value problem once an initial data set is given. An initial data set consists in a 3 dimensional Riemanian hypersurface $\Sigma_{0}$ together with a 3 dimensional Riemanian metric $g$ and a second fundamental form $k$. 

The initial hypersurface has moreover to satisfy constraint equations which appear when writing the equation $(1)$ in term of a 3+1 splitting. In the case of a maximal foliation, these equations take the following form :
\begin{eqnarray}
&& tr k = 0,\\
&& \nabla^{j}k_{i,j}=0,\\
&&
 R=|k|^2
\end{eqnarray}
where here $R$ stands for the Riemann curvature of the metric $g$. 
The other equations coming from the reformulation of equation (1.1) are the following evolutions equations : 

\begin{eqnarray}
&&\partial_t g_{ij}=-2nk_{ij},\\
&&\partial_t k_{ij}=-\nabla_i\nabla_j n + n(R_{ij}-2k_{ia}k^{a}_j)
\end{eqnarray}

together with the lapse equation :

\begin{equation}
\begin{aligned}
\Delta n = |k|^2 n
\end{aligned}
\end{equation}
with condition $n \rightarrow 1$ at spatial infinity on $\Sigma_0$.
The result \cite{krb} as well as what we will prove here can apply to maximal foliations as well as constant mean curvature foliations (CMC). In this case the leaves are compact and the mean curvature can be taken as being the time function. We will consider here the case of maximal foliation, the one with CMC foliation being simpler due to the compacity of the leaves. We will consider the case of asymptotic flatness. $\Sigma_0$ is said to be asymptotically flat if the complement of a compact set $K \in \Sigma_0$ is diffeormorphic to the complement of a $3$-sphere and that there exists a system of coordinate in which :

\begin{eqnarray}
g_{ij}&=&(1+\frac{M}{r})\delta_{ij} +o_4(r^{-3/2}) \\
k_{ij}&=& o_3(r^{-5/2})
\end{eqnarray}

The system of equations (4) to (9) is a closed system of equations. We are naturally led to ask the question of well-posedness for this system. The first result of this theory is the local well-posedness result of Choquet-Bruhat  \cite{cb}, here as stated in \cite{ck}:

\begin{thm} (Local existence theorem) Let $(\Sigma_0,g_0,k_0)$ be an initial data set verifying the following conditions :

1) $(\Sigma_0,g_0)$ is a complete Riemannian manifold diffeormorphic to $\mathbb{R}^3$.

2)The isoperimetric constant of $(\Sigma_0,g_0)$ is finite.

3) $Ric(g_0) \in {H}_{2,1}(\Sigma_0,g_0)$ and $k_0 \in {H}_{3,1}(\Sigma_0,g_0)$.
 
4) The initial slice satisfies the constraint equations.

Then there exists a unique, local-in-time smooth development, foliated by a normal, maximal time foliation $t$ with range in some $\left[0,T\right]$ and with $t=0$ corresponding to the initial slice $\Sigma$. Moreover :
 
 a) $g(t)-g \in \mathcal{C}^1([0,T],H_{4,1})$
 
 b) $k(t)\in \mathcal{C}^0([0,T],H_{3,1})$
 
 c) $Ric(g)(t)\in \mathcal{C}^0([0,T],{H}_{2,1})$
 \end{thm}
where for a given tensorfield $h$ on $\Sigma$, $\|h\|_{H_{n,s}}$ denotes the norm :
$$\|h\|_{H_{n,s}}= ( \sum_{i=0}^n\int_\sigma (1+d_0^2)^{s+i}\abs{\nabla^i h}^2 )^\demi$$
\vspace{5 mm}

where $d_0$ stands for the geodesic distance to a basepoint $O$.

 The natural question for such a local result is the minimum regularity of the inital data set for having well-posedness. The scaling property of (1.1) could lead to think that the critical sobolev exponent for $g$ should be $3/2$. The local theorem can be easily extended to $5/2 + \e $. An important improvement led to the proof for $2 + \e $, see \cite{kr2}. The so-called $L^2$-conjecture asserts that the optimal regularity for the initial data set is $Ric(g) \in L^2$ corresponding to an initial metric $g$ in $H^2$, the proof is however not yet avable.

\vspace{5 mm}
Consistant with this conjecture, we can expect to prove continuation arguments with $g(t)$ having this critical regularity, which is precisely the purpose of \cite{krb} and of the continuation argument of this paper.
We are now ready to state the main result of  \cite{krb} :

\begin{thm}
Let  $\mathbf{(M,g)}$ be a globally hyperbolic development of an asymptotically flat initial data set $(\Sigma_{0},g_0,k_0)$ satisfying the assumptions of the local in time result and globally foliated by a normal, maximal foliation given by the level sets of a smooth time function t such that $\Sigma_{0}$ corresponds to $t=0$. Suppose moreover that $\Sigma_{0}$ satisfies the geometric condition {\bf (G1)} (see below). 
We suppose that :
\begin{eqnarray}
&a)& \int_{\Sigma_{0}} \abs{\bo{R}}^2(x)d\mu(x) \ls \Delta_0\\
&b)& \normf{n^{-1}}_{L^\infty([0,t_1[,L^\infty(\Sigma_{t}))} \ls\Delta_1\\
&c)& \normf{k}_{L^\infty([0,t_1[,L^\infty(\Sigma_{t}))} + \normf{\nabla(log(n))}_{L^\infty([0,t_1[,L^\infty(\Sigma_{t}))} \ls \Delta_2
\end{eqnarray}
than the space-time together with the maximal foliation can be extended beyond time $t_1$.
\end{thm}

\vspace{5 mm}

This theorem implies directly the following breakdown criterion :

\begin{thm}{\bf [Breakdown Criterion]}
With the notations of theorem (1.1.2), the first time $T*$ with respect to the $t$-foliation of a breakdown is charecterized by the condition :

$$\limsup_{t \to T*} {\big( \normf{k}_{L^\infty(\Sigma_{t})}+ \normf{\nabla(log(n))}_{L^\infty(\Sigma_{t})}\big)} =+ \infty$$
\end{thm}

\vspace{10 mm}
A lot of commentaries should be done on this theorem. But before entering the remarks and elements of the proof of  \cite{krb}, let us state what will be the result of this paper.

\begin{thm} {\bf [Main theorem]}
We keep here the notations of theorem (1.1.2). We suppose that there exists $\Delta_0, \Delta_1$ and $\Delta_2$ such that : 
\begin{eqnarray}
&a)& \int_{\Sigma_{0}} \abs{\bo{R}}^2(x)d\mu(x) \ls \Delta_0\\
&b)& \normf{n^{-1}}_{L^\infty([0,t_1[,L^\infty(\Sigma_{t}))} \ls\Delta_1\\
&c)& \int_0^{t_1}(\normf{k}_{L^\infty(\Sigma_{t})}+ \normf{\nabla(log(n))}_{L^\infty(\Sigma_{t})})^2ndt \ls \Delta_2
 \end{eqnarray}

\vspace{5 mm}

than the space-time together with the maximal foliation can be extended beyond time $t_1$.
\end{thm}

\vspace{5 mm} This theorem implies of course the corresponding breakdown criterion.
Before doing some remarks on the theorems, we will describe the geometric condition {\bf (G1)} :

\begin{thm}{\bf [Geometric assumption \bf (G1)]} The initial surface $\Sigma_0$ is said to satisfy {\bf (G1)} if and only if there exists a convering of $\Sigma_0$ by a finite number of charts $U$ such that for any fixed chart, the induced metric $g$ verifies :

$$C^{-1} \abs{\xi}^2 \ls g_{ij}(x)\xi_i \xi_j \ls C\abs{\xi}^2 , \hs \forall x \in U$$

with $C$ a fixed number.
\end{thm}
\subsection{Remarks on the theorems :}

\hspace{5mm}   The first remark is that theorem (1.1.4) is trivially better than theorem (1.1.2). One has to remark that even in the case of theorem (1.1.2), one has to add pointwise conditions on the lapse $n$ because a lower bound is not implied by (1.1.9) as it was the case in \cite{krb} in which the corresponding elliptic equation : $\Delta n = \abs{k}^2n-1$ for CMC foliation led to such lower bound. By the maximum principle, the lapse takes its value in $[0,1]$. With hypothesis $b)$, the lapse is thus supposed to be uniformly bounded from above and from below.
   
The fact that working with asymptotic flatness does not change the proof could be understood when thinking to global non-linear stability of Minkowski Space. We will see that thanks to \cite{ck} and \cite{kn}, our space-time will be uniformly asymptotically flat outside a compact set. The proof is thus concerned with the control of the geometry of $\bo{M}$ in a compact set. Moreover, the additional difficulty created by the weights in the case of maximal foliation will not be a problem as from the one hand they don't play any role in the estimates for the compact region and from another hand the exterior is very regular.
  
      Let us now make some remarks on the main hypothesis of the two theorems. Before theorem (1.1.2) was proved, the best breakdown criterion avaible was a result of M.Anderson which requires uniform bounds in space and time on the curvature $\bo{R}$. Consistant with the $L^2$ conjecture the theorems (1.1.2) and (1.1.4) only requires $L^2$ bounds on the curvature. Contrary to Anderson's result, the  \cite{krb} result requires a deep understanding of the causal structure of null cones which is the purpose of the article \cite{krc}. 
     
     However a condition on the $L^2$ norm of the curvature on the initial slice is not sufficient to obtain a global existence result. The heuristic idea is that the minimum requirement (even if a $L^2$ local result is avaible) is :
     
    $\al$ ) Having a uniform control on the geometry of the leafs $\Sigma_{t}$ 

    $\bet$ ) Having uniform $L^2(\Sigma_{t})$ control on the curvature and its derivatives.

The purpose of the conditions $b)$ and $c)$ of theorem (1.1.2) (resp. (1.1.4)) is heuristically to assure these two very minimum requirements.
The conditions $b)$ and $c)$ can be more clearly understood when looking at the deformation tensor of $\bo{T}$ denoted by $\pi={}^{(\bo{T})}\pi=\mathcal{L}_{\bo{T}}\bo{g}$.

When expressing the components of this tensor relative to an orthonormal frame $\bo{T},e_1,e_2,e_3$, we obtain the following expressions :

\begin{equation}
\pi_{00}=0, \hspace{3mm}\pi_{0i}=\nabla_i log(n), \hspace{3mm}  \pi_{ij}=-2k_{ij}.
\end{equation}
where we've used the standard conventions about latin and greek indices in General Relativity.
Thus, the additional conditions are nothing else than uniform controls on the "non-killingness" of the vector field $\bo{T}$. In the case of a $L^\infty$ in space and time on $\pi$ or equivalently condition $c)$ of theorem (1.1.2), the vector field $\bo{T}$ is said to be almost Killing. The idea beyond this notion of almost-killingness is quite clear if we recall that a standard approach to get a continuation result in PDE theory is to use a local existence result together with the existence of a conserved quantity. As it will be shown later on, energy estimates in general relativity typically lead to $L^2$ estimates on the curvature. The idea is that Killing vectorfields give birth to conserved quantities. By controlling the non-killingness of a vectorfield, we can hope to get control on the associate quantity.

We will now review the main elements of the proof of theorem (1.1.2). We would like to insist in this part on the global structure of the proof of \cite{krb}, precisely on how the other articles of the sequence are used in the proof of the breakdown criterion. As we will see, part of the work to obtain theorem (1.1.4) from the proof of theorem (1.1.2) is just checking that this weaker assumption is in fact sufficient. However we will see that a key element has to be redone : the control of the causal structure of the light cones with the integral condition c). This will be the subject of most of this paper (the section 2).

Let us sketch the proof of theorem (1.1.2). According to the local existence result, a continuation result requires from the one hand a geometric control of the leaves and roughly a $H^2$ control on $R$. A first remark (see below) is that the hypothesis of theorem (1.1.4) are sufficient to control basic aspects of the geometry of the leaves, this will be the first step of the proof. As an hypothesis, we have the control of the $L^2$ norm of the curvature. The purpose of the work will thus be to obtain $L^2$ control on the two first derivatives. If energy estimates are sufficient to get $L^2$ bounds from the initial data hypothesis and $b)$ and $c)$, these techniques applied to the derivatives of the curvature will lead to a $L^2$ control of $\bo{DR}$ and $\bo{D^2R}$ involving not only the initial condition but also $\normf{\bo{R}}_{L^\infty(\Sigma_{t})}$. 

There are not a lot of techniques avaible to prove uniform pointwise bounds in PDE theory. The first one consists in using Sobolev embeddings which lead typically to a loss. The second one is by using a representation formula. In fact, the curvature $\bo{R}$ satifies a non linear wave equation of the form :

$$\square_{\bo{g}} \bo{R}=\bo{R}\star\bo{R}$$

where $\star$ is a bilinear symmetric operator from the four times covariant tensorfields into itself and from
Weyl tensor fields into itself.

Thus, we can hope to find a parametrix for this equation as it is the case with flat background.  This parametrix has been derived in \cite{krk}. One of its main properties is that both the integral and the error terms can be expressed as integrals along a null cone. Now, having an integral representation formula, it seems possible to derive $L^\infty$ bounds on $\bo{R}$ using $L^2$ bounds. It remains to be able to control the causal structure of the null cones and in particular its radii of conjugacy and injectivity (see \cite{krc} and \cite{kri}).

Moreover, we can remark that the asymptotical flatness as stated here (that is consistantly with \cite{ck}) is the natural requirement for physical quantities to be defined but we do not need so much regularity. Let us list the elements of spatial asymptotics which will play a role in the proof :

\begin{itemize}
\item $L^{\infty}$ bound on $R$ outside a compact set (for the control of the radius of injectivity of null cones).
\item  $g_{ij}-\delta_{ij}$ being $o_2(r^{-1})$ which is essential to control the weighted estimates needed to apply the local-in-time result.
\end{itemize}

 and of course the regularity $Ric(g(0)) \in H_{2,1}(O,\Sigma_0)$ and $k(0) \in H_{3,1}(O,\Sigma_0)$. In fact a development with this regularity which is more a asymptotically euclidian regularity would be sufficient. 

We shall also make a remark on the link between $n$ and $k$. In view of (1.19), one can ask if an integral control on $k$ would imply the same on $\nabla log(n)$. More precisely does a pointwise control on $n$ and a bound on $\int_0^{t_\star} \norme{k}_{L^\infty(\Sigma_t)}ndt$ implies a bound on $\int_0^{t_\star} \norme{\nabla log(n)}_{L^\infty(\Sigma_t)}ndt$. In view of (1.9), the question underlying this problem is the control of the constant in the following elliptic estimate : $\norme{\nabla n}_{L^\infty(\Sigma_t)} \lesssim \norme{n}_{L^\infty(\Sigma_t)} + \norme{\Delta n}_{L^\infty(\Sigma_t)}$. However the constant in this inequality depends on the harmonic radius on $\Sigma_t$. But to control this radius we need at the minimum a $L^2$ control on $R$, but the way we derive such a control requires itself bounds on $\nabla n$ via the energy estimates as we will see later on. So it seems to me that the control of all the components of ${}^{(\bo{T})}\pi$ has to be kept.

\vspace{5mm}
 Let us now begin with the proof of the very minimum control that we mentioned as requirements $\al)$ and $\bet)$.
\subsection{Basic geometric control and energy estimates}

\hspace{5mm}In this section we will consider a space-time such as in theorem (1.1.4) satisfying the hypothesis on $[0,t_\star[$. 

\subsubsection{Geometric control :}
Contrary of the case of a CMC foliation we will not of course get control on the volume of the leaves which is infinite, but we keep the following geometric control:

\begin{thm}
Under the assumptions of Theorem (1.1.4), the leaves $\Sigma_t$ of the $t$ foliation satisfy uniformly the hypothesis {\bf (G1)} for $t \in [0,t_\star[$.
\end{thm}

{\bf Proof :}
Consider the good coordinate chart $U$ given by {\bf (G1)} and consider the transported coordinates $t,x_1,x_2,x_3$ on $I \times U$. Let $X$ a time independant vector tangent on {\bf M} tangent to $\Sigma_t$.
Then :

\begin{equation}
\partial_t g(X,X)= -\demi n k(X,X)
\end{equation}
This implies :
\begin{equation}
-\demi \ns{nk}{\infty} \abs{X}_g ^2 \ls \partial_t ( \abs{X}_g ^2) \ls \demi \ns{nk}{\infty} \abs{X}_g ^2
\end{equation}
which leads to :
\begin{equation}
\abs{X}_{g(0)}^2 e^{-\int_0^t\ns{nk}{\infty}(s)ds}  \ls \abs{X}_{g(t)} ^2 \ls \abs{X}_{g(0)}^2 e^{\int_0^t\ns{nk}{\infty}(s)ds}\end{equation}

and thus using hypothesis $c)$ of the main theorem together with {\bf (G1)}, we get (1.3.1).

\vspace{5 mm}
This uniform control on the metric allows us to get uniformity in the constants of some of the Sobolev embeddings on the leaves $\Sigma_t$ (see \cite{krb}).

\begin{thm}
Under the assumptions of Theorem (1.1.4), there exists a constant independant of  $t \in [0,t_\star[$ and depending only on $\Delta_1$, $\Delta_2$ and $t_\star$ such that for every smooth scalar function on $\Sigma_t$ and any smooth tensorfield $F$ on $\Sigma_t$ and any $2\ls p \ls6$ , we have :
\begin{eqnarray}
\ns{f}{\frac{3}{2}} &\ls& C(\ns{\nabla f}{1} + \ns{ f}{1})\\
\ns{F}{p}&\ls&C(\ns{\nabla F}{2}^{3/2-3/p}\ns{F}{2}^{3/p-1/2}+\ns{\abs{F}^{2p/3}}{1}
\end{eqnarray}
\end{thm}

We also have the following result :

\begin{lem}
Let F be a tensor field on $\Sigma_t$, then :
\begin{equation}
\ns{F}{4}^2\ls \ns{\nabla^2F}{2} \ns{F}{\infty}
\end{equation}
\end{lem}

and the less trivial following (this theorem require the existence of a good local system of coordinates, see the part on the radius of injectivity) :
\begin{lem}
Let f be a scalar function on $\Sigma_t$, then :
\begin{equation}
\ns{f}{\infty}\ls C(\ns{\nabla^2f}{2} + \ns{f}{2})
\end{equation}
\end{lem}

\vspace{5mm}

\begin{thm}
Under the assumptions of Theorem (1.1.4), the leaves $\Sigma_t$ of the $t$ foliation are uniformly asymptotically flat for $t \in [0,t_\star[$.
\end{thm}

{\bf Proof :}
The proof of the uniformity of the asymptotic flatness relies on the proof of non-linear stability of Minkowski Space. With the notation of \cite{kn}, there exists a sufficiently large compact $K$ such that the initial data satisfies the global smallness assumption $J_K(\Sigma_0,g_0,k_0) \ls \e^2$ for a sufficiently small $\e>0$. Then applying the main theorem of \cite{kn} as stated in page 126, we get in particular control on the $\mathcal{O}$, $\mathcal{D}$ and $\mathcal{R}$ norms. This imply in particular that outside the domain of influence of $K$, the curvature in uniformly bounded and tends uniformly to zero on the slices $\Sigma_t$.
\vspace{5mm}

A primary thing which has also to be proved is that the hypothesis $a)$, $b)$ and $c)$ are sufficient to control the $L^2$ flux along the leaves of the foliation from the one hand and along the null cones from the other hand (the two coming in fact together). This basic result is a consequence of the energy estimates for EVE.

\subsubsection{Energy estimates :}

\hspace{5mm}Let us briefly describe the procedure used to derive energy estimates for the curvature tensor $\bold{R}$. This procedure can be applied to the whole class of Weyl tensor field that is the class of four covariant traceless tensor $\bold{W}_{\al \bet \ga \de}$ satisfying the symmetries of the curvature tensor. From a Weyl tensor field $W$, we define its so-called Bel-Robinson tensor :

$$\bold{Q}[W]_{\al \bet \ga \de}=W_{\al \vr \ga \s}{{{W_{\bet}}^{\vr}}_{\delta}}^{\s}+{}^{\star}W_{\al \vr \ga \s}{}^{\star}{{{W_{\bet}}^{\vr}}_{\delta}}^{\s}$$

where we have used the standard notation $\star$ for Hodge duality.

This tensor satisfies the following properties (see \cite{kn} or \cite{ck}) :

\begin{prop}

Let $Q=Q[W]$ be the Bel-Robinson tensor relative to a Weyl tensor W then :

i) $Q$ is symmetric and traceless relative to all pairs of indices.

ii) $Q$ satisfies the following condition : for any non-spacelike future directed vector fields $X_1, X_2, X_3, X_4$ the quantity $Q(X_1,X_2,X_3,X_4)$ is nonnegative.

iii) if $W$ satisfies the Bianchi equations then $Q$ is divergence free
\end{prop}

Property $i)$ comes directly from the algebraic properties of $W$. Properties $ii)$ will be of great use when applied to our vectorfield $\bold{T}$. The property $iii)$ is naturally the infinitesimal formulation of a conservation law. 

The following proposition is also fundamental :

\begin{prop}

Let $Q=Q[W]$ be the Bel-Robinson tensor relative to a Weyl tensor W and $X_1, X_2, X_3$ a triplet of vector fields and let P the following covariant vectorfield :

$$P_{\al}=Q_{\al \bet \ga \de}X_1^{\bet}X_2^{\ga}X_3^{\de}.$$

then we have :

$$Div P = Div Q _{\bet \ga \de}X_1^{\bet}X_2^{\ga}X_3^{\de}+\frac{1}{2}Q_{\al \bet \ga \de} \big( {}^{(X_1)} \pi^{\al \bet}X_2^{\ga}X_3^{\de}+ {}^{(X_2)} \pi^{\al \ga}X_1^{\bet}X_3^{\de}+ {}^{(X_3)} \pi^{\al \de}X_1^{\bet}X_2^{\ga} \big)$$

\end{prop}

The result of this proposition as well as the positivity result applied to $P_{\al}=Q[R]_{\al \bet \ga \de}T^{\bet}T^{\ga}T^{\de}$ gives raise to the following theorem :

\begin{thm}
Let $\bo{R}$ be the Riemann curvature tensor of a solution of the EVE and satisfying the hypothesis $a), b) $ and $c)$ of the Main theorem, then for all t in $[0,t_\star[$:

\begin{eqnarray}
&a)& \int_{\Sigma_{t}} \abs{\bo{R}}^2(x)d\mu(x) \ls C_1(\Delta_0,\Delta_1,t_\star)\\
&b)& \int_{\mathcal{N}^{-}(p)} \bo{Q[R]}(\bo{T},\bo{T},\bo{T},L) \ls C_2(\Delta_0,\Delta_1,t_\star)
 \end{eqnarray}

where $\mathcal{N}^{-}(p)$ stands for the past null cone initiating at the point p that is the boundary of the causal past of $p$ and the integral is taken in time between $0$ and $t^{\star}$ and $L$ for the null geodesic generator of $\mathcal{N}^{-}(p)$, that is the vectorfield on $\mathcal{N}^{-}(p)$ satisfying :
\begin{equation}
 \bo{D}_L L=0 \hs <L,L>=0
\end{equation}
and the normalization condition at $p$ :
$$<L,T>(p)=1$$

\end{thm}
 
 Following again the notation of \cite{krb}, we will note this flux : 
 
 $$\mathcal{F}(p)= \int_{\mathcal{N}^{-}(p)} \bo{Q[R]}(\bo{T},\bo{T},\bo{T},L)= \int_0^{t(p)}n \int_{\Sigma_{t}}\bo{Q[R]}(\bo{T},\bo{T},\bo{T},L)dA_t$$.
 
 If the integral appearing in $a)$ involves all the components of the curvature tensor, it is not the case of the one involved in $\mathcal{F}(p)$.
 
 We recall the definition of the null components of a Weyl Field $W$ relative to a null pair $(e_3,e_4)$ :
 
 \begin{eqnarray*}
 &&\al(W)(X,Y)=W(X,e_4,Y,e_4)\\
 &&\bet(W)(X)=\frac{1}{2}W(X,e_4,e_3,e_4)\\
 &&\rho(W)=\frac{1}{4}W(e_3,e_4,e_3,e_4)\\
&&\sigma(W)=\frac{1}{4}{}^{\star}W(e_3,e_4,e_3,e_4)\\
&&\underline{\al}(W)(X,Y)=W(X,e_3,Y,e_3)\\
&&\underline{\bet}(W)(X,Y)=W(X,e_3,e_3,e_3)
\end{eqnarray*}
 
$\al(W)$ and $\underline{\al}(W)$ are symmetric traceless tensors. We have here a decomposition of the ten independant coefficients of the Weyl field $W$. Writing $\bold{N}$ the exterior unit normal to $S_t=\Sigma_t \cap \mathcal{N}^{-}(p)$ in $\Sigma_t$:
 \begin{eqnarray*}
&&L=e_4=b(\bold{T}+\bold{N})\\
&&\underline{L}=e_3=b^{-1}(\bold{T}-\bold{N})
 \end{eqnarray*}

If we denote by $s$ the affine parameter relative to $L$, we have the following simple lemma (see \cite{krb}) :

\begin{lem}
There exists C($\Delta_1$) such that :
\begin{equation}
C^{-1}\ls \abs{\frac{dt}{ds}}\ls C
\end{equation}
\end{lem}

We have moreover :

$$ \bo{Q}(\bo{T},\bo{T},\bo{T},L)=  b^3 |\al|  ^2 + 3  b^2 | \bet |^2 + 3 b( \rho^2+ \sigma^2) +|\underline{\bet} |^2$$.

The two components of the "bad" tensor $\underline{\al}$ are not controled by our hypothesis. Thus, the control provided by the previous theorem is only a control of a reduced curvature flux :

$$\mathcal{R}(p)= \int_{\mathcal{N}^{-}(p)}| \al |^2 +|\bet|^2 + |\underline{\bet}|^2 + \abs{\rho}^2+ \abs{\sigma}^2$$.

\subsubsection{Energy estimates for higher derivatives }

\hspace{5 mm}As explained after the presentation of the local existence theorem that we will use, we will have to control higher derivatives of $\bold{R}$. These energy estimates make use of the wave structure of $\bold{R}$ :

\begin{equation}
\square_{\bo{g}} \bo{R}=\bo{R}\star\bo{R}
\end{equation}

where the bilinear operator $\star$ has the following definition :

\begin{equation*}
\bo{R}\star\bo{R}_{\al \bet \ga \de}=-\bo{R}_{\mu \s \ga \de}(\bo{R}^ {\hspace{2mm}\mu \s}_{\bet \hspace{4mm} \al} -\bo{R}^{\hspace{2mm}\mu \s}_{\al \hspace{4mm} \bet})+\bo{R}_{\al \s \ga \mu}(\bo{R}^ {\hspace{2mm}\mu \s}_{\delta \hspace{4mm} \bet} -\bo{R}^{\hspace{2mm}\mu \s}_{\bet \hspace{4mm} \delta})+\bo{R}_{\al \s \mu \de}(\bo{R}^{\hspace{2mm}\s \mu}_{\bet \hspace{4mm} \delta} -\bo{R}^ {\hspace{2mm}\mu \s}_{\ga\hspace{4mm} \delta})
\end{equation*}

It is natural according to the standard hyperbolic theory to associate to the covariant d'Alembertian $\square_{\bo{g}}$ the following Energy-momentum tensor :

$$\bo{Q}[U]_{\al \bet} = h^{IJ}\bo{D}_\al U_I\bo{D}_\bet U_J - \frac{1}{2}\bo{g}_{\al \bet} h^{IJ}\bo{g}^{\mu \nu}\bo{D}_\mu U_I\bo{D}_\nu U_J$$
where $h$ stands for the natural Riemaniann metric associated to the Lorentzian metric $\bo{g}$. As it is the case for standard wave equations in flat background, the computation of the covariant derivatives of this tensor leads to simplifications and we finally obtain the following expression :

\begin{eqnarray*}
\bo{D}^\bet \bo{Q}[U]_{\al \bet}&=&h^{IJ}\bo{D}_\mu U_I \square U_J + h^{IJ}(\bo{D}_\bet \bo{D}_\al U_I -\bo{D}_\al \bo{D}_\bet U_I )\bo{D}^\bet U_J\\
&&+ \bo{D}^\bet h^{IJ}(\bo{D}_\al U_I\bo{D}_\bet U_J-\frac{1}{2}\bo{g}_{\al \bet} \bo{g}^{\mu \nu}\bo{D}_\mu U_I\bo{D}_\nu U_J)
\end{eqnarray*}

therefore :

$$|\bo{D}^\bet (\bo{Q}[U]_{\al \bet} \bo{Q}^\al)|\lesssim|\bo{D}U||\square U| +|\bo{R}| |\bo{D}U||\square U| + |\bo{D}U|^2$$

where the constant in the $\lesssim$ depends on all the constants of the Main Theorem.

Using moreover the fact that :
$$\bo{Q}[U](\bo{T},\bo{T})=\frac{1}{2}|\bo{D}U|^2$$

we have, 

$$ \int_{\Sigma_{t}} |\bo{D}U|^2 \lesssim \int_{\Sigma_{0}} \abs{\bo{D}U}^2 + \int_0^t n  \int_{\Sigma_{s}} (\abs{\bo{D}U}|\square U| +|\bo{R}| |\bo{D}U||\square U| + |\bo{D}U|^2)dA_s ds$$

Thus, as explained in the presentation of the main structure of \cite{krb}, we need pointwise bounds on the curvature tensor $\bo{R}$ to control $L^2$ norms of derivatives of order one of this tensor. More precisely the previous formula leads directly to the following theorem :

\begin{thm}
With the hypothesis of our Main theorem, there exist a constant $C=C(t_\star,\Delta_0,\Delta_1,\Delta_2)$ such that for all $t$ in the time slab $[0,t_\star[$ :

$$\normf{\bo{DR}}_{L^2(\Sigma_{t})}^2\leqslant C \big(\normf{\bo{DR}}_{L^2(\Sigma_{0})}^2 +  \int_0^t\normf{\bo{R}}_{L^\infty(\Sigma_{s})}^2ds\big)$$
\end{thm}

Note that due to hypothesis $b)$ of the Main Theorem, we can omit to put the lapse $n$ into the time integral if we add the dependance on $\Delta_1$ in the constant C.

Exactly the same procedure but applied to $U=\bo{DR}$ leads to the corresponding estimates for the second derivatives of the curvature tensor :

\begin{thm}
With the hypothesis of our Main theorem, there exist a constant $C'=C'(t_\star,\Delta_0,\Delta_1,\Delta_2)$ such that for all $t$ in the time slab $[0,t_\star[$ :

$$\normf{\bo{D^2R}}_{L^2(\Sigma_{t})}^2\leqslant C' \big(\normf{\bo{D^2R}}_{L^2(\Sigma_{0})}^2 +  \int_0^t\normf{\bo{R}}_{L^\infty(\Sigma_{s})}^2\normf{\bo{DR}}_{L^2(\Sigma_{s})}^2ds\big)$$
\end{thm}

\subsection{Estimate for $\|\bo{R}\|_{H_{2,1}(\Sigma_t,O_t)}$ :}

\hspace{5mm}The generalized energy type estimates of the previous subsection would be sufficient for the case of the $CMC$ foliation. However as we choose to present the theorem here in the case of a maximal foliation, we have to deal with weighted Sobolev norms in the local existence result. We will see that in fact, it causes few difficultes. 

Let's begin with our base point $O$ on $\Sigma_0$, following the integral curves of $\bo{T}$, we get the points $O_t$. Let us denote by $\mathcal{J}^{-}(E)$ the causal past of a subset $E \subset \Sigma_{t_1}$ for a $t \in [0,t_\star[$. We note $E(t,G,H):= \{ p \in \Sigma_t, d_e(p,O_t) \in [G,H] \}$ the annulus relative to the euclidian norm and $E'(t,G,H):= \{ p \in \Sigma_t, d_t(p,O_t) \in [G,H] \}$ relative to the geodesic distance on $\Sigma_t$. As a consequence of condition $b)$ (the pointwise bound on $n$) and theorem (1.3.1), we have that there exists a constant $C:=C(t_\star, \Delta_1,\Delta_2)$ such that :

\begin{lem}
For any $t \in [0,t_1]$, any $0\ls G \ls H < \infty$,
\begin{eqnarray*}
E(t, G,H)  &\subset& \mathcal{J}^{-}(E(t_1,G,H))\cap \Sigma_t\\
\mathcal{J}^{-}(E(t_1,G,H)) \cap \Sigma_t&\subset& E(t, G - C(t_1-t),H + (t_1-t)) 
\end{eqnarray*}
\end{lem}

{\bf Proof :}
Following the integral curves of n$\bo{T}$ and using the definition of transported coordinates give the first inclusion. 
For the second one, let us denote by $D$ the constant of theorem (1.3.1) and we recall that $\Delta_1$ is the constant controlling $n$ and $n^{-1}$.
Let $c(\tau)=(\tau,x(\tau))$ be an arbitrary causal curve ending in $\mathcal{I}^{-}(E(t_1,G,H))\cap \Sigma_t$ then by definition :
\begin{equation}
0 \rs \bo{g}(\dot{l}(\tau),(\dot{l}(\tau))
\end{equation}

and using the bounds :
\begin{equation}
\Delta_1^2 \rs n^2(\tau,x(\tau)) \rs D^{-1}\abs{\dot{x}}^2
\end{equation}

which implies the preceding lemma with a constant $C=D \Delta_1^2$. 

Moreover outside a compact set of $\mathcal{M}$, $E'(t,(1+\e)G,(1-\e)H) \subset E(t,G,H)$
Let us consider for a positive integer $m$ the following set $E_{(t,m)}=E(t,2^m,2^{m+1})$ and $F_{m}=F(0,2^m-C t_\star ,2^{m+1} + C t_\star )$.We will now apply the proposition (1.3.2) to $\mathcal{J}^{-}(E(t,m))$ which together with the positivity of the flux give that there exists $D:=D(t_\star, \Delta_1,\Delta_2)$ such that :

\begin{equation}
\int_{E_{(t,m)}} \abs{\bo{R}}^2 d\mu_t \ls D \int_{F_{m}} \abs{\bo{R}}^2d\mu_0
\end{equation} 

than we remark that there exits $H(t_\star, \Delta_1,\Delta_2)$ such that on $\Sigma_0$ :
\begin{equation}
\sum_{m \in \mathbb{N}} \bo{1}_{F_{m}} \ls H
\end{equation}

Then, we have :
\begin{eqnarray}
\|\bo{R}\|_{L_{2,1}(\Sigma_t,O_t)} &\ls& \sum_{m \in \mathbb{N}} (1+2^{2m+2})\int_{E_{(t,m)}} \abs{\bo{R}}^2 d\mu_t \\
& \ls& D \sum_{m \in \mathbb{N}} (1+2^{2m+2})\int_{F_{m}} \abs{\bo{R}}^2 d\mu_0\\
& \ls& D \sum_{m \in \mathbb{N}} \frac{1+2^{2m+2}}{1+(2^m - C t_\star)^2)} \int_{F_{m}} (1+d_0^2)\abs{\bo{R}}^2 d\mu_0\\
& \ls&D \cdot H \cdot \int_{\Sigma_0} (1+d_0^2)\abs{\bo{R}}^2 d\mu_0\\
&\lesssim& \|\bo{R}\|_{L_{2,1}(\Sigma_0,O)} 
\end{eqnarray}

We apply the same method to prove that :

\begin{thm}
The theorems (1.3.3), (1.3.4) and (1.3.5) hold with $\ns{\bo{R}}{2}$, $\ns{\bo{DR}}{2}$ and $\ns{\bo{D^2R}}{2}$ replaced respectivly by $\|\bo{R}\|_{L_{2,1}(\Sigma_t,O_t)}$, $\|\bo{DR}\|_{L_{2,2}(\Sigma_t,O_t)}$ and $\|\bo{D^2R}\|_{L_{2,3}(\Sigma_t,O_t)}$
\end{thm}

where we have used the following weighted norms :
\begin{equation}
\|f\||_{L_{2,i}(\Sigma_t,O_t)}=\int_{\Sigma_t} (1+d_t^2)^{1+i}\abs{f}^2 d\mu_t
\end{equation}
We will also need estimates on $k$ to apply the local in time result. The standard $L^2$ estimates for $k$
 are easy to get from the $3D$ Hodge systems satisfied by $k$ on $\Sigma_t$ (see \cite{ck} and subsection (1.9)). However, proving the weighted estimates with the method we used for $\bo{R}$ would require the control of the flux of $k$ and its derivatives along null hypersurfaces. We could deal with such a quantity if we were able to derive  a wave equation on $k$, more precisely an equation of the form :
 \begin{equation}
 \square {}^{(\bo{T})}\pi = [\square, \mathcal{L}_{\bo{T}}] \bo{g}=F( {}^{(\bo{T})}\pi)
  \end{equation}

However when doing the calculation, we do not obtain a good structure similar to the one obtained for $\bo{R}$, that is a $F$ depending only on $ {}^{(\bo{T})}\p$. The way we find to avoid this difficulty is by using a modified $3D$ Hodge systems on the $E(m,t)$ to prove that there exists $J(t_\star, \Delta_1,\Delta_2)$ such that :

\begin{equation}
\|\bo{k}\|_{H_{3,1}(\Sigma_t,O_t)} \ls J \|\bo{R}\|_{H_{2,1}(\Sigma_t,O_t)}
\end{equation}

This will be done in subsection 1.9.

\vspace{1 cm}

So at this step of the proof, the problem consists mainly in deriving pointwise estimates for the curvature involving $L^2$ estimates of $\bo{R}$ and its derivatives which is in a sense the heart of the proof of \cite{krb}. This is the Kirchoff-Sobolev type parametrix proved in \cite{krk}.

\subsection{Geometry of null cones :}

%With the preceeding formula we have a usefull too obtain pointwise control on the curvature $\bo{R}$. Essentially the main ingredient missing is the control of the geometry of the null cones which is a crucial thing in comparison with the preceeding breakdown criterion of Anderson or even with the proof of of global nonlinear stability of Minkovsky space (see \cite{ck} and \cite{kn}. The control of the Ricci coefficients of the radius of conjugacy and the estimates on the Ricci coefficients of the $(t,u)$-foliation will be the subject of the second section of this paper. It is the part of theorem 3 in which the most differences appear in comparison with the one of theorem 2. 

\subsubsection{Geometrical definitions :}

\hspace{5mm}Let us begin by recalling the definitions associated with the geometry of the null cones. We denote by $\mathcal{N}^{-}(p)$ the null boundary of the causal past of the point $p$, $\mathcal{J}^{-}(p)$. In general it's an achronal, Lipschitz hypersurface ruled by the set of past null geodesics initiating in $p$. We use the following parametrization at the vertex $p$ for the geodesics. For every direction $\omega \in \bo{S^2}$, we consider the null vector $l_\omega$ in $T_pM$ such that :

$$\bo{g}(l_\omega, \bo{T}_p)=1,$$

We consider associated to it the past null geodesic $\gamma_\omega(s)$ with initial data $p$ and initial speed $\dot{\gamma}(0)=l_\omega$. We the define a null vectorfield $L$ on $\mathcal{N}^{-}(p)$ by setting :

$$L(\gamma(p))=\dot{\gamma}(p)$$

When two geodesics intersect, we have a multi-valued vectorfield. We associate to $L$ its affine parameter $s$ id est $L(s)=1$ and $s=0$ at $p$.

For a sufficiently small $\delta>0$ the exponential map $\mathcal{G}:(s,\omega) \rightarrow \gamma_\omega(s)$ is a diffeomophism from $(0,\delta) \times \bo{S^2}$ to its image in $\mathcal{N}^{-}(p)$. For each $\omega \in \bo{S^2}$ either $\gamma_\omega(s)$ can be continued for all positive values of $s$ or there exists a terminal point after which the $\gamma_\omega(s)$ are not longer in  $\mathcal{N}^{-}(p)$ but in its interior $\mathcal{I}^{-}(p)$. If the exponential is map is singular (resp. non-singular) the terminal point is said to be a conjugate (resp. cut-locus) terminal point. $\mathcal{N}^{-}(p)$ is thus regular except in its vertex and terminal points. Its regular part will be denoted by $\dot{\mathcal{N}}^{-}(p)$. The set $\mathcal{G}^{-1}(\mathcal{N}^{-}(p) \backslash  \dot{\mathcal{N}}^{-}(p))$ has $dA_{\mathcal{N}}^{-}(p):=dsdA^2$ mesure zero.

\begin{defi}
Given a point $p \in \mathcal{M}$ we note $i_{\star}^{-}(p)$ (resp $c_{\star}^{-}(p)$)  the supremum over all values $s>0$ for which the exponential map $\mathcal{G}$ is a global (resp. local) diffeormorphism, it is the past null radius of injectivity (resp. conjugacy) at p relative to the geodesic foliation.
Consistantly we note $i_{\star}^{-}(p,t)$ the supremum over all values $t(p)-t>0$ for which the exponential map $\mathcal{G}$ is a global diffeormorphism, it is the past null radius of injectivity at $p$ relative to the $t$-foliation.
\end{defi}

If we denote by $d^{-}(p,t)$ the $t$ distance to the past boundary of $\mathcal{M}$, we have the following :

\begin{defi}
Under the assumption of theorem (1.1.4) there exists a positive number $i_{\star}$ depending only on the constants of $\Delta_0$, $\Delta_1$ and $\Delta_2$ and $t_\star$ such that for all $p \in \mathcal{M}$ we have : 
\begin{equation}
i_{\star}^{-}(p,t)>min(i_{\star},d^{-}(p,t))
\end{equation}
\end{defi}

{\it Remarks :} Contrary to the situation in \cite{krb}, the preceding theorem is not just an application of \cite{kri}, because this article makes the assumptions of theorem (1.1.2) which are stronger than ours. So this whole work has to be checked to see that the integral condition on $k$ and $n^{-1}\nabla n$ is in fact sufficient. We will not reproduce here the proof, it would be pure copy. The main idea is that the heart of the proof relies on a argument of Cheeger-Gromov type which depend only on the uniform control of the $L^2$ norm of $R$. This theorem is the following :

\begin{thm}
Given $\Lambda>0$, $v>0$ and $\e >0$ there exists an $r_0>0$ such that on any 3D complete Riemannian manifold $(M,g)$ with $\|\bo{R}\|_{L^2}\ls \Lambda$ and volume radius, at scales $\ls 1$ bounded from below by $v$, verifies the following :

Every geodesic ball $B_r(p)$ with $p \in M$ and $r\ls r_0$ admits a system of harmonic coordinates $x=(x^1, x^2, x^3)$ relative to which we have,

\begin{eqnarray}
(1+\e)^{-1}\de_{ij}\ls g_{ij} &\ls& (1+\e)\de_{ij}\\
r \int_{B_r(p)}\abs{\partial^2} g_{ij}dv_g &\ls& \e
\end{eqnarray}
\end{thm}

Thus once we have control on  $\|R\|_{L^2}$ (see the subsection on energy estimates and the subsection on the control of $k$) and the control on the metric (see the subsection on the geometric control), we can apply this theorem.

In fact we have to mention that the controls on the raddii of conjugacy and injectivity come {\it together}. So in fact the main argument on which we have to work is the control of the radius of conjugacy. But this one is a consequence from the control of the causal structure which is the part of the breakdown criterion where the largest number of differences appear in the proof of theorem (1.1.4) in comparison with the proof of theorem (1.1.2) : it will be the subject of the section 2 of this paper.

\vspace{.5 mm}

We have also to remark that the fact that we do not work with CMC foliation but with maximal foliation does not change the statement of the previous result. We define $l_\star(p)$ the smallest value of $s$ for which there exists two distinct null geodesics initiating at $p$ and intersecting for the affine parameter (the smallest value for the two geodesics) $s_\star=l_\star(p)$. The theorem of \cite{kri} requires that there exists a future compact set $\mathcal{C} \subset \mathcal{M}$ and $\de_\star > 0$ such that $ \forall$ $p \in \mathcal{C}^c$, $l_\star(p)>\de_\star$. 

This condition is trivially satisfied in the case of CMC foliation in view af the compacity of the leaves. In our case the assymptotic flatness is sufficient to assure that there exists  $\mathcal{C} \subset \mathcal{M}$ outside which we have in particular pointwise bound on $R$ which is sufficient to assure the control of the geodesic loops. In fact in comparison with the same situation in Riemannian geometry, the surprising fact is that an integral condition on the curvature is sufficient in the compact case (see the preliminary discussion in \cite{kri}).

\subsubsection{Extension of the derivation :}

\hspace{5mm}We can associate on any point of $\mathcal{N}^{-}(p) \textbackslash \{p\}$ a conjugate null vector $\Lb$ such that $\bo{g}(L,\Lb)=-2$ and $\Lb$ is orthogonal to the leafs $S_t$. Endowed with $\gamma$, the restriction of $\bo{g}$, $S_t$ is a two dimentional compact surface. We choose an orthonormal base $(e_1,e_2)$ to complete the null frame $(e_1,e_2,\Lb,L)$. We will denote by $e_3=\bo{T}+\bo{N}$ and  $e_4=\bo{T}-\bo{N}$. The null frame $(e_1,e_2,e_3,e_4)$ is called the canonical null frame associated to the $t$ foliation. We define as usual the Ricci coefficients associated to these frames (see section 2 for the details) and extend the angular derivation to arbitrary covariant $S_t$ tangent tensor $F$ by setting : 
\begin{equation}
\nablab_X F(Y_1,......,Y_k)= X(F((Y_1,......,Y_k)) - F(\nablab_X Y_1,.....,Y_k) -F(Y_1,.....,\nablab_XY_k) 
\end{equation}
where $X,Y_1,......,Y_k$ $S$ tangent. Given an $S$-tangent vector-field $X$ we note $\nablab_LX$ by :

\begin{equation}
\nablab_LX= \bo{D}_L + \demi \bo{g}(\bo{D}_LX, \Lb)L
\end{equation}

and extend it to any covariant $S$-tangent tensor $F$ by setting :

\begin{equation}
\nablab_L F(Y_1,......,Y_k)= L(F((Y_1,......,Y_k)) - F(\nablab_L Y_1,.....,Y_k) -F(Y_1,.....,\nablab_LY_k) 
\end{equation}

The covariant derivatives can be extended to space-time not necessarily $S$-tangent along a fixed surface $S_t$  (see \cite{ck}).

\subsubsection{Construction of the optical function :}
\hspace{5mm}An optical function is necessary to make sense both of the Kirchoff-Sobolev parametrix and the trace theorems of subsection (1.10). Let p an an arbitrary point of $\mathcal{M}$. We want to construct a function $u$ in a neighborhood of $\mathcal{N}^{-}(p, \delta)$ (with $0<\delta < i_\star(p,t)$, vanishing on $\mathcal{N}^{-}(p, \delta)$. We consider the timelike geodesic initiating at $p$ with initial speed $\bo{T}_p$. We can define such a geodesic $\Gamma_{\e_1,\e_2} : (1-\e_1,1+\e_2) \rightarrow \mathcal{M}$ with $\Gamma_{\e_1,\e_2}(1)=p$ and $\Gamma'_{\e_1,\e_2}(1)=\bo{T}_p$. We remark that there exists a constant $r>0$ such that, for every $p$ we can construct such a geodesic with $r<max(\e_1,\e_2)$. 
Now $u$ is defined to be constant on each  $\mathcal{N}^{-}(q, \delta)$ for $q=\Gamma_{\e_1,\e_2}(t)$ and $t \in (1-\e_1,1+\e_2)$ and :

\begin{equation}
u_{\mathcal{N}^{-}(q, \delta)}=t-1
\end{equation}

This defines a smooth function on a neighborhood of $\mathcal{N}^{-}(p, \delta)$ verifying :
\begin{equation}
\bo{g}^{\al \bet}\partial_\al u \partial_\bet u =0
\end{equation} 

{\it Remark :} We remark that by the same argument as above and the control on $max(\e_1,\e_2)$ the constructed neighborhood of $\mathcal{N}^{-}(p, \delta)$ contains a $\kappa$-geodesic neighborhood of $p$ in $\Sigma_t$ with $\kappa$ independently bounded from below.
\subsection{Causal structure of null cones :}

\hspace{5mm}This part is the one which has to be redone in comparison with \cite{krb}. In fact even in this article, the authors use the result of \cite{krc} in which the setting is geodesic foliation. The control on the lapse together with hypothesis $b)$ is sufficient (see lemma (1.3.3) ) to prove that on every null cone the affine geodesic parameter is uniformly controlled by the time $t$. However even if we have this control, the control of the different norms involved on the $S_t$ is not implied by the one on $S_s$, the work has to be redone. In particular, the role of the control on the derivative of $n$ is not a priori clear. Note that the following estimate hold both for the Ricci coefficients relative to the canonical null frame and the associated affine parameter $s$ and the  Ricci coefficients relative relative to the renormalized null frame and its affine parameter $s'$. 

\begin{thm}
For any $t\in(t(p)-\delta,t(p))$ with $\delta < i_\star(p,t)$ the Ricci coefficients relative to the renormalized null frame :
\begin{eqnarray}
\sup_t \abs{tr \chi - \frac{2}{s(t)}} & \lesssim & D(t_\star, \Delta_i) \\
\| \sup_t (t(p)-t) \abs{\nablab tr\chi} \|_{L^2_\omega}& \lesssim & D(t_\star, \Delta_i) \\
\sup_{\omega \in \bo{S^2}} \int_{t(p)-\delta}^{t(p)} \big(  \abs{\hat{\chi}}^2 + \abs{\zeta}^2 + \abs{\hb}^2\big)(t,\omega) dt & \lesssim & D(t_\star, \Delta_i) \\
\| \mu \|_{L^2(\mathcal{N}^{-}(p,\delta)} & \lesssim & D(t_\star, \Delta_i)
\end{eqnarray}
The points on the null cone are here parametrized by the coordinate $(t,\omega)$ given by the exponential map.
\end{thm}

{\bf Proof :} The proof is the subject of the whole section 2.

\subsection{The Kirchoff-Sobolev parametrix :}
\hspace{5mm}Let us briefly describe without proof the Kirchoff-Sobolev parametrix of \cite{krk}. We begin with the standard formula in flat background. 

We consider the solutions to the covariant tensorial wave equation in the $3+1$ flat Minkovsky space-time :

\begin{equation}
\square_m \Psi = F
\end{equation}

then it's well known that the solution of this equation with a given initial data set can be written as the sum of two terms :

\begin{equation}
\Psi = \Psi_h+ \Psi_F
\end{equation}

where $\Psi_h$ has flat d'Alembertan and the same initial data set and $ \Psi_F$ can be written :

\begin{equation}
\Psi_F(t,x)=\frac{1}{4\pi}\int_{\mathbb{R}^{3+1}}H(t-s)\delta\big(-(t-s)^2+\abs{x-y}^2 \big)f(s,y)dsdy
\end{equation}

where $H$ denotes the standard Heaviside function and $\delta$ the standard Dsirac. Some remarks can be immediately made.  The first one is of course the speed of propagation and the other one of great importance for us is that this integral involves only $F$ on the past null cone initiating in the point $x$. However the situation is more difficult in curved background. One can hope to have the same kind of structure that is :

\begin{equation}
 \Psi_F(p)=\frac{1}{4\pi}\int_{\mathcal{I}^{-}(p)}r(p,q)\delta\big(d^2(p,q) \big)f(q)dv(q)
 \end{equation}

where $\mathcal{I}^{-}(p)$, $d$, $r$ stand respectively for the causal past of the point $p$, the Lorentzian distance defined by the metric $g$ and a correction factor satisfying a transport equation along the past null cone ${\mathcal{N}^{-}(p)}$. Of course a minimum requirement for such a formula is to work on a part of the causal past of $p$ in which the Lorentzian distance is smooth enough.

Let us consider the tensorial equation :
\begin{equation}
\square \Psi = F
\end{equation}
 with $\Psi$ a k covariant tensor field. Let $p\in\mathcal{M}$. Let us consider $\mathcal{N}_{-}(p,\delta)$ for $\delta<i_\star(p,t)$. Consider now the solution of the following transport equation with blowing up initial data :

\begin{equation}
\bo{D}_L\bo{A}+\frac{1}{2}\bo{A}tr \chi =0, \hspace{3 mm} \bo{A}(p)=\bo{J_0} \hspace{3 mm} \text{on }\mathcal{N}_{-}(p,\delta)
\end{equation}
where $\bo{J_0}$ is a fixed k-tensor in $p$ such that $\abs{\bo{J_0}} \leq 1$, then we have the following representation formula.

\begin{thm}{\bf Representation Theorem}
The solution $\Psi$ of the equation (59) has the following form

\begin{equation}
\bo{\Psi}(p)\cdot\bo{J}_0=-\int_{\mathcal{N}^{-}(p)}\bo{A}\cdot \square \Psi+\int_{\mathcal{N}^{-}(p)}\bo{g}(\mathcal{E},\Psi)
\end{equation}
where the error term $\mathcal{E}$ has the following representation formula :

\begin{equation}
\mathcal{E}_a=\frac{1}{2}\bo{R}_{a \lambda \gamma \delta} \bo{A}^\lambda \bo{\underline L}^\gamma \bo{L}^\delta + \big( \Delta \bo{A} + \xi_a\bo{D}_a\bo{A} \big) _\alpha + \frac{\mu}{2}\bo{A}_\alpha\end{equation}
\end{thm} 

\subsection{Closure of the estimates :}

\subsubsection{Consequence of the parametrix :}

\begin{prop}
The following estimate holds for all $p \in \mathcal{M}$ and $0<\delta<i_\star^{-}(p,t)$ :
\begin{eqnarray*}
\abs{\bo{R}(p)}& \lesssim& \mathcal{E} + \nnc{\bo{R}}{\infty} \nnc{\bo{A}}{2}\big(\mathcal{R}(p, \delta)+ \nnc{\mu}{2}\big)  \\
&&+ \nnc{\nablab \bo{A}}{2}\big(  \nnc{\nablab \bo{R}}{2} +  \nnc{\bo{R}}{\infty} \nnc{\zeta}{2}\big)
\end{eqnarray*}
\end{prop}

{\bf Proof :}

Once we have proved the estimates for the Ricci coefficients relative to the $t$ foliation, the proof is pure copy of the one in \cite{krb}. In particular this part being local, the difference between the CMC and the maximal foliations setting plays no role. We reproduce it here. 

Let's introduce a smooth cut-off function $f$ taking its values in $[0,1]$ supported in $[t(p),t(p)-\delta]$ and identically equal to $1$ in $[t(p),t(p)-\delta/2]$.  We have :

\begin{eqnarray}
f\bo{R}(p)&=&\bo{R}(p)\\
\square(f \bo{R})&=& f \bo{R} \star \bo{R} + (\square f) \bo{R} + 2 \bo{D}^\al f \bo{D}_\al \bo{R}
\end{eqnarray}

then applying the representation formula :

\begin{eqnarray}
4 \pi \bo{R}(p) \cdot \bo{J_0}&=& I(p) + J(p)+K(p)+L(p)+E(p)\\
I(p)&=&\int_{\mathcal{N}^{-}(p,\delta)} \bo{A}\cdot f (\bo{R} \star \bo{R}) \\
J(p)&=&- \demi \int_{\mathcal{N}^{-}(p,\delta)} \bo{A} \cdot \bo{R}(.,., \bar{L},L) \cdot f \bo{R} \\
K(p)&=& \int_{\mathcal{N}^{-}(p,\delta)} (\Deltab \bo{A} + \zeta^a \nablab_a \bo{A}) \cdot f \bo{R} \\
L(p)&=& \demi \int_{\mathcal{N}^{-}(p,\delta)} \mu \bo{A}\cdot f \bo{R} \\
E(p)&=& \int_{\mathcal{N}^{-}(p,\delta)} \big( \square f (\bo{A} \cdot \bo{R}) + 2 \bo{D}^\al f ((\bo{A} \cdot \bo{D_\al R}) \big)
\end{eqnarray}

{\bf Estimate of I(p) :}

\hspace{5mm}The important remark to estimate $I(p)$ is that $W=\bo{R} \star \bo{R}$ is also a Weyl tensorfield. So the natural question is the one of the null components of $W$ as a function of the null components of $\bo{R}$. Due to the formula giving $W$, the null components are quadratic expressions of the one for $\bo{R}$. However recall that we have no a priori control of the $L^2$ norm of $\underline{\al}(\bo{R})$ on the null cone, so it's important to understand the presence of this term in the null components of $W$. 
Following \cite{ck}, we can assign the natural signature to the null components, id est the power of $a$ in the transformation $\al \rightarrow \al'$ etc .... when performing the base transformation $ L \rightarrow a L$ and $ \underline{L} \rightarrow a^{-1}\underline{L}$. Due to this scaling property and the definition of $W$, we can deduce that the only terms in which $\underline{\al}$ appears are quadratic terms of the form $\mathcal{Q}(\underline{\al}, R_0)$ where $R_0$ has signature $0$, $1$ or $2$. In particular we will not have quadratic  terms of the form $\mathcal{Q}(\underline{\al},\underline{\al})$.

So if we note :

\begin{equation}
\abs{\bo{R}}'^2= \abs{\al}^2 +  \abs{\bet}^2 +  \abs{\underline{\al}}^2 + \abs{\rho}^2 +  \abs{\sigma}^2
\end{equation}

we have :

\begin{equation}
\abs{\bo{R}\star\bo{R} } \lesssim \abs{\bo{R}} \cdot \abs{\bo{R}}'
\end{equation}

and also :

\begin{eqnarray}
\abs{I(p)}&\lesssim&\int_{\mathcal{N}^{-}(p,\delta)}   \abs{\bo{A}} \cdot \abs{\bo{R}} \cdot \abs{\bo{R}}' \\
&\lesssim&\big(\int_{\mathcal{N}^{-}(p,\delta)} \abs{\bo{R}}'^2 \big)^\demi \big(\int_{\mathcal{N}^{-}(p,\delta)}  \abs{\bo{A}}^2 \cdot \abs{\bo{R}}^2 \big)^\demi\\
&\lesssim&\mathcal{R}(p, \delta) \big( \int_{t(p)-\delta}^{t(p)} \ns{\bo{R}}{\infty}^2 \ndleb{\bo{A}}{2}^2 dt \big)\\
&\lesssim&\mathcal{R}(p, \delta) \big( \int_{t(p)-\delta}^{t(p)} \ndleb{\bo{A}}{2}^2 dt \big)^\demi \sup_{t' \in [t(p)-\delta, t(p)]} \ndleb{\bo{R}}{\infty}
\end{eqnarray}

{\bf Estimate for J(p) :}

The same method leads to :
\begin{equation}
\abs{J(p)}\lesssim \mathcal{R}(p, \delta) \big( \int_{t(p)-\delta}^{t(p)} \ndleb{\bo{A}}{2}^2 dt \big)^\demi \sup_{t' \in [t(p)-\delta, t(p)]} \ndleb{\bo{R}}{\infty}
\end{equation}

{\bf Estimate for L(p) :}

\begin{equation}
\abs{L(p)}\lesssim    \big( \int_{t(p)-\delta}^{t(p)} \ndleb{\mu}{2}^2 dt \big)^\demi\big( \int_{t(p)-\delta}^{t(p)} \ndleb{\bo{A}}{2}^2 dt \big)^\demi \sup_{t' \in [t(p)-\delta, t(p)]} \ndleb{\bo{R}}{\infty}
\end{equation}

{\bf Estimate for K(p) :}

\begin{eqnarray*}
\abs{K(p)}&\lesssim& \int_{\mathcal{N}^{-}(p,\delta)} \big( \abs{\nablab \bo{A}}  \abs{ \bo{R}}+ \abs{\zeta}\abs{\nablab \bo{A}}  \abs{\bo{R}} \big)\\
&\lesssim& \nnc{\nablab \bo{A}}{2} \big( \nnc{\nablab \bo{R}}{2}  +  \nnc{\bo{R}}{\infty} \nnc{\zeta}{2} \big)
\end{eqnarray*}

{\bf Estimate for E(p) :}

We have :

\begin{equation}
\abs{\bo{D}t} \lesssim 1
\end{equation}

but to contrary to \cite{krb}, we have only :

\begin{equation}
\int\abs{\square t}^2n dt \lesssim 1
\end{equation}

therefore , we have the following estimates :

\begin{equation}
\|f'\|_{L^\infty}\lesssim \delta^{-1},  \hspace{5mm}  \int_{t(p)-\delta}^{t(p)-\delta/2} \|\square f\|_{L^\infty(\Sigma_t)}^2 dt \lesssim \delta^{-4}
\end{equation}

\begin{eqnarray*}
\abs{E(p)}&\lesssim&\sup_{t' \in [t(p)-\delta, t(p)-\delta/2]} \big( \delta^{-1} \ndleb{\bo{R}}{2} + \ndleb{ \bo{DR}}{2} \big) \\
&\lesssim&\delta^{-1}  \sup_{t' \in [t(p)-\delta, t(p)-\delta/2]} \big( \ns{\bo{R}}{2} + \ns{ \bo{DR}}{2} + \ns{ \bo{D^2R}}{2} \big) 
\end{eqnarray*}

Moreover we have the following estimates on $\bo{A}$:

There exists $C(t_\star, \Delta_i)$ such that :
\begin{eqnarray}
\nnc{(t(p)-t)\bo{A}}{\infty} &\ls& C\\
\nnc{\nablab \bo{A}}{2} &\ls& C\\
\nnc{\bo{A}}{2} &\ls& \delta^\demi C\\
\end{eqnarray}

(Little changes based mainly on the use of Gronwall type arguments have to made to obtain the same estimates obtained in \cite{krb} for $\bo{A}$. Thus we refer directly to this article).

Then putting everything together and choosing $\delta$ small enough but depending only on $t_\star$ and the $\Delta_i$, we get :

\begin{equation}
\|\bo{R}(t)\|_{L^\infty} \lesssim C \delta^{-1}\sup_{t' \in [t-\delta, t-\delta/2]}\|\bo{R}(t)\|_{H^2}
\end{equation}

\subsubsection{Control on $\bo{R}$ :}

%\begin{equation}
%\|\bo{R}(t)\|_{L^\infty} \lesssim \mathcal{E}+ \delta^\demi \sup_{p \in \Sigma_t} \nnc{\bo{R}}{\infty} +  \sup_{p \in \Sigma_t}\nnc{\nablab \bo{R}}{2}
%\end{equation}
Now returning to the energy estimates, we get :

\begin{equation}
\|\bo{DR}(t)\|_{L_{2,2}}^2 \lesssim \|\bo{DR}(t-\delta/2)\|_{L_{2,2}}^2 +  \delta^{-2}\sup_{t' \in [t-\delta, t-\delta/2]}\|\bo{R}(t)\|_{H^2}^2
\end{equation}

\begin{equation}
\|\bo{D^2R}(t)\|^2_{L_{2,3}} \lesssim  \|\bo{D^2R}(t-\delta/2)\|^2_{L_{2,3}}  +\delta^{-2}\sup_{t' \in [t-\delta, t-\delta/2]}\|\bo{R}(t)\|_{H^2}^2
\end{equation}

Finally :
\begin{equation}
\|\bo{R}(t)\|_{H_{2,1}} \lesssim \delta^{-1}\sup_{t' \in [t-\delta, t-\delta/2]}\|\bo{R}(t)\|_{H_{2,1}}
\end{equation}

then iterating by step of size $\delta/2$, we get :

\begin{thm}
With our hypothesis, there exits $C(t_\star,\Delta_i)$ such that for every $t \in [0,t_\star[$ :
\begin{equation}
\|\bo{R}(t)\|_{H_{2,1}(\Sigma_t,O_t)} \ls C\|\bo{R}(0)\|_{H_{2,1}(\Sigma_0,O)}
\end{equation}
\end{thm}

Now we have to derive estimates and weighted estimates of $k$ and that will imply estimates on $R$.

\subsection{3D Hodge systems and  estimates :}

\hspace{5mm}As explained above the difference between our approach of the analysis of the causal geometry of our space time and the one in \cite{krc} lies in the use of a priori estimates of $k$ and $n$ on the 2 dimensional surfaces $S_{t,u}$. The idea underlying this approach is to use the $L^\infty([0,t_\star[,L^2(\Sigma_{t}))$ control on $\bo{R}$ to get a control on $\nabla k$ and $\nabla^2n$. 

{\bf First estimates for $k$ and $n$ :}

Let us begin by recalling a result concerning rank 2 symmetric Hodge systems on a 3 dimensional Riemannian manifold $\Sigma$ :

\begin{lem}
Let $V$ a 2 symmetric traceless 2-tensor on a 3 dimensional Riemannian manifold $\Sigma$  satisfying :
\begin{eqnarray*}
\nabla \cdot V = D, \hspace{5mm} \nabla \wedge V = C,
\end{eqnarray*}
Then,
$$\int_{\Sigma} \big( \abs{\nabla V}^2 + 3R_{mn}V^{im}{V_i}^n-\frac{1}{2}R\abs{V}^2 \big)= \int_{\Sigma}  \big(\abs{C}^2 + \frac{1}{2}\abs{D}^2\big)$$

For a scalar, $\phi$,

$$\int_{\Sigma}  \abs{\nabla^2 \phi}^2 + \int_{\Sigma} R^{ij}\nabla_{i}\phi\nabla_{j}\phi=\int_{\Sigma} \abs{\Delta \phi}^2$$

\end{lem}

We will now apply the first part of the lemma to $k$ and the second to $n$.

\vspace{5mm}
$k$ is a rank 2 symmetric tensor verifying :

$$\nabla \cdot k =0, \hspace{5mm} \nabla \wedge k = H$$

where $H$ stands for the magnetic-part of the electric-magnetic decomposition of the tensor of the curvature of the leaf $\Sigma$ :

$$H(X,Y)=\langle{}^{\star}\bo{R}(X,\bo{T})\bo{T},Y\rangle$$
$$E(X,Y)=\langle \bo{R}(X,\bo{T})\bo{T},Y\rangle$$

Thus, the previous theorem applied to $k$ on a leaf $\Sigma_t$ for a fixed $t \in [0,t_\star[$ and taking into account the constraint equation gives us :

$$\int_{\Sigma_t} \big( \abs{\nabla k}^2 + 3(k^2)^{mn}(E_{mn}(k^2)_{mn}-\demi\abs{k}^4 \big)= \int_{\Sigma_t} \abs{H}^2$$

This leads to the following :
\begin{equation}
\int_{\Sigma_t} \big( \abs{\nabla k}^2+ \frac{1}{4}\abs{k}^4 \big) \leq \int_{\Sigma_t} \abs{\bo{R}}^2
\end{equation}

Then applying the second part of the lemma above to the lapse and using the elliptic equation :
\begin{equation}
\begin{aligned}
\Delta n = |k|^2 n
\end{aligned}
\end{equation}

gives :
\begin{equation}
\int_{\Sigma_t}  \abs{\nabla^2 n}^2 + \int_{\Sigma_t} R^{ij}\nabla_{i}n\nabla_{j}n=\int_{\Sigma_t} |k|^4 n^2
\end{equation}
then using the pointwise estimates for the lapse $n$ and (26) gives :
\begin{equation}
\int_{\Sigma_t}  \abs{\nabla^2 n}^2 + \int_{\Sigma_t} R^{ij}\nabla_{i}n\nabla_{j}n \lesssim \int_{\Sigma_t} \abs{\bo{R}}^2
\end{equation}
now for all $\e$, there exists a $C_\e$ such that :
\begin{eqnarray*}
\int_{\Sigma_t} R^{ij}\nabla_{i}n\nabla_{j}n &\leq& C_\e \int_{\Sigma_t} \abs{R}^2 + \e \int_{\Sigma_t}\abs{\nabla n}^4\\
&\leq&C_\e \int_{\Sigma_t} \abs{R}^2 + D\e \big(\int_{\Sigma_t}\abs{\nabla^2 n}^2 \big)\sup_{\Sigma_t}n^2
\end{eqnarray*}

where $D$ can be taken independant of the time $t \in [0,t_\star[$.

Moreover as we have the following constraint equation :

$$R_{ij}-k_{ia}k^{a}_j+trk k_{ij}= E_{ij}$$

we have using (1.17) uniformly in $t$ :

$$\int_{\Sigma_t} \abs{R}^2 \lesssim  \int_{\Sigma_t} \abs{\bo{R}}^2$$

thus using the pointwise bound on $n$ and taking $\e$ small enough we get :
\begin{equation}
\int_{\Sigma_t}  \abs{\nabla^2 n}^2 \lesssim \int_{\Sigma_t} \abs{\bo{R}}^2
\end{equation}

{\bf Higher order derivatives of $k$ and $n$ :}

From the constraint equations in the case of maximal foliations we have :

\begin{equation}
R_{ij}-k_{ia}k^a_j=E_{ij}
\end{equation}

moreover differentiating the curl equation of $k$, we get :

\begin{equation}
\Delta k_{jm}-\nabla^i \nabla_j k_{im}= \in^s_{ij}\nabla^iH_{sm}
\end{equation}

thus using that $k$ is divergence free and commuting the derivatives, we get that $\Delta k_{jm}- \in^s_{ij}\nabla^iH_{sm}
$ is a quadratic expression in $R$ and $k$.

Integration over $\Sigma_t$,

\begin{equation}
\int_{\Sigma_t} \abs{\Delta k}^2 \ls \int_{\Sigma_t} \big( \abs{k}^6 + \abs{E}^2\abs{k}^2 + \abs{\nabla H}^2 \big)
\end{equation}

as a consequence :

\begin{equation}
\int_{\Sigma_t} \abs{\nabla^2 k}^2 \ls \int_{\Sigma_t} \big( \abs{k}^6 + \abs{E}^2\abs{k}^2 + \abs{\nabla H}^2 \big) + \int_{\Sigma_t} \abs{R}^2\abs{\nabla k}^2  
\end{equation}

here we have to be a little more careful that in \cite{krb} as we do not have a priori control on $\ns{k}{\infty}$.
In particular, the $L^\infty$ bounds implied by the control on $\bo{R}$ do not imply $L^\infty$ bounds on $R$, nevertheless :

\begin{eqnarray*}
\ns{\nabla^2 k}{2}^2& \ls & C(t_\star, \Delta_1,\Delta_2) +  \int_{\Sigma_t} \abs{R}^2\abs{\nabla k}^2\\
&\ls& C(t_\star, \Delta_1,\Delta_2)+ \ns{\Delta k}{2} \ns{\nabla^2 k}{2}\ns{\nabla k}{2}^4\\
&\ls& C(t_\star, \Delta_1,\Delta_2)(1+ \ns{\nabla k}{2}^4\ns{\nabla^2 k}{2})
\end{eqnarray*}

which implies the following : 
\begin{thm}
There exists a constant $J(t_\star, \Delta_i)$ such that for all $t \in [0,t_\star[$ :
\begin{equation}
\ns{\nabla^2 k}{2}^2 \ls J
\end{equation}
\end{thm}

At this step of the proof, we thus have pointwise bounds for $k$. The proof for the derivatives of order 3 is similar. We begin by differentiating two times the curl equation and so on.

\vspace{3mm}

As for the energy estimates, we have now to face the problem of the weighted Sobolev spaces. As mentioned before, we do not have a sufficiently good wave equation for $k$ that's why we have to use an other idea. We need to control the weighted Sobolev norms in a neighborhood of spatial infinity. To do that we use the asymptotic flatness property. Considering that $O$ is the origin of the system of coordinate given by the theorem, we have that the geodesic distance is controlled by the euclidian norm relative to the transported coordinates $r=\sqrt{x_1^2+x_2^2+x_3^2}$. Now that we have the weighted estimates for $\bo{R}$, we derive a 3D Hodge system for $fk:=(1+r^2)k$ and it's easy to control lower order terms in the divergence and curl. Let us consider a smooth scalar function $f$. Let $\tilde{k}=fk$, $\tilde{k}$ satisfies the following $3D$ Hodge :

\begin{eqnarray}
tr \tilde{k} &=&0 \\
\nabla^{j} \tilde{k}_{ij} &=& k_{ij}\nabla^j f \\
\nabla \wedge  \tilde{k}_{ij}&=& f H_{ij} + k_{bj} \in_i^{ab} \nabla_a f + k_{bi}  \in_j^{ab} \nabla_a f
\end{eqnarray} 

remark that $\abs{\nabla f}\leq 2min(f,f/r)$.
The idea of using modified Hodge systems is taken from \cite{ck}, work in which due to the differences between an exterior region and an interior region, the authors have to work on a modified $k$.

We thus can apply to $\tilde{k}$ the same treatment that we did on $k$ using the fact that the additional terms are low order and can be controlled. We thus obtain :

\begin{thm}
There exists a constant $C(t_\star,\Delta_i)$ such that :
\begin{equation}
\|k\|_{H_{3,1}(\Sigma_t,O_t)}  \ls C\|\bo{R}\|_{H_{2,1}(\Sigma_t,O_t)}
\end{equation}
\end{thm}

{\bf Proof :}

Applying the same calculations that previously we get :

\begin{eqnarray}
\int_{\Sigma_t} \big( \abs{\nabla \tilde{k}}^2+ \frac{1}{4}\abs{\tilde{k}}^4 \big) &\lesssim& \int_{\Sigma_t} \abs{\bo{R}}^2 +  \int_{\Sigma_t} \abs{\nabla f}^2 \abs{k}^2 \\
& \lesssim & \int_{\Sigma_t} f^2\abs{\bo{R}}^2 +  \int_{\Sigma_t} \abs{min(f,f/r)}^2 \abs{k}^2 \\
& \lesssim & \int_{\Sigma_t}f^2 \abs{\bo{R}}^2 +  \int_{\Sigma_t} \abs{\tilde{k}}^2\abs{min(1,1/r)} ^2 \\
& \lesssim & \int_{\Sigma_t} f^2\abs{\bo{R}}^2 + \e \int_{\Sigma_t} \abs{\tilde{k}}^4 + C_\e
\end{eqnarray}

taking $\e$ small enough we get :

\begin{equation}
\int_{\Sigma_t} \big( \abs{\nabla \tilde{k}}^2+ \frac{1}{4}\abs{\tilde{k}}^4 \big) \lesssim \int_{\Sigma_t}f^2 \abs{\bo{R}}^2
\end{equation}

The same modification could be applied to the estimates on the derivatives of $k$.

\vspace{5mm}

The previous theorem together with the control on $\|\bo{R}\|_{H_{3,1}(\Sigma_t,O_t)}$ and the constraint equation :

\begin{equation}
R_{ij}-k_{ia}k^a_j=E_{ij}
\end{equation}

gives :

\begin{thm}
There exists a constant $C(t_\star,\Delta_i)$ such that :
\begin{equation}
\|R\|_{H_{2,1}(\Sigma_t,O_t)}  \ls C\|\bo{R}\|_{H_{2,1}(\Sigma_t,O_t)}
\end{equation}
\end{thm}

\subsection{Conclusion :}

To conclude we have :

 \begin{equation}
\|R\|_{H_{2,1}(\Sigma_t,O_t)}  + \|k\|_{H_{3,1}(\Sigma_t,O_t)} \ls C(t_\star, \Delta_i)
\end{equation}

It remains to check that we have a uniform control on the isoperimetric constant $(\Sigma_t,g_t)$. This is an implication of the uniform control on $\|R\|_{L^\infty(\Sigma_t)}$.

\subsection{Trace theorem with finite curvature flux in 2D geometry :}

Now let us consider a family $S_{t,u}$ for $t\in{[t_0,t_1]}$ and a fixed $u$ of smooth 2 dimensional surfaces in a fixed 3 dimensional Riemannian manifold. We suppose moreover that the constant of the trace theorem can be uniformly controlled, then we have :
\begin{eqnarray*}
\|k\|_{L^\infty_a(L^4(S_a))}^2& \lesssim&  \int_{\Sigma_t} \abs{\bo{R}}^2\\
\|\nabla n\|_{L^\infty_a(L^4(S_a))}^2& \lesssim&  \int_{\Sigma_t} \abs{\bo{R}}^2
\end{eqnarray*}

Moreover we can at the minimum heuristically control half a derivative in $L^2(S_{t,u})$ norm of $k$ and $3/2$ derivatives of the lapse $n$ restricted to the same family of 2 dimensional surfaces. We would like to apply this theorem to the surfaces $S_{t,u}$. However the standard theory requires pointwise bound of the curvature of the surfaces $S_{t,u}$ that we will denote by $K_{t,u}$. 

Let us recall the following relation :
\begin{equation*}
K=-\frac{1}{4}tr\chi tr \chiu+\frac{1}{2}\hat{\chi}\cdot \hat{\chiu}-\rho
\end{equation*}
but we only have in our case of  $\ntsl{\rho}{2}{2}<+\infty$ as assumption of our theorem due to the hypothesis of finite curvature flux $a)$. Thus there is no chance to get the required bound and directly the trace theorem.

Let us consider a fixed leaf $\Sigma_t$. Suppose moreover that we have constructed an optical function $u$ in a neighborhood $M$ of a truncated null cone $\mathcal{N}=\mathcal{N}^{-}(p,\delta)$ such that $u=0$ on it. We suppose moreover that $M \cap \Sigma_t = M_t$ contains a $r/2$-neighborhood of our reference surface $S_{t,0}$. We consider on $M_t$ the induced foliation by the sets of $u$ with lapse $b$ and second fundamental form $\theta$. For a point $q \in M_t$, we consider the $2D$ surface $S_{t,u(q)}$ and note $r(q)=\frac{1}{4 \pi}\big(\int_{S_{t,u(q)}} 1)^\demi$.

Suppose moreover that :

\begin{equation*}
\| tr \theta - \frac{2}{r} \|_{L^\infty (M_t) + L^3(M_t)} \lesssim 1 \hspace{5 mm} {\bf (H)}
\end{equation*}
and that we have pointwise bounds for $b$ on $M_t$:
\begin{equation*}
b \sim 1 \hspace{5 mm} {\bf (P)}
\end{equation*}

\subsubsection{Lebesgue estimates :}

\begin{thm}{\bf Trace theorem}
Let $F$ be an arbitrary tensor tangent to $S_{t,0}$ at each point of $M_t$ then we have the following estimate :

\begin{eqnarray*}
\int_{S_{t,0}} \abs{F}^4 &\lesssim &\big( \nm{F}{4}^4 + \nm{F}{6}^4  \big)+ \nm{F}{6}^3\nm{\nabla F}{2} \\
&\lesssim & \nm{\nabla F}{2}^4 +  \nm{r^{-1} F}{2}^4
\end{eqnarray*}
\end{thm}

{\bf Proof :}

Let us consider a smooth function $\phi$ from $M_t$ to $[0,1]$ such that : $\phi=1$ in a neighborhood of $S_{t,0}$ in $M_t$, $\phi(q)=0$ for $q \in M_S$ and $d_t(p,q) \rs \frac{r}{2}$ and $\abs{\nabla \phi} \ls \frac{Cf}{r}$ on $M_t$, where $f=1_{d_t(p,q) \ls \frac{r}{2}}$.

\vspace{3mm}
Let  $G= \phi F$. we have :

\begin{equation*}
tr\theta = div \hspace{2mm}N
\end{equation*}

thus we have :
\begin{eqnarray*}
\int_{S_{t,0}} \abs{F}^4 &=& - \int_{Ext S_{t,0}} div (\abs{G}^4 N)\\
&=&  - \int_{Ext S_{t,0}} \big( tr \theta \abs{G}^4 + 4 \abs{G}^2G\cdot \nabla_N G \big)\\
&\ls&  -  \int_{Ext S_{t,0}} \abs{G}^4 (tr \theta - \frac{2}{r}) - \int_{Ext S_{t,0}} \abs{G}^4\frac{2}{r} -  \int_{Ext S_{t,0}} 4 \abs{G}^2G\cdot \nabla_N G\\
&\ls&\big( \nm{G}{4}^4 + \nm{G}{6}^4  \big) \big(\| tr \theta - \frac{2}{r} \|_{L^\infty (M_t) + L^3(M_t)} +  \|\frac{2}{r}\|_{L^{3,\infty}(Ext S_{t,0})} \big)\\
&& + \nm{G}{6}^3 \nm{\nabla G}{2}\\
& \lesssim& \big( \nm{G}{4}^4 + \nm{G}{6}^4  \big) + \nm{G}{6}^3\nm{\nabla G}{2} \\
 &\lesssim& \big( \nm{F}{4}^4 + \nm{F}{6}^4  \big) + \nm{F}{6}^3\nm{\nabla F}{2}+ \nm{F}{6}^4\nm{f/r}{3}\\
 &\lesssim& \big( \nm{F}{4}^4 + \nm{F}{6}^4  \big) + \nm{F}{6}^3\nm{\nabla F}{2}\\
 &\lesssim & \nm{\nabla F}{2}^4 +  \nm{r^{-1} F}{2}^4
  \end{eqnarray*}

\begin{thm}
Suppose that we can construct an optical function in the neighborhood of every truncated null cone. Suppose moreover that these foliations satisfy {\bf (H)} uniformly. Let us denote $S$ a generic $S_{t,u}$ and $M_S$ its corresponding neighborhood in $\Sigma$. We have uniformly :

\begin{eqnarray*}
\|k\|_{L^4(S)}^2& \lesssim&  \int_{M_S} \abs{\bo{R}}^2\\
\|\nabla n\|_{L^4(S)}^2&\lesssim& \int_{M_S} \abs{\bo{R}}^2
\end{eqnarray*}

\end{thm}

{\bf Proof :}
It's sufficient to remark that $k \in L^4(M_t)$ and  $\nabla k \in L^2(M_t)$ which implies that $k \in L^p(M_t)$ for every $4\ls p\ls6$ and in particular for $p=6$.
For the lapse $n$ we have $n \in  L^\infty(M_t)$ and $\nabla^2 n  \in  L^2(M_t)$, thus by integration by part :
\begin{equation*}
\nm{\nabla n}{4}^2\ls\nm{n}{\infty}\nm{\Delta n}{2}
\end{equation*}

thus we can apply the trace theorem to $\nabla n$. We have also to remark that we can take $\kappa$ uniformly bounded from bellow (see the Remark in subsection (1.5.3)).

{\it Remark :} In our application (see section 2), we will make large use of these estimates.  The hypothesis {\bf (H)} will be quite natural in this context because we will derive $L^\infty$ bound for $tr \chi$ and as we have :

\begin{equation*}
\chi_{AB}=\theta_{AB} - k_{AB}
\end{equation*}

and by maximality $trk=0$, it gives :

\begin{equation*}
tr \theta =tr \chi - k_{NN}
\end{equation*}
thus as we have $k_{NN}\in L^p(M_t)$  for all $p$, $4\ls p\ls6$ just by using $3D$-Hodge systems, we will be able to apply the trace theorem.   

Moreover we have using the same idea :

\begin{equation*}
\int_{S_{t,0}} \abs{F}^2 \lesssim\big( \nm{F}{2}^2 + \nm{F}{3}^2  \big) \ +\nm{F}{2}\nm{\nabla F}{2}
\end{equation*}

\subsubsection{Sobolev estimates :}

The proof of Sobolev type trace theorems under assumption {\bf (H)} requires the use of a geometric version of Littlewood-Paley theorem (see \cite{krg} and the appendix below). The proof of such theorem requires an additional assumption $\bo{H2}$ :

there exists C such that :

\begin{equation}
\norm{\theta}_{L^\infty_u(L^4(S_{t,u})}+ \norm{\nablab b}_{L^\infty_u(L^4(S_{t,u})}
\end{equation}
We will use the notation $\tilde{H}^s(M_t)$ for $s\in[\frac{1}{2},\frac{3}{2}]$ for Sobolev-type spaces of order s defined as the subspace of $\mathcal{D}'(M_t)$ of finite $\mathcal{N}_s (M_t)$ norm wher we denote by $\Lambda:=(Id-\Deltab)^{\frac{1}{2}}$ and :

\begin{equation}
\norm{A}_{\mathcal{N}_1 (M_t)}:=\norm{r^{-s}A}_{L^2(M_t)} + \norm{\Lambda^s A}_{ L^2(M_t)} + \norm{\Lambda^{s-1} \nablab_N A}_{L^2(M_t)}
\end{equation}

With our definition we have by complex interpolation that :

\begin{equation}
\tilde{H}^s(M_t)=[\tilde{H}^{\frac{1}{2}}(M_t),\tilde{H}^{\frac{3}{2}}(M_t)]
\end{equation}

Moreover, we have that $\norm{\Lambda A}_{L^2(M_t)}= (\norm{ A}_{L^2(M_t)}^2 + \norm{\nablab A}_{L^2(M_t)}^2)^{\frac{1}{2}}$ thus :

\begin{equation}
\norm{A}_{\mathcal{N}_1 (M_t)}\approx \norm{ r^{-1}A}_{L^2(M_t)} + \norm{\nablab A}_{L^2(M_t)} + \norm{\nablab_N A}_{L^2(M_t)}
\end{equation}

For a $S_t$-tangent tensor-field, we define for $s\rs 0$ : $\norm{F}_{H^s(S_t)}=\norm{r^{-s}F}_{L^2(S_t)}+\norm{\Lambda^{s}F}_{L^2(S_t)}$.

With the same notations we need to prove the following :
\begin{thm}
Let $F$ be an arbitrary tensor tangent to $S_{t,u}$ at each point of $M_t$. Suppose that the $(t,u)$ foliation verifies {\bf (H)} {\bf (H2)}and{\bf (P)} in $M_t$, then we have the following estimate :

$$\|F\|_{H^\demi_{S_{t,0}}}  \lesssim \|F\|_{\tilde{H}^1(M_t)} $$
\end{thm}

{\bf Proof :}
By interpolation it is sufficient to prove the continuity of the trace operator from $\tilde{H}^{\frac{3}{2}}(M_t)$ to $H^1(S_{t,u)})$ and from $\tilde{H}^{\frac{1}{2}}(M_t)$ to $L^2(S_{t,u)})$. For the second mapping, we write :

\begin{eqnarray*}
\int_{S_{t,0}} \abs{F}^2 &\lesssim& \big( \nm{F}{2}^2 + \nm{F}{3}^2 +\norm{F}^2_{\mathcal{N}_\demi(M_t)}\big)\\
 &\lesssim&\norm{F}^2_{\mathcal{N}_\demi(M_t)}\\
\end{eqnarray*}

The continuity of the other mapping (from $\tilde{H}^{\frac{3}{2}}(M_t)$ to $H^1(S_{t,u)})$) requires a commutator lemma proved below :

\begin{lem}
We have the following commutator Lemma :
\begin{equation}
| \int_{M_t} \nablab F \cdot [\nablab_N,\nablab]Fd\mu_Sbdu | \lesssim \norm{F}^2_{\mathcal{N}_{\frac{3}{2}}(M_t)}
\end{equation}
\end{lem}
But controlling these terms by the $\mathcal{N}_{\frac{3}{2}}(M_t)$ norm of F is a direct consequence of the commutator lemma for the second term and by interpolation and definition of our norms for the first one.

Once we have this lemma we recall that :
\begin{equation*}
\int_{S_{t,0}} \abs{\nablab F}^2 \lesssim    \big( \nm{\nablab F}{2}^2 + \nm{\nablab F}{3}^2 +|\int_{M_t}\nablab_N \nablab F \cdot  \nablab F| +\int_{M_t}\nablab_N \nablab F\cdot \nablab F|\big)
\end{equation*}

The only difficulty to prove the continuity of the mapping relies in proving :

\begin{equation}
|\int_{M_t}\nablab_N \nablab F \cdot  \nablab F|\lesssim \norm{F}^2_{\mathcal{N}_{\frac{3}{2}}(M_t)}
\end{equation}

but :
\begin{eqnarray}
|\int_{M_t}\nablab_N \nablab F \cdot  \nablab F|&\lesssim& |\int_{M_t}\nablab \nablab_N F \cdot  \nablab F| + | \int_{M_t} \nablab F \cdot [\nablab_N,\nablab]F |
\end{eqnarray}

As a corollary, we have the following proposition :

\begin{prop}
Suppose that we have proved that the $(t,u)$ foliations locally defined satisfy the hypothesis  {\bf (H)} and  {\bf (P)} uniformly. Let $S$ be a generic $2D$ surface obtained as the subsection of a null cone $\mathcal{N}$ and a leaf $\Sigma_t$. Let $M_S$ its canonical neighborhood in $\Sigma_t$. Then we have :
\begin{eqnarray*}
\|k\|^2_{H^\demi(S)}& \lesssim& \int_{M_S} \abs{\bo{R}}^2\\
\|n\|^2_{H^{\frac{3}{2}}(S)}& \lesssim&  \int_{M_S} \abs{\bo{R}}^2
\end{eqnarray*}
\end{prop}

\subsubsection{Other trace estimates :}

\hs We will later need $\mathcal{B}^0$ estimates for $k$ and $ \nabla log n$. So let us consider a truncated null cone $\Hyp = \cup_{[0,1]} S_{(t,u)}$ where the  $S_{(t,u)}$ are uniformly weakly regular and of radius comparable to $1$. Suppose moreover that these surfaces verify {\bf (H)} and {\bf (P)} uniformly, then we have :

\begin{prop}[Besov estimates]
Let $a$, $0\ls a < \demi$ :
\begin{eqnarray}
\|k\|^2_{\mathcal{B}^a}& \lesssim& \sup_{t \in[0,1]}\int_{M_{S_{t,u}}} \abs{\bo{R}}^2 \\
\|\nabla log n\|^2_{\mathcal{B}^a} & \lesssim&\sup_{t \in[0,1]} \int_{M_{S_{t,u}}} \abs{\bo{R}}^2
\end{eqnarray}
\end{prop} 

{\bf Proof :}

Following the notations of previous subsection , we have :
\begin{eqnarray}
\|k\|_{\mathcal{B}^a}& \lesssim&C_\kappa \sum_m 2^{am}\sup_{t \in[0,1]}\big(\int_u \int_{S_{t,u}} 2^m\abs{P_m k}^2\big)^\demi\\
& \lesssim&C_\kappa \sum_m 2^{m(a-\demi)}\sup_{t \in[0,1]}\big(\int_u \int_{S_{t,u}} 2^{2m} \abs{P_m k}^2\big)^\demi
\end{eqnarray}

but :
\begin{equation}
\sup_{t \in[0,1]}\big(\int_u \int_{S_{t,u}} 2^{2m} \abs{P_m k}^2\big) \lesssim \sup_{t \in[0,1]}\big(\int_u \int_{S_{t,u}} \abs{\nablab P_m k}^2\big) \lesssim \sup_{t \in[0,1]} \int_{M_{S_{t,u}}} \abs{\nablab k}^2
\end{equation}

and idem for $\nabla log n$.

\subsection{Appendix : Some results from a geometric version of LP theory }

This theory is based on the heat flow and the ideas can be traced back to E.Stein. The geometrical context of this theory is $2D$ weakly regular surfaces :

\begin{defi}
Let $S_0 \subset \Sigma_0$ be a compact 2-surface diffeomorphic to $\bo{S^2}$ with $\gamma$ a Riemannian metric on it. We say that $\Sigma_0$ is weakly regular if it can be covered by a finite number of coordinate charts $U$ with coordinates $\omega^1$, $\omega^2$ relative to which,

$$C^{-1}\abs{\xi}^2 \leqslant \gamma_{ab}(p) \xi^a\xi^b \leqslant C\abs{\xi}^2, \hspace{5mm}\text{uniformly for all }p \in U$$

$$\sum_{a,b,c}\int_U \abs{\partial_c \gamma_{ab}}^2dx^1dx^2 \leqslant C^{-1}$$

We say moreover that $\gamma$ is W-weakly spherical if in addition :

$$\sum_{a,b,c}\int_U \abs{\partial_c \gamma_{ab}-R^2\tilde{\gamma}_{ab}}^2 dx^1dx^2 \leqslant I_0^2$$

where $\tilde{\gamma}$ is the standard metric on $\bo{S^2}$ and $R$ is a constant, $W^{-1} \leqslant R  \leqslant W$. Here $I_0$ is a sufficiently small constant. 
\end{defi}

For the properties of such surfaces see \cite{krg}. We say that a family of surfaces is uniformly weakly spherical if the elements of this family are W-weakly spherical with constants bounded independently of the elements.

Following the article \cite{krg}, we introduce a family of LP projectors $(P_k)_{k\in\mathbb{Z}}$ verifying $\sum_k P_k^2=Id$, these operators are based on the heat flow on $S$ whose evolution semi-group will be denoted $U(\tau)$ and on a  function $m$ of the family $\mathcal{M}$ defined as the family of smooth functions on $[0,\infty[$  and verifying the following vanishing moment property at order N :
\begin{equation}
\int_0^\infty \tau^{k_1}\partial_\tau^{k2}m(\tau)d\tau=0, \hs \abs{k_1}+\abs{k_2}\leq N
\end{equation}

From the heat flow and such a function, we define the projection operators as follows :
\begin{equation} 
P_k=\int_0^\infty 2^{2k}m(2^{2k}\tau)U(\tau)Rd\tau
\end{equation}

This family also satisfies the following additional properties that they share with the usual LP projectors :

\begin{prop}{\bf LP type properties} The LP projectors defined in \cite{krg} verify the following properties :

i) $L^p$ boundness :  For any $1\ls p < \infty$ and any interval $I\subset \mathbb{Z},$
\begin{equation}
\|P_I F\|_{L^p(S)} \lesssim \|F\|_{L^p(S)}
\end{equation}

ii) $L^p$ almost orthogonality, for any $(k,k')$ and families of LP projector (P,P') corresponding to the base functions (m,m'):
\begin{equation} 
\|P_kP'_{k'} F\|_{L^p(S)} \lesssim 2^{-2\abs{k-k'}}\|F\|_{L^p(S)}
\end{equation}

iii) Bessel inequality 
\begin{equation} 
\sum_{k \in \mathbb{Z}} \|P_k F\|_{L^2(S)}^2 \lesssim \|F\|_{L^2(S)}^2
\end{equation}

iv) Reproducing property 

Given any integer $l\ls2$ and $\bar{m} \in \mathcal{M}_l$, there exists $m \in  \mathcal{M}$ such that :
\begin{equation}
{}^{(\bar{m})}Pk ={}^{(m)}P_k \cdot {}^{(m)}P_k
\end{equation}

v) Other estimates :  For any $1\ls p\ls\infty$ and any $k \in \mathbb{Z}$ :
\begin{eqnarray}
\|\Deltab P_k F\|_{L^p(S)} &\lesssim& 2^{2k}\|F\|_{L^p(S)}\\
\|  P_kF\| _{L^p(S)} &\lesssim& 2^{-2k} \| \Deltab F\| _{L^p(S)}
\end{eqnarray}
Given a $m\in \mathcal{M}$, it's possible to find an $m'\in \mathcal{M}$ and the LP projectors family associated to it $P'$ such that $\Deltab P_k=2^{2k}P'_k$

Moreover we have the following estimates :

\begin{eqnarray}
\| \nabla P_k F\|_{L^2(S)} &\lesssim& 2^{k}\|F\|_{L^2(S)}\\
\|  P_kF \| _{L^2(S)} &\lesssim& 2^{-k} \|  \nablab F \| _{L^2(S)}\\
\| P_k \nablab F\| _{L^2(S)} &\lesssim& 2^{k} \|   F\| _{L^2(S)}
\end{eqnarray}

vi) Weak Bernstein inequalities : For any $2\ls p<\infty$
\begin{eqnarray}
\|  P_kF\| _{L^p(S)} &\lesssim& (2^{(1-\frac{2}{p})k}+1) \| F\| _{L^2(S)}\\
\|  P_{<0}F\| _{L^p(S)} &\lesssim&\| F\| _{L^2(S)}
\end{eqnarray}
and their dual estimates
\begin{eqnarray}
\| P_kF\| _{L^2(S)} &\lesssim& (2^{(1-\frac{2}{p})k}+1) \| F\| _{L^{p'}(S)}\\
\|  P_{<0}F\| _{L^2(S)} &\lesssim&\| F\| _{L^{p'}(S)}
\end{eqnarray}

vii) Commutator estimates . : Given two tensorfields F, G and denoting by $F \cdot G$ any contraction of their tensorial product. We have the following estimates for the commutator $[P_k,F]\cdot G$ :
\begin{eqnarray}
\|  [P_k,F] \cdot G|_{L^2(S)} &\lesssim& 2^{-k} \| \nablab F\| _{L^{\infty}(S)} \| \nablab G\| _{L^{2}(S)}\\
\|  [P_k,F]\cdot G|_{L^2(S)} &\lesssim&\big(2^{-2k} \| \Deltab F\| _{L^{\infty}(S)} +2^{-k} \| \nablab F\|_{L^{\infty}(S)}\| \nablab G\| _{L^{2}{S}}\big)
\end{eqnarray}

\end{prop}
With the help of this family of operator, we can define the fractional Sobolev norms as we usually do on numerical spaces with the standard LP theory based on Fourier transform.

\begin{defi}
Given an arbitrary tensor F on a fixed surface $S=S_t$ for some $t$, $0 \ls t\ls T \ls1$ and $a\in \mathbb{R}$
$$\| F \|_{H^a(S)}=\| \Lambda^aF \|_{L^2(S)}:=\big(\sum_{k\rs0}2^{2ka}\|P_k F\|^2_{L^2(S)}\big)^\demi + \|P_{<0}F\|_{L^2(S)}$$

where $\Lambda=(Id-\Delta)^\demi$. Remark that these Sobolev type spaces are inhomogoneous.
We follow again KR in defining the corresponding Besov type norms for an arbitrary S-tangent vectorfield F on $\Hyp$ :

\begin{eqnarray*}
\|F\|_{\mathcal{B}^a}&=&\sum_{k\rs0}2^{ka}\sup_{t\in[0,1]}\|P_k F\|_{L^2(S)} + \sup_{t\in[0,1]}\|P_{<0}F\|_{L^2(S)}\\
\|F\|_{\mathcal{P}^a}&=&\sum_{k\rs0}2^{ka}\|P_k F\|_{L^2(\Hyp)} + \|P_{<0}F\|_{L^2(\Hyp)}
\end{eqnarray*}
\end{defi}

\subsection{A commutator Lemma :}

In the proof of Sobolev type trace theorems we have used the following lemma :
\begin{lem}
Commutator Lemma :
\begin{equation}
A:=| \int_{M_t} \nablab F \cdot [\nablab_N,\nablab]F| \lesssim \norm{F}^2_{\mathcal{N}_{\frac{3}{2}}(M_t)}
\end{equation}
\end{lem}

We have for any arbitrary k-tensor on $\Sigma_t$ tangent to $S_{t,u}$ at any point,
\begin{equation}
| [\nablab_N,\nablab]F| \leq |\nablab(log b)||\nablab_N F| + |\theta||\nablab F|+ |R||F| +|\nablab(log b)||F||\theta|
\end{equation}
  
 With our assumptions {\bf (H)} and {\bf (H2)} which will be verified in the bootstrap of section 2, we will have :
 \begin{eqnarray*}
 A &\lesssim& \norm{F}_{\mathcal{N}_{\frac{3}{2}}(M_t)}^2( \norm{\nablab b}_{L^\infty_u(L^4(S_{t,u})} +  \norm{\nablab b}_{L^\infty_u(L^4(S_{t,u})} \norm{\theta}_{L^\infty_u(L^4(S_{t,u})} +  \norm{\theta}_{L^\infty_u(L^4(S_{t,u})} + \\
 &&\norm{R}_{L^2(M_t)})
 \end{eqnarray*}
 
 where we have used we the notation $\nabla=(\nablab, \nablab_N)$ :
 \begin{eqnarray*}
 \norm{\nabla F}_{L^2_u(L^4(S_{t,u})} &\lesssim &  \norm{F}_{\mathcal{N}_{\frac{3}{2}}(M_t)}\\
 \norm{F}_{L^6(M_t)} &\lesssim & \norm{F}_{\mathcal{N}_1(M_t)}
 \end{eqnarray*}
 
 The second one is a consequence of Sobolev imbedding once we have made the remark that the $\mathcal{N}_1$ norm is equivalent to the standard $H^1$ norm. For the first one, remark that :
 
 \begin{equation}
 \norm{\nabla F}_{L^2_u(L^4(S_{t,u})}^2\lesssim \big(  \int_u \norm{\nabla F}_{L^2(S_{t,u})} ^2+ \norm{\Lambda^\demi \nabla F}_{L^2(S_{t,u})}^2 \big) \lesssim  \norm{F}_{\mathcal{N}_{\frac{3}{2}}(M_t)}^2
 \end{equation}

 \cleardoublepage

\section{Causal Structure of null Hypersurfaces :}

\subsection{Definitions :}

\hspace{5 mm}We will consider in this section a null hypersurface $\Hyp$ typically a troncated null cone initiating in a point $p$ of a part of our space-time foliated by a regular time function $t$. Thus $\Hyp$ itself is foliated by the level surfaces $S_t$ of $t$ restricted to $\Hyp$. We will denote by $S_0$ the initial 2D compact surface that is $\Hyp\cap \Sigma_{t_0}$. We will work on a truncated hypersurface in subsection 1 to 14 and we will describe the modifications to consider the null cone till the vertex to conclude in subsection 15 .

We will moreover assume that we have constructed an optical function in the neighborhood of $\Hyp$ vanishing on $\Hyp$ and verifying the eikonal equation :

\begin{equation}
g^{\al \bet} \partial_\al u\partial_\bet u =0
\end{equation}

Following the standard formulation, we say that two vectors $(\Lb,L)$ defined in this neighborhood of $\Hyp$ form a null pair compatible with $t$ and $u$ iff  $\langle \Lb,L \rangle= -2$ and at every point there are normal to the sphere $S_{t,u}$ passing through that point. Moreover, an arbitrary orthonormal frame on $S_{t,u}$ will be denoted by $(e_a)_{a=1,2}$. $(e_1,e_2,\Lb,L)$ is called a null frame compatible with the $t$-foliation. We have by definition :
\begin{equation}
<e_a,L>=<e_a, \Lb>=0, \hspace{5mm}<e_a,e_b>=\delta_{a,b}.
\end{equation}

Associated with this frame, we define the following geometric quantities :

\begin{thm}{\bf Ricci coefficients}
The Ricci coefficients relative to an adapted frame $(e_A,e_B,L,\Lb)$ are defined as follows :
\begin{eqnarray*}
\chi_{AB}=\langle \bo{D_A}L,e_B \rangle &\hspace{5mm}& \underline{\chi}_{AB}=\langle\bo{D_A}\Lb,e_B \rangle \\
 \xi_{A}=\frac{1}{2}\langle \bo{D_L}L,e_A \rangle  &\hspace{5mm}& \underline{\xi}_{A}=\frac{1}{2}\langle \bo{D_{\Lb}}\Lb,e_A \rangle\\
\h_A=\frac{1}{2}\langle \bo{D_{\Lb}}L,e_A \rangle&\hspace{5mm}&  \underline{\h}_{A}=\frac{1}{2}\langle \bo{D_L}\Lb,e_A\rangle \\
\Omega =\frac{1}{4}\langle \bo{D_L}L,\Lb \rangle&\hspace{5mm}&\underline{\Omega}=\frac{1}{4}\langle \bo{D_{\Lb}}\Lb,L\rangle \\
V_A=\frac{1}{2}\langle \bo{D_A}L,\Lb\rangle
\end{eqnarray*}
\end{thm}

If we denote by {\bf (T)} the unit future-directed vector normal to $\Sigma_t$ and  by {\bf (N)} the unit vector normal to  $S_{t,u}$ in $\Sigma_t$, two among the compatible null frames play an important role :
\begin{defi}
The null pair $(L,\Lb)$ with $L= \bo{T}-\bo{N}$ and  $\Lb= \bo{T}+\bo{N}$ is called the standard null pair.
The null pair $(L',\Lb')$ with $L'= b^{-1}(\bo{T}-\bo{N}) = -\bo{g}^{\mu \nu} \partial_\nu u \partial_\mu$ and  $\Lb'= b(\bo{T}+\bo{N})$ is called the renormalized null pair.
\end{defi}
We will use the adjectives standard and renormalized for the corresponding null frames. The renormalized null pair correspond to the null pair $(L, \Lb)$ of section 1, that is $L$ is the null geodesic generator of our null cone. We denote by  $s$ the affine parameter corresponding to $L$ that is $L(s)=1$ and $s(p)=0$ and  $s'$ the corresponding affine parameter for $L'$.

As we want to make comparisons with the original proof in \cite{krc}, we introduce the following null frame adapted to the geodesic foliation id est the foliation given by the level sets of $s'$. From our normalized null frame $(e_1,e_2,\Lb',L')$ we obtain such a frame by the following transformation :

\begin{eqnarray*}
e'_a&=&e_a-e_a(s')L'\\
e'_3&=&\Lb'-2e_a(s')e_a-2 \abs{\nabla s'}^2 L'\\
e'_4&=&L'
\end{eqnarray*}

Then we have the following relations :

\begin{thm}{\bf Ricci coefficients (case of the geodesic foliation :)}
In the case of the geodesic foliation and relative to the null frame $(e'_1,e'_2,e'_3,e'_4)$ we have in particular the following cancelations:

\begin{eqnarray*}
\xi^{(g)}_A&=&\frac{1}{2}\langle \bo{D_{e'_4}}e'_4,e'_A\rangle=0\\
\Omega^{(g)}&=&\frac{1}{4}\langle \bo{D_{e'_4}}e'_4,e'_3 \rangle=0\\
V^{(g)}_A&=&\frac{1}{4}\langle \bo{D_{e'_A}}e'_4,e'_3 \rangle=-\underline{\h}^{(g)}_A
\end{eqnarray*}
\end{thm}

If we denote by $b$ the lapse function of the foliation induced by $u$ on each $\Sigma_t$ and by $\theta$ the second fundamental form of the surfaces $S_{t,u}$ relative to $\Sigma_t$, we have :

\begin{thm}{\bf Ricci coefficients (case of the (t,u) foliation :)}
In the case of the $(t,u)$ foliation and relative to the standard $(t,u)$ compatible null frame $(e_1,e_2,\Lb,L)$, we have the following relations :
\begin{eqnarray*}
\chi_{AB}&=&\tht_{AB}-{}^{(2)}k_{AB}\\
\underline{\chi}_{AB}&=& -\tht_{AB}-{}^{(2)}k_{AB}\\
\underline{\x}_A&=&n^{-1}\nablab_An-b^{-1}\nablab_Ab\\
\underline{\h}_A&=&n^{-1}\nablab_An-\e_A=-\ebaa_A\\
\h_A&=&b^{-1}\nablab_Ab+\e_A\\
\Omega&=&\frac{1}{2}(-n^{-1}\nablab_Nn+\delta)\\
\underline{\Omega}&=&\frac{1}{2}(n^{-1}\nablab_Nn+\delta)\\
V_A&=&\e_A
\end{eqnarray*}
 \end{thm}

where we have used the following $2+1$ splitting of the second fundamental form $k$ :
\begin{eqnarray*}
{}^{(2)}k_{AB}=k_{AB}\\
\e_A=k_{AN}\\
\delta=k_{NN}
\end{eqnarray*}

The proof of this proposition can be found in \cite{ck} p. 171. When there is no risk of confusion between $k$ and ${}^{(2)}k$, we will omit the $(2)$
 
We will also make use of the following notations : 

\begin{eqnarray*}
\deb=k_{NN}-n^{-1}\nablab_Nn\\
\bar{\e}_A=k_{AN}-n^{-1}\nablab_An\\
\end{eqnarray*}

We have seen using 3D Hodge systems and trace theorems that if we have a basic control on $tr\theta$, we get estimates for $k, \e, \bar{\e},\de$ and $\deb$ on $\Hyp$, we will make use of course of these estimates. As there is no risk of confusion in this section between the different derivations, {\bf we will omit the bars and note $\nabla$ for $\nablab$, etc ...}

\subsection{Structure equations :}

\hspace{5mm}As explained above, the global structure of this proof will be the same as in \cite{krc}. As we would like to insist here on the differences between our case and their, we will present first the structure equations relative to the geodesic foliation which is the setting in  \cite{krc} and in comparison the structure in our case.

\subsubsection{Structure equations relative to the geodesic foliation}

The structure equations relative to the geodesic foliation are the following  :

\begin{eqnarray}
e'_4(tr \chi ^{(g)})&=&-|\hat{\chi}^{(g)}|^2-\frac{1}{2}(tr\chi^{(g)})^2- {\bo R}_{44} \\
\nabla_{e'_4}\hat{\chi}^{(g)}&=&- \chi^{(g)} \cdot \hat{\chi}^{(g)} - \al^{(g)} \\
\nabla \cdot \hat{\chi}^{(g)}&=&-\bet^{(g)}+\frac{1}{2}\nabla tr \chi^{(g)} + \frac{1}{2} tr \chi^{(g)} \z^{(g)} - \z^{(g)} \cdot \hat{\chi}^{(g)} + {\bo R}_{4a} \\
e'_4(tr \underline{\chi}^{(g)} )&=&2 \nabla \cdot \underline{\eta}^{(g)}+ 2  \rho ^{(g)}-\demi(tr\chi^{(g)}) tr \underline{\chi}^{(g)} - \hat{\chi}^{(g)}  \cdot \underline{\hat{\chi}}^{(g)} + 2 |\underline{\eta}^{(g)}|^2-{\bo R}_{43}\\
\nabla_{e'_4}\underline{\hat{\chi}}^{(g)}&=&\nabla \hat{\otimes}  \underline{\eta}^{(g)} -  \frac{1}{2}tr \chi^{(g)}  \cdot \hat{\underline{\chi}}^{(g)} - \frac{1}{2}tr \underline{\chi}^{(g)}  \cdot\hat{\chi}^{(g)}+\underline{\eta}^{(g)} \hat{\otimes}  \underline{\eta}^{(g)}+{\bo R}_{ab} - \demi {\bo R}_c^c\gamma_{ab}\\
\nabla_{e'_4}\underline{\z}^{(g)}&=&-\bet^{(g)} + \chi^{(g)} \cdot \big( - \z^{(g)} + \underline{\eta}^{(g)} \big) \\
\nabla \wedge \underline{\eta}^{(g)}&=&\frac{1}{2}\hat{\chi}^{(g)} \wedge \underline{\hat{\chi}}^{(g)} - \sigma^{(g)}\\
K^{(g)}&=&-\frac{1}{4}tr\chi tr \chiu^{(g)}+\frac{1}{2}\hat{\chi}^{(g)}\cdot \hat{\chiu}^{(g)}-\rho^{(g)}+\demi {\bo R}_a^a-\demi {\bo R}_{43}\\
\nabla \cdot  \underline{\hat{\chi}}^{(g)}&=& \frac{1}{2} \nabla tr \underline{\chi}^{(g)}-\frac{1}{2}  tr \underline{\chi}^{(g)} \z^{(g)} + \z^{(g)} \cdot \underline{\hat{\chi}}^{(g)}+\underline{\bet}^{(g)}+{\bo R}_{3a}
\end{eqnarray}

\subsubsection{Structure equations relative to the $(t,u)$ foliation :}

Relative to the standard null pair, we have :

\begin{eqnarray}
L(b)&=&-\deb b\\
L(tr \chi )+ \deb tr\chi&=&-|\hat{\chi}|^2-\frac{1}{2}(tr\chi)^2- {\bo R}_{44}\\
\nabla_L\hat{\chi} + \deb \hat{\chi} &=&-tr \chi  \cdot \hat{\chi} - \al \\
\nabla \cdot \hat{\chi} + \hat{\chi} \cdot \e  &=&-\bet+\frac{1}{2}\nabla tr \chi + \frac{1}{2} tr \chi \e + {\bo R}_{4.}  \\
\nabla_L\h&=&-\bet - \chi \cdot \big(  \h + \bar{\e} \big) \\
\nabla \wedge \eta&=&\frac{1}{2}\underline{\hat{\chi}} \wedge \hat{\chi} + \sigma\\
L(tr \underline{\chi} )- \deb tr \underline{\chi}&=&2 \nabla \cdot \underline{\eta}+ 2  \rho -\frac{1}{2}(tr\chi) tr \underline{\chi} - \hat{\chi}  \cdot \underline{\hat{\chi}} + 2 |\underline{\eta}|^2 -{\bo R}_{43}\\
\nabla_L\underline{\hat{\chi}} - \deb\underline{\hat{\chi}} &=&\nabla \hat{\otimes}  \underline{\eta} -  \frac{1}{2}tr \chi  \cdot \hat{\underline{\chi}}- \frac{1}{2}tr \underline{\chi}  \cdot\hat{\chi}+\underline{\eta} \hat{\otimes}  \underline{\eta} +{\bo R}_{ab}- \demi {\bo R}_c^c\gamma_{ab}\\
\nabla \cdot  \underline{\hat{\chi}} -  \underline{\hat{\chi}}   \cdot \e&=& \frac{1}{2} \nabla tr \underline{\chi}-\frac{1}{2}  \e\hspace{.5mm} tr \underline{\chi}  +\underline{\bet}+ {\bo R}_{3.}\\
K&=&-\frac{1}{4}tr\chi tr \chiu+\frac{1}{2}\hat{\chi}\cdot \hat{\chiu}-\rho +\demi {\bo R}_a^a-\demi {\bo R}_{43}
\end{eqnarray}

We could have written these equations relative to the renormalized null pair. Sometimes it will be easier to deal with them, we of course have the corresponding equations :
\begin{eqnarray*}
L'(b)&=&- \deb\\
L'(tr \chi' )&=&-|\hat{\chi'}|^2-\frac{1}{2}(tr\chi')^2-{\bo R}_{44}\\
etc ...
\end{eqnarray*}
\hspace{5mm}In fact as we remark on this two first equations, the structure equations look simpler relative to the renormalized null pair, but as we have in mind to use a priori estimates coming from trace theorems we prefer the relations involving directly the components of $k$. For instance the estimates on the derivatives of $\chi$ together with the a priori estimates give estimates on $\chiu$. To pass from estimates on derivatives of $\chi'$ to estimates on $\chiu'$, we would have to prove a prori estimates on $bk$ that we don't have. (Recall that we have an a priori control on half  derivative of $k$ as soon as we will have some basic control on $tr \theta$).

\subsection{Remarks :}

We have the following relations between the geometric quantities relative to the standard null frame and the renormalized ones :
\begin{eqnarray}
\chi'&=&b^{-1}\chi\\
\chiu'&=&b\chiu\\
\xi'&=&b^{-2}\xi\\
\underline{\xi}'&=&b^2\underline{\xi}\\
\h'&=&\h\\
\hb'&=&\hb\\
\Omega'&=&0=b^{-2}\Omega -2b^{-1}L(b^{-1})\\
\underline{\Omega}'&=&b^{2}\underline{\Omega}-2b\Lb(b)\\
V'_A&=& V_A+b^{-1}\nabla_Ab
\end{eqnarray}

The relations between the null components is directly given by :
\begin{equation} 
nullccomponents'=b^{-signature(nullccomponents)}nullccomponents
\end{equation} 

where $nullccomponents$ stands for a null component among $(\al, \underline{\al}, \bet, \underline{\bet}, \sigma, \rho)$ and $signature(nullccomponents)$ stands for its signature.

We have moreover the following relations :
\begin{eqnarray}
\chi^{(g)}&=& \chi' \\
\al^{(g)}&=& \al'
\end{eqnarray}

Let us now do some remarks about the differences between the formulation of the structure equations relative to the $(t,u)$-foliation and the one relative to the geodesic foliation. The first remark is that these equations seem a little bit more complicated due to the presence of additional terms. Moreover we have to keep trace of the differences between the three quantities : $V$, $\h$ and $-\underline{\h}$. This was not the case in the formulation of \cite{krc}. But as soon as we will have some control on the geometry, we will have :

\begin{itemize}
\item Estimates for the following quantities : $\e$, $\hb$, the restriction of $k$, $\de$ and $\deb$.
\item The estimates on $\chi$ imply some estimates on $\underline{\chi}$ due to the relation :

$$\chi + \chiu=-2k$$

\end{itemize}

\subsection{Bianchi equations and mass equations :}

Following \cite{ck}, we have the following commutation formulas : 

\begin{thm}

We have the following commutations formulas for an arbitrary $m$-covariant $S$ tangent vector-field $\bo{\pi}$ :

$$\nabla_B \nabla_L \bo{\pi}_{A_1,....A_m} -   \nabla_L \nabla_B \bo{\pi}_{A_1,....A_m} = \chi_{BC} \nabla_C \bo{\pi}_{A_1,....A_m}-n^{-1}\nabla_B n \nabla_L\bo{\pi}_{A_1,....A_m}$$
$$+ \sum_i \big(\chi_{A_i B} \ebaa_C - \chi_{BC} \ebaa_{A_i} + \bo{R}_{C A_i 4 B} \bo{\pi}_{A_1,.... C,....A_m} \big)$$

In particular for a scalar $f$ :

$$\nabla_B \nabla_L f -   \nabla_L \nabla_B f = \chi_{BC} \nablab_C f-n^{-1}\nabla_B n \nabla_L f$$

For a one-form $F$ :
$$L(div F)-div(\nabla_L F)= - \chi \cdot\nabla F -n^{-1} \nablab n \cdot \nabla_L F - \frac{1}{2}tr \chi \cdot \ebaa \cdot F - \hat{\chi} \cdot \ebaa F - \beta F$$

\end{thm}

$\mathit{Remark :}$ Contrary to the case of geodesic foliation we do not have :

$$\hb^{(g)} + \zeta^{(g)} =0$$

which implies in the preceeding formulas the presence of  a term involving $\nabla_L F_{A_1,....A_m}$ which itself implies the presence of additional terms in every formula or calculation that we transpose from \cite{krc}. 

Let us now see the additional terms in the following equations which are special forms of Bianchi equations :

\begin{thm} {\bf Bianchi equations} We have the following Bianchi equations :

\begin{eqnarray*}
\nabla_L \bet + 2 tr \chi \bet&=& div \alpha - \deb \beta + \alpha \cdot (2 \e + \hb)+{\bo D}_4{\bo R}_{a4}-{\bo D}_a{\bo R}_{44}+ \demi tr(\chi){\bo R}_{4.}- \hat{\chi}_{.}^b{\bo R}_{4b} \\
L(\rho) + \frac{3}{2} tr \chi \rho &=& div \bet -\frac{1}{2}\underline{\hat{\chi}} \cdot \alpha + \e \cdot \bet + 2 \hb \cdot \beta+\demi {\bo D}_{4}{\bo R}_{34}-\demi {\bo D}_{3}{\bo R}_{44}-\demi \hat{\chi}^{ab}{\bo R}_{ab} + \frac{1}{4} tr\chi{\bo R}_{43}\\
&&-\frac{1}{4}tr\underline{\chi}{\bo R}_{44}\\
L(\sigma) +  \frac{3}{2} tr \chi \sigma &=& - curl \bet + \frac{1}{2} \hat{\underline{\chi}} \cdot \alpha - \e \wedge \beta-2\hb \wedge \bet-\demi \hat{\chi}\wedge {\bo R} + \hb \wedge {\bo R}_{4.}\\
\nabla_L \underline{\bet} + tr \chi \underline{\bet}&=& -\nabla \rho + {}^{*}\nabla \sigma + 2  \hat{\underline{\chi}} \cdot \bet + \deb \cdot \underline{\beta}-3\hb \cdot \rho +3 \hb{}^{*}\sigma + {\bo D}_{3}{\bo R}_{4.} + \frac{1}{4}tr\underline{\chi}{\bo R}_{4.}  \\
&&-\frac{1}{4}tr\chi{\bo R}_{3.}-\demi\underline{\hat{\chi}}\cdot {\bo R}_{4.}+\demi \hat{\chi}\cdot {\bo R}_{3.}-\hb^b{\bo R}_{b.}+ \demi \hb {\bo R}+ \hb {\bo R}_{34}
\end{eqnarray*}
\end{thm}

$\mathit{Remark :}$ Let us remark that in comparison to \cite{krc}, we have additional terms. These ones come from (in \cite{ck} notations) :

$$\Omega=\frac{1}{4}\langle \bo{D_L}L,\Lb \rangle=\frac{1}{2}\deb$$

This term is of course vanishing in the case of the geodesic foliation. Again the additional terms are not so bad because of our control on $\deb$.

{\it Remark :} Contrary to \cite{krc}, we do not need to renormalize Bianchi equation because of the better control that we have on every term involving $\chiu$.

We will moreover use the reduced mass $\mub$ :

\begin{equation}
\mub=-div \hspace{.5mm}\h + \frac{1}{2}\hat{\chi}\cdot \underline{\hat{\chi}} -\rho + \frac{1}{2} \hspace{.2mm} \delta \hspace{.2mm} tr\chi - \abs{\h}^2 -\demi{\bo R}_{43} 
\end{equation}

The goal of studying such a quantity is as we will see later to close the 2D Hodge system on $\h$ for which we do not have any a priori estimates (due to the dependance on $b$) contrary to $\hb$. We can derive a transport equation for such a quantity (see \cite{ck} p.363).  

\begin{thm}{\bf Transport equation for the reduced mass :} The reduced mass $\mub$ satisfies the following transport equation :
\begin{eqnarray*}
L(\mub) + tr\chi \mub&=&  2 \hat{\chi} \cdot \nabla \h + (\h-\hb)(b\nabla(b^{-1} tr\chi) + tr \chi \h)\\
&& -\demi tr \chi (\hat{\chi}\cdot \hat{\chiu}-2 \rho + 2 \hb \cdot \h) + 2 \h \cdot \hat{\chi}\cdot \ \h -  \delta\abs{\hat{\chi}}^2 -\demi \deb (tr \chi)^2 - \demi {\bo D}_{3}{\bo R }_{44} \\
&&+2 \zeta^a{\bo R}_{a b} - \frac{1}{4}tr\chi{\bo R}_{34}-tr{\underline \chi}{\bo R}_{44}
\end{eqnarray*}
\end{thm}

{\it Remark :} We do not use the same mass as in \cite{krc}. The main problem is the presence of a term involving $div \hspace{.2mm} \underline{\h}$ for which we do not have a priori estimate (see the trace theorems of part 1).

In order to estimate the full derivative of $\chi$ we will need in addition the following transport equation :

\begin{eqnarray*}
\nabla_L(\nabla tr \chi) &=& -\frac{3}{2}tr \chi \nabla tr\chi-\hat{\chi} \cdot \nabla tr \chi - 2 \hat{\chi}\cdot \nabla\hat{\chi}-\deb \nabla tr \chi - \deb tr \chi \\
&&- n^{-1}\nabla n \big(\abs{\hat{\chi}}^2+ \demi (tr\chi)^2 + \deb tr \chi)- {\bo D}_{.}{\bo R}_{44}-2 \chi_{.}^b{\bo R}_{4b}-\nabla (log n){\bo R}_{44}
\end{eqnarray*}

Again the second line of this transport does not vanish contrary to the case of the geodesic foliation and this fact creates additional problems.

Let us now define the appropriate norms and quantities needed to state our result.

\subsection{Definitions :}

We will introduce the appropriate norms for our problem but before we need to recall the following lemma :

\begin{lem}
Relative to the $(t,u)$ foliation on $\Hyp$, we have the following relation for a scalar function $f$,

$$\frac{d}{dt}\int_{S_{t,u}} f dA_{t,u} = \int_{S_{t,u}} \big(\frac{d}{dt}f  + n tr \chi f\big)dA_{t,u}$$

where $dA_{t,u}=dA_{\gamma_s}$ denotes the volume element on $S_{t,u}$, with $\gamma_{t,u}$ being the induced metric on $S_{t,u}$. In particular, if we introduce the area $A(t,u)$ of the 2-surface and the radius $r(t,u)$ defined by :
$$A(t,u)=4\pi r^2(t,u)$$
we have along a given $C_u$ :
$$\frac{d A}{d t}= \int_{S_{t,u}} n tr \chi dA_{t,u}$$

and along $C_u$
$$\frac{dr}{d t}=\frac{r}{2} \overline{ntr\chi}$$
\end{lem}

where we have used the following notation for the mean value on a surface $S_{t,u}$ 

$$ \overline{f(t,u)}=\frac{1}{A(t,u)}\int_{S_{t,u}} f dA_{t,u}$$.

The presence of the lapse $n$ has of course no influence due to the pointwise control on the lapse that we will have.

We will also make use of the corresponding lemmas with respect to the affine parameter $s$ relative to $L=\bo{T}+ \bo{N}$ and $s'$ the geodesic affine parameter (thus relative to $L'=b^{-1}L$).

Throughout this paper we will consider a null hypersurface which for our application (see section 1) will be a truncated null cone initiating on a initial 2-surface which will be supposed to be weakly spherical (see section 1) and with a radius $r(0)\geqslant 1$.

\subsection{Assumptions :}

We will consider in this section a truncated null cone $\Hyp$ foliated by our $t$-function into 2 dimensional compact surfaces. We will thus suppose that :

$$\Hyp=\bigcup_{t\in[t_0,t_0+T]}S_t$$.

(We will now omit the subscript $u$ as we work on a fixed $C_u$).

Let us now describe what will be our assumptions on the initial surface and its geometry :

\vspace{5 mm}

{\bf Assumptions on the initial surface :}

We suppose that the inital surface $S_0$ satisfies the following properties :

\begin{itemize}
\item {\bf (HI1)} $S_0$ W-is weakly spherical with constants smaller that  $\mathcal{I}_0$ and $r(0) \rs1$ 

\item {\bf (HI2)} $\nli{tr \chi -\frac{2}{r}}{\infty} + \| \nabla tr \chi \|_{B_{2,1}^0(S_0)} +  \| \tilde{\mu} \|_{B_{2,1}^0(S_0)}\ls \mathcal{I}_1$
\item {\bf (HI3)} there exists a constant $C$ such that :

$$C^{-1}\ls b(\omega)\ls C \hs \text{for all } \omega \in S_0$$

\end{itemize}

\vspace{5 mm}
{\it Remarks :} 

$\al$) The only true hypothesis is {\bf HI2}. Up to a scaling argument {\bf HI1} is no longer restrictive. {\bf HI3} is just a matter of normalization of $b$. As we choose to be consistent with section 1 where $b(p)$ is normalized to $n(p)$ at the vertex, we have to add this hypothesis which of course disapears when adapting the proof from the truncated null cone case to the null cone with vertex case.

$\bet$) In comparison with \cite{krc}, we have to put a condition of W-weakly spherical which has no importance, it is just due to the relation along null geodesics :

$$\frac{dt}{ds'}=(nb)^{-1}$$

which with the pointwise bound on $n$ (see {\bf (HST1)}) force us to change the range allowed for $R$ in the definition of weakly spherical surfaces.

$\gamma$) Of course the main difference in the main hypothesis {\bf HI2} lies in the fact that we do not need initial assumption on $\underline{\chi}$ which are in fact automatic due to {\bf (HST2)} and the relation between $\chi$ and $\underline{\chi}$.

Moreover we assume the following space-time controls :

\vspace{5 mm}

{\bf Space-time assumptions :} 

\begin{itemize}

\item {\bf (HST1)} $\normf{n}_{L^\infty([0,T],L^\infty(\Sigma_{t}))}+\normf{n^{-1}}_{L^\infty([0,T],L^\infty(\Sigma_{t}))} \ls \Delta_1$

\item {\bf (HST2)}   $\int_0^{t_1}(\normf{k}_{L^\infty(\Sigma_{t})}+ \normf{\nabla(log(n))}_{L^\infty(\Sigma_{t})})^2ndt \ls \Delta_2$

\end{itemize}

{\it Remark :} We have seen that in the case of our application for the breakdown criterion, if we suppose that :
\begin{equation}
\int_{\Sigma_{0}} \abs{\bo{R}}^2(x)d\mu(x) \ls \Delta_0
\end{equation}

then the following reduced flux is also finite :
\begin{equation}
\mathcal{R}(T)=\big( \ntsl{\al}{2}{2}^2 +  \ntsl{\bet}{2}{2}^2 +  \ntsl{\sigma}{2}{2}^2 +  \ntsl{\rho}{2}{2}^2 +  \ntsl{\underline{\bet}}{2}{2}^2\big)^\demi
\end{equation}

We recall that two of the null components out of ten are missing, the two coefficients "coded" by $\underline{\al}$. 

We will also note :

\begin{equation}
\mathcal{R'}(0)=\int_{M_{S_0}}\abs{\bo{R}}^2d\mu_0
\end{equation}

this integral is taken over $M_{S_0}$ as defined in section 1. Let us also define :

\begin{equation}
\mathcal{R''}(T)=\int_{t_0}^{t_0+T} (\normf{k}_{L^\infty(\Sigma_{t})}+ \normf{\nabla(log(n))}_{L^\infty(\Sigma_{t})})^2dt 
\end{equation}

In the formulation of the theorem, we will make the assumption that $T=1$. That is :
$$\Hyp=\bigcup_{t\in[t_0,t_0+1]}S_t$$

We will note $\mathcal{R}= max (\mathcal{R}(1),\mathcal{R'}(0),\mathcal{R'''}(1))$ and  $\mathcal{I}:= max (\mathcal{I}_0,\mathcal{I}_1)$.

\vspace{3mm}
We will prove that if $\mathcal{R}$ and $\mathcal{I}$ is small enough, we can control the geometry of $\Hyp$.
In our application this quantity is not small, but only finite. A reformulation a our theorem states that we can control geometry of the troncated null cone for $t-t_0$ small enough and depending only on the fundamental constants $t_\star$ and $\Delta_i$.

\subsection{Statement of the main theorem }

\begin{thm}{\bf Main theorem}
Consider a troncated null cone $\Hyp$ verifying ${\bf (HST1)}$ and ${\bf (HST2)}$. Suppose that $\Hyp$ is foliated by our time function $t$ and that it  initiates on a 2-surface $S_0$ (corresponding to $t=t_0$) diffeomorphic to $\bo{S^2}$ and verifying the hypothesis ${\bf (HI1)}$, ${\bf (HI2)}$, ${\bf (HI3)}$.  Suppose moreover that  $\mathcal{R}$ and $\mathcal{I}$ are small enough, then we have the following estimates :

Primary estimates :
\begin{eqnarray}
\ntsl{tr \chi - \frac{2}{s}}{\infty}{\infty} & \lesssim & \mathcal{R} + \mathcal{I} \\
\|\nabla tr \chi\|_{\mathcal{B}^0}+ \| \mu \|_{\mathcal{B}^0} & \lesssim & \mathcal{R} + \mathcal{I}  \\
\nli{\int_0^1\abs{\hat{\chi}}^2}{\infty} + \nli{\int_0^1\abs{\h}^2}{\infty} & \lesssim & \mathcal{R} + \mathcal{I} \\
\nli{\sup_{t\in[0,1]}\abs{\hat{\chi}}}{2} + \nli{\sup_{t\in[0,1]}\abs{\mu}}{2} &\lesssim& \mathcal{R} + \mathcal{I} \\
\mathcal{N}_1(\hat{\chi})+\mathcal{N}_1(\h)+\mathcal{N}_1(tr\chi-\frac{2}{s})& \lesssim &  \mathcal{R} + \mathcal{I} \\
\end{eqnarray}
Secondary estimates :
\begin{eqnarray}
\left\| tr \underline{\chi} + \frac{2}{s}\right\| _{L^{\infty}_{[0,1]}L^\infty_x +L^{\infty}_{[0,1]}H^\demi_x}  & \lesssim &  \mathcal{R} + \mathcal{I}\\
\|\hat{\chiu}\|_{L^{\infty}_{[0,1]}H^\demi_x} & \lesssim &  \mathcal{R} + \mathcal{I}\\
%\ntsl{\nabla tr \chiu}{2}{2}&\lesssim &  \mathcal{R} + \mathcal{I}%
\end{eqnarray}
\end{thm}

{\it Remarks :} 

\begin{itemize}
\item These results should be compared to the result in \cite{krc}. In our case, we obtain better estimates for $\chiu$. In fact, the estimates on $\chi$, $\h$ and $\m$ come exactly as in \cite{krc} and the estimates on $\chiu$ come as consequences of the global structure properties and the primary estimates. 
\item The goal is clearly to use as often as possible the a priori estimates. However, as it was mentioned in the end of section 1, we need the a priori estimates in their "trace formulation". But we need a minimum control on $tr \theta$ to apply these results. And we do not have this control a priori. So we will have to apply the trace theorems into the bootstrap argument, as it is the case for the sharp trace theorems of \cite{krs}.
\end{itemize}

\subsection{Strategy of the proof :}
\subsubsection{Derivation of the estimates :}

\hspace{5mm}Starting with the transport equation on $\chi$, we see that if we want to control $\chi$ in $L^\infty$, we have to control the trace term :
\begin{equation}
\int_0^t \abs{\hat{\chi}}^2n dt + \int_0^t \abs{{\bo R}_{44}}
\end{equation}

The first idea consists in using a trace theorem on $\Hyp$. By comparison with the standard trace theorems, we could expect :

\begin{equation}
\int_0^t \abs{\hat{\chi}}^2n dt \lesssim \norme{\nabla \hat{\chi}}^2_{L^2(\Hyp)}
\end{equation}

Such a trace theorem together with the Codazzi equation on $div \hspace{.7mm}\hat{\chi}$ could lead with the help of elliptic estimates for Hodge systems on the leaves to the control of $\norme{\nabla \hat{\chi}}^2_{L^2(\Hyp)}$ and 
$\int_0^t \abs{\hat{\chi}}^2n dt $. However, this trace theorem fails to be true if we do not add structure conditions on $\hat{\chi}$. Here is the main idea of \cite{krc}. If $\hat{\chi}$ can be written under the following form, then we have the type of trace theorem mentioned before :
\begin{equation}
\hat{\chi}= \nabla_LQ +E
\end{equation}
where we control roughly two derivatives of $Q$ and $E$ is an error term.
The next step consists in proving that $\hat{\chi}$ is of this form. To do that we recall that we have :

\begin{eqnarray*}
L(\rho) + \frac{3}{2} tr \chi \rho &=& div \bet -\frac{1}{2}\underline{\hat{\chi}} \cdot \alpha + \e \cdot \bet + 2 \hb \cdot \beta + F({\bo R}, {\bo D}{\bo R})\\
L(\sigma) +  \frac{3}{2} tr \chi \sigma &=& - curl \bet + \frac{1}{2} \hat{\underline{\chi}} \cdot \alpha - \e \wedge \beta-2\hb \wedge \bet + G({\bo R}, {\bo D}{\bo R})
\end{eqnarray*}

It implies that $\beta = \dope_1^{-1}L(\rho, -\sigma)$ + other terms where $\dope_1$ stands for an Hodge operator (see below).

So $\dope_2(\hat{\chi}):= div \hspace{.7mm} \hat{\chi} = -L \dope_1^{-1}(\rho, -\sigma) + [L,\dope_1^{-1}](\rho, -\sigma) + ....$. It follows from this approach of the problem that we have to control commutators. Moreover a typical problem will come from terms of the type : 

\begin{equation}
\int_0^t \abs{\dope_2^{-1}\nabla tr \chi}^2 n dt
\end{equation}

The difficulty comes here from the non-locality of $\dope_2^{-1}\nabla$ which leads to consider the besov space $B^1_{2,1}(S)$ instead of $L^\infty(S)$.

To close the estimates for $\chi$, we need to control simultaneously $\h$. In fact $(\mu,\h)$ satisfies a coupled transport-Hodge system similar to the one satisfied by $(\nabla tr \chi, \hat{\chi})$. The structure of the derivation of the control on these quantities is quite parallel to the one we have just described. 

\subsubsection{Differences with \cite{krc} :}

\hspace{5mm}We will follow step by step the original paper \cite{krc}. Let us list however the differences :
\begin{itemize}
\item {\bf Simplifications :} The main source of simplification comes from the link between $\chi$ and $\chiu$. It implies in particular that we do not need bootstrap assumptions on $\chiu$. The second source of simplifications is the estimates that we will get once we will have a minimal control on the geometry of the surfaces. However, these simplifications are needed in the sense that in comparison with the structure equations relative to the geodesic foliations, we have more unknown in our setting.

\item {\bf Difficulties :} This number of unknown in the structure equations is the first additional difficulty. From another hand, we mentioned that we will have to prove estimates for commutators. But in every commutator calculation, our setting produces more terms that in the case of the geodesic foliation. We will thus have additional terms almost everywhere some of them requiring special estimates. Moreover as explained before, the estimates on some of the terms (see end of section 1) will in fact come into the bootstrap as it is the case for the application of the sharp trace theorems of \cite{krs}. This idea is something new in comparison to \cite{krc}.
\end{itemize}

\subsection{Hodge operators :}
\hs We have already used Hode systems in section 1 to derive estimates for $k$ on the leafs $\Sigma_t$. The use of 2 dimensional Hodge systems on the $S_{t,u}$ will be of great use in the proof of the theorem. For instance (2.17) and (2.34) form a 2 dimensional system for $\h$.

 Following for this part exactly \cite{krc}, we introduce the following notations :

\vspace{3mm}

$\mathbf{Notations}.$ We consider the following Hodge operators acting on a $S_{t}$ :
\begin{itemize}
\item The Hodge operator $\dope_1$ takes an 1-Form F into the pair of scalar functions (div F, curl F).
\item The operator $\dope_2$  takes any $S$-tangent symmetric, traceless tensor $F$ into the $S$ tangent one form div F.
\item ${}^{\star}\dope_1$ takes the pair of scalar functions $(\rho,\sigma)$ into the $S$-tangent one form $-\nabla \rho + (\nabla \sigma)^{\star}$.
\item ${}^{\star}\dope_2$ takes  1 form $F$ on $S$ into the 2-covariant, symmetric traceless tensor $-\demi\hat{\mathcal{L}}_F\gamma$ the traceless part of the Lie derivative of the metric $\gamma$ relative to $F$.
\end{itemize}
We recall the following properties of these operators :

\begin{itemize}
\item ${}^{\star}\dope_1$ is the $L^2$ adjoint of $\dope_1$ and ${}^{\star}\dope_2$ is the $L^2$ adjoint of $\dope_2$.
\item The kernels of $\dope_1$ and $\dope_2$ in $L^2$ are trivial.
\item The kernel of ${}^{\star}\dope_1$ consists in the pairs of constant functions.
\item The kernel of ${}^{\star}\dope_2$ consists in the conformal Killing vector fields on $S$
\item The $L^2$ range of $\dope_1$  consists of all pairs of functions on $S$ with vanishing mean curvature.
\item The $L^2$ range of $\dope_2$  consists of all pairs of all $L^2$ integrable 1-form on $S$ orthogonal to the Lie algebra of all conformal Killing vector fields on $S$.
\end{itemize}

Moreover we have the following relations where K stands for the Gauss curvature of $S_t$ :
\begin{eqnarray*}
{}^{\star}\dope_1 \cdot \dope_1&=& - \Delta + K \\
{}^{\star}\dope_2 \cdot \dope_2&=& - \demi \Delta + K \\
\dope_1{}^{\star}  \cdot  \dope_1&=& - \Delta\\
\dope_2{}^{\star}  \cdot  \dope_2&=& - \demi \big(\Delta+K \big)
\end{eqnarray*}

\subsection{Bootstrap assumptions and consequences :}
We assume that there exists of sufficiently small constant $M_0$, $0<M_0<\demi$ such that :

Assumptions {\bf (B1)} :

\begin{equation*}
\sup_\Hyp r \abs{\overline{b^{-1}tr \chi}- \frac{2}{r}}\ls M_0, \hs \sup_\Hyp r \abs{b^{-1} tr \chi -\overline{b^{-1}tr \chi} }\ls M_0
\end{equation*}

Assumptions {\bf (B2)} :

\begin{eqnarray*}
\nl{\hat{\chi}}{\infty}{2}, \nl{\h}{\infty}{2} &\ls& M_0\\
\nl{\mu}{2}{\infty},\nl{\nabla tr \chi}{2}{\infty} &\ls& M_0\\
\mathcal{N}_1(\hat{\chi}), \hspace{2 mm} \mathcal{N}_1(\hat{\h}) &\ls& M_0
\end{eqnarray*}

{\it Remark :} There are different ways of writing {\bf (B1)} depending on the application we have in mind. For instance the first part could be written as a control on $\abs{\overline{tr\chi}-\frac{2}{s}}$ is we want to use an equation in $s$. Here we use this formulation because we have in mind an equation in $s'$. The reader as to keep in mind that we will write equations along null geodesics and that due to the following relations :
\begin{eqnarray}
\frac{dt}{ds}&=&n^{-1}\\
\frac{dt}{ds'}&=&(nb)^{-1}
\end{eqnarray}

equations can be rewritten in the formulation the most useful for us. Moreover, we will have uniform from above and below for the two lapses $n$ and $b$ thus the constants in the transport lemmas can be taken uniformly bounded.

We have the following comparison result :

\begin{thm}{\bf Comparison theorem}
We have the following estimates on $\Hyp$ :

\begin{eqnarray}
\abs{b^{-1}r-s}&\lesssim& \mathcal{R} + M_0 \\
\abs{n-b}&\lesssim& \mathcal{R} 
\end{eqnarray}

And as a consequence we have :

\begin{eqnarray*}
\sup_\Hyp r \abs{\overline{tr \chi}- \frac{2}{s}}\ls M_0 + C \mathcal{R}, \hs \sup_\Hyp r \abs{ tr \chi -\overline{tr \chi} }\ls M_0 + C \mathcal{R}
\end{eqnarray*}
\end{thm}

{\bf Proof :}
Let us consider :
\begin{equation}
L'(b^{-1}r-s)=-\deb b^{-2}+b^{-1}\overline{b^{-1}tr\chi}\frac{r}{2} -b^{-1}=  -\deb b^{-2}+b^{-1}\frac{r}{2} \big(\overline{b^{-1}tr\chi}-\frac{2}{r} \big)
\end{equation}

Thus integrating in $s'$ from the vertex and using the pointwise bounds on $b$ gives (203).

For (204) just remark that $n(p)=b(p)$ by normalization, moreover on $\Hyp$ :

\begin{eqnarray}
\abs{n(q)-n(p)}&\lesssim& \mathcal{R}\\
\abs{b(q)-b(p)}&=&\abs{b(p)(e^{-\int_p^q - \deb ds}-1)} \lesssim \mathcal{R}
\end{eqnarray}

where the integral is taken on a null geodesic from $p$ to $q$.

\vspace{5mm}

We have also the following :

\begin{thm}
As a consequence of {\bf (B1)} and {\bf (B2)}, we have that for every $t_0 \ls t\ls T \ls t_0 + 1$, the volume elements $\abs{\gamma(t)}$ and $\abs{\gamma(0)}$ stay comparable, moreover, the $S_{t}$ are uniformly weakly spherical.
\end{thm}

{\bf Proof :}

We have the following relation :
\begin{equation}
\frac{dr}{ds'}=1+ \frac{r}{2}(\overline{b^{-1}tr \chi}-\frac{2}{r})
\end{equation}

this implies ,

\begin{equation}
\abs{r(s')-r_0-s'} \ls s' M_0
\end{equation}

thus :
\begin{equation}
r_0+ (1-\demi M_0)s' \ls r \ls r_0+ (1+ \demi M_0)s' 
\end{equation}

but along $C_u$:

\begin{equation}
\frac{ds'}{dt}=nb
\end{equation}
which implies that $s'$ is bounded in $\Hyp$ by a constant. If we suppose  without loss of generality $0\ls M_0 \ls \demi$, we get :
\begin{equation}
r_0+ \frac{3}{4}s' \ls r \ls r_0+ \frac{5}{4}s' 
\end{equation}

From another hand :
\begin{equation}
\abs{b^{-1}tr\chi-\frac{2}{r}}\ls 2 \frac{M_0}{r}
\end{equation}

This implies that for all $t \in [0,1]$ :
\begin{equation}
\frac{1}{r_0 + 2 s'}\ls b^{-1}tr \chi \ls  \frac{3}{r_0 + \demi s'}
\end{equation}
 but we have $Q^{-1}(t-t_0)\ls s' \ls Q (t-t_0)$, thus :
 \begin{equation}
\frac{1}{r_0 + 2Q(t-t_0)}\ls b^{-1}tr \chi \ls  \frac{3}{r_0 + \demi Q^{-1} (t-t_0)}
\end{equation}
 
but $\frac{d}{dt}(log \sqrt{\abs{\gamma}})=2n tr\chi$ and thus using pointwise bounds for $n$ and $b$, there exists $G$ such that :

\begin{equation}
G^{-1}\ls \frac{\abs{\gamma(t)}}{\abs{\gamma(0)}} \ls G
\end{equation}

We have now to prove that the surfaces $S_t$ stay uniformly weakly spherical.
We introduce the transported local coordinates $(s',\omega)$.
We have :

\begin{equation} 
\frac{d}{ds'}\gamma_{ab}=2\chi_{ab}
\end{equation}
then :

\begin{eqnarray*}
\frac{d}{ds'}\partial_c \gamma_{ab}&=& \partial_c(tr \chi\cdot\gamma_{ab}+2 \hat{\chi}_{ab})\\
&=& \nabla_c tr \chi \gamma_{ab}+ tr \chi \cdot \partial_c \gamma_{ab}+2  \partial_c \hat{\chi}_{ab}\\
&=& \nabla_c tr \chi \gamma_{ab}+ tr \chi \cdot \partial_c \gamma_{ab}+2  \nabla_c \hat{\chi}_{ab}+ \partial \gamma \cdot\hat{\chi}
\end{eqnarray*}
denoting schematically  $g:= \nabla \chi+ \partial \gamma \cdot\hat{\chi}$ and $v_t=\frac{\sqrt{\abs{\gamma_t}}}{\sqrt{\abs{\gamma_0}}}$ :

\begin{equation}
\partial\gamma(s',\omega)=v_s\big( \partial\gamma(0,\omega) + \int_0^s v_r^{-1}g(r,\omega)dr \big)
\end{equation}
then using the uniform control on $v_s$ and the fact that $s'$ is bounded : 
\begin{eqnarray*}
\nl{\partial\gamma - \partial v_s \gamma(0)}{2}{\infty} &\lesssim& \nl{\nabla \chi }{2}{1}+  \nl{\partial \gamma \hat{\chi} }{2}{1}\\
&\lesssim& M_0+M_0 \nl{\partial\gamma - \partial v_s \gamma(0)}{2}{\infty}
\end{eqnarray*}

with our assumption on $M_0$ and on $S_0$ together with the control on $v_s$ gives :
\begin{equation}
\nl{\partial\gamma - (1+s')\gamma(0)}{2}{\infty}\lesssim\mathcal{I}+M_0
\end{equation}

This estimate together with the uniform control on the components of $\gamma_t$ that we have proved gives the uniformity in the weakly spherical property.
\vspace{5mm}

As a consequence given  an arbitrary $S$-tensor on $\Hyp$ the following quantities are equivalent :

$$\big(\int_0^T n dt\int_{S_{t,u}}\abs{F}^2dA_{t,u} \big)^{\demi} \hs \text{and} \hs \big(\int_0^T  dt\int_{S_{0}}\abs{F(t,\omega}^2dA_{0} \big)^{\demi}$$

We can of course put or omit the lapse in the definition of this quantity later on referred as $\ntsl{F}{2}{2}$. The same remark can of course been made for the other Lebesgue norms. We also define for $1\ls p < \infty$ :

\begin{eqnarray*}
\ntsl{F}{\infty}{p}&=&\sup_{t\in[0,T]} \nli{F(t)}{p}\\
\ntsl{F}{2}{p}&=&\big( \int_0^T \nli{F(t)}{p}ndt\big)^{\demi}\\
\nl{F}{\infty}{2}&=&\sup_{\omega \in S_0}\big( \int_0^T \abs{F(t,\omega)}{2}ndt\big)^{\demi}\\
\nl{F}{2}{\infty}&=&\nli{\sup_{t\in[0,T]} \abs{F(t,\omega)}}{2}
\end{eqnarray*}

Sadly we will also have to use Sobolev and Besov norms due to the non locality of some crucial operators. We have to recall that as we are working with 2D compact surfaces with low regularity, the standard theory using local coordinates does not work and these type of utilisations deeply requires the use of the geometric Littlewood-Paley theory developed in \cite{krg}. The results of this article useful in the rest of this section can be found in the appendix of section 1.

We introduce also the following notations :

\begin{nota}
For an arbitrary $S$-tangent tensor $F$ on $\Hyp$ :
$$\nd{\overline{\nabla}F}=\nd{\nabla F}+\nd{\nabla_L F}$$
$$\mathcal{N}_1(F)=\nd{r^{-1}F}+\nd{\overline{\nabla}F}$$
$$\mathcal{N}_2(F)=\nd{r^{-2}F}+\nd{r^{-1}\overline{\nabla}F}+\nd{\nabla \overline{\nabla}F}$$
\end{nota}

Let us recall the following lemma :

\begin{lem}Let us consider the following transport equation $\nabla_L F+k tr\chi F = G$ for S-tangent tensors $F,G$ on $\Hyp$ and $\lim_{t\rightarrow 0}v(t)^kF=0$. Then, for any $1\ls p \ls \infty$,
$$\nl{F}{p}{\infty} \lesssim \nli{F(0)}{p} + \nl{G}{p}{1}$$
\end{lem}

{\bf Proof :}
we have :
\begin{equation}
\frac{d}{dt}\abs{F}^2+kntr\chi=2nG\cdot F
\end{equation}

If we denote by $v_t=\frac{\sqrt{\abs{\gamma_t}}}{\sqrt{\abs{\gamma_0}}}$, we have :

\begin{equation}
\abs{F}^2(t,\omega)=\lim_{s \rightarrow 0}v_t(\omega)^{-2k}\big(\abs{F}^2(s,\omega) + \int_s^t v_{t'}(\omega)^{2k}G \cdot F n dt' \big)
\end{equation}

using the bounds previously proved :

\begin{equation}
\abs{F}^2(t,\omega)\lesssim \abs{F}^2(0,\omega) + \int_0^t \abs{G} \cdot \abs{F} n dt'
\end{equation}

whence :
\begin{equation}
\abs{F}(t,\omega)\lesssim \abs{F}(0,\omega) + \int_0^t \abs{G}  dt'
\end{equation}

whence :
\begin{equation}
\sup_{[0,1]} \abs{F}(t,\omega)\lesssim \abs{F}(0,\omega) + \int_0^1 \abs{G}  dt'
\end{equation}

\vspace{5mm}

From another hand integrating (161) along the null geodesics and using the initial assumptions on $S_0$ there exists $D(C,\Delta_2)$ such that :

$$D(C,\Delta_2)^{-1}\ls b(p)\ls D(C,\Delta_2)\hs \text{for all } p\in \Hyp$$

Then we will also need the following lemma :

\begin{lem}Let us consider the following transport equation $\nabla_L F+k tr\chi F \pm \deb F= G$ for S-tangent tensors $F,G$ on $\Hyp$ and  $\lim_{t\rightarrow 0}v(t)^kF=0$. Then, for any $1\lesssim p \lesssim \infty$,
$$\nl{F}{p}{\infty}\lesssim  \nl{G}{p}{1}$$
\end{lem}
$\bo{Proof :}$ To prove Lemma 2 it is sufficient to apply Lemma (2.10.1) to $b^{\mp 1}L$ and using (2.12).

The theorem (2.10.2) allows us to apply the geometric Littlewood-Paley of \cite{krg} and in particular deduce the following form of Sobolev inequalities.

\begin{defi}
We will use the following notations :

\begin{itemize}
\item we denote by $R$ the full collection of null curvature components $\al$, $\bet$, $\rho$, $\sigma$, $\underline{\al}$, $\underline{\bet}$.
\item we denote by $R_0$ the collection of null curvature components $\bet$, $\rho$, $\sigma$, $\underline{\al}$, $\underline{\bet}$.
\item we denote by $A$ the collection  $tr\chi-\frac{2}{s}$, $\hat{\chi}$, $\h$
\item we denote by $C$ the collection $\nabla n$, $k_{(2)}$, $\e$, $\hb$, $\de$, $\deb$.
\item we denote by $\underline{A}$ the collection $tr\chi-\frac{2}{s}$, $\hat{\chi}$, $\h$ and $tr\underline{\chi}+\frac{2}{s}$, $\hat{\underline{\chi}}$
\item we denote by $M$ the collection  $\mu$, $\nabla tr \chi$.
\item we denote by $\nabla A$ the collection  $\nabla(tr\chi-\frac{2}{s})$, $\nabla \hat{\chi}$, $\nabla \h$
\end{itemize}
\end{defi}

{\it Remarks :} 

1) The elements of type $C$ are the ones for which we we will get estimates when we will have control on $tr \theta$ by application of the trace theorems.

2) contrary to \cite{krc} the coefficients $\underline{A}$ are not so bad as sums of terms of type $A$ and terms of type $C$.

\begin{thm}{\bf Relations on the surfaces $S_t$}
We have the following relations on the surfaces $S_t$ with constants not depending on $t\in [0,T]$, $T\ls1$. For an arbitrary scalar function $f$ and  an arbitrary tensorfield $F$, we have :

\begin{eqnarray}
\nlsl{f}{2}&\lesssim&\nlsl{ \nabla f}{1}+\nlsl{r^{-1}f}{1}\\
\nlsl{f}{\infty}&\lesssim&\nlsl{\nabla^2 f}{1}+\nlsl{r^{-2}f}{1}\\
\nlsl{F}{p}&\lesssim&\nlsl{\nabla F}{2}^{1-\frac{2}{p}}\nlsl{F}{2}^{\frac{2}{p}}+\nlsl{r^{-1+\frac{2}{p}}F}{2} \text{ $2\ls p < \infty$}\\
\nlsl{F}{\infty}&\lesssim&\nlsl{\nabla^2 F}{2}^{\demi}\nlsl{F}{2}^{\demi}+\nlsl{r^{-1}F}{2}
\end{eqnarray}
\end{thm}

On the  cone we have the following Sobolev type relations :

\begin{thm}{\bf Relations on the surface $\Hyp$} 
For an arbitrary S tangent tensorfield F and for $T \in [0,1]$, we have :
\begin{eqnarray}
\ntsl{F}{\infty}{4}&\lesssim&T^{-\demi}\ntsl{F}{2}{4} + \nl{F}{\infty}{2}\cdot \ntsl{r^{\demi}\nabla_L F}{2}{2}\\
\ntsl{F}{\infty}{4}&\lesssim&T^{-\demi}\big( \ntsl{r^{\demi}F}{2}{2} + \ntsl{r^{-\demi}\nabla F}{2}{2} \big) + \ntsl{r^{\demi}\nabla_L F}{2}{2}\\
\ntsl{F}{6}{6}&\lesssim&T^{-\frac{1}{3}}\big( \ntsl{r^{\frac{2}{3}}F}{2}{2} + \ntsl{r^{\frac{5}{3}}\nabla F}{2}{2} \big) +T^{\frac{2}{3}}\ntsl{r^{-\frac{2}{3}}\nabla_L F}{2}{2}\\
\ntsl{F}{\infty}{\infty}&\lesssim&T^{-\demi}\big( \ntsl{r^{-1}F}{2}{2} + \ntsl{r \nabla^2 F}{2}{2} \big)\\
\nonumber&&+T^\demi \big( \nl{\nabla_L F}{\infty}{2}+\ntsl{\nabla \nabla_L F}{2}{2}\big)
\end{eqnarray}
\end{thm}

As a consequence, we have the following :

\begin{thm}
As a consequence of the bootstrap assumptions, the elements of type $A$ verify the following estimates for $2\ls p<\infty$ :

$$ \nl{A}{\infty}{4}, \hs \nl{A}{6}{6} \lesssim M_0$$
and 
$$\nl{A}{2}{p} \lesssim M_0$$
\end{thm}

we also have :

\begin{thm}
The Gauss curvature $K_t$ of $S_t$ satisfies :
\begin{equation}
\ntsl{K_t-\frac{1}{s(t)^2}}{2}{2}\lesssim M_0 + \mathcal{R}
\end{equation}
\end{thm}

{\bf Proof :}

Let's begin by recalling the Gauss curvature equation :

\begin{equation}
K=-\frac{1}{4}tr\chi tr \chiu+\frac{1}{2}\hat{\chi}\cdot \hat{\chiu}-\rho +{\bo {Ric}}
\end{equation}
Due to the curvature term, we cannot hope better than a $L^2_x(L^2_t)$ estimate.

\begin{equation}
K-\frac{1}{s^2(t)}= \frac{1}{2r}A -\frac{1}{2r}\underline{A} - \frac{1}{4}A \cdot \underline{A} + R + {\bo {Ric}}
\end{equation}

with the previously defined notations.
We thus have :

\begin{eqnarray}
\ntsl{K-\frac{1}{s^2(t)}}{2}{2}&\lesssim& \ntsl{\underline{A}}{2}{2} +  \ntsl{A}{\infty}{4}\cdot \ntsl{\underline{A}}{2}{4} \\
&&+  \ntsl{R}{2}{2} +\ntsl{ {\bo {Ric}}}{2}{2} \\
&\lesssim& M_0+ M_0 \mathcal{R} + \mathcal{R}
\end{eqnarray}

\vspace{5 mm}

\begin{thm}
The Gauss curvature $K_t$ of $S_t$ satisfies for $\demi < a<1$ :
\begin{equation}
\ntsl{\Lambda^{-a}(K_t-\frac{1}{s(t)^2})}{\infty}{2}\lesssim M_0 + \mathcal{R}
\end{equation}
\end{thm}

{\bf Proof :}

We use the following non-sharp Sobolev embedding :

\begin{lem}

The following inequality holds for an arbitrary tensorfield $F$ uniformly on the $S_t$ with $2<p<\infty$, $s>1-\frac{2}{p}$ :
\begin{equation}
\|F\|_{L^p(S_t)}\lesssim \|\Lambda^s F\|_{L^2(S_t)}
\end{equation}
\end{lem}

Then we have for $\demi<a<1$ :
\begin{equation}
\ntsl{\Lambda^{-a}(K_t-\frac{1}{s(t)^2})}{\infty}{2}\lesssim \ntsl{\Lambda^{-a}\underline{A}}{\infty}{2} + \ntsl{\Lambda^{-a}A \cdot \underline{A}}{\infty}{2} + \ntsl{\Lambda^{-a}\rho}{\infty}{2}
\end{equation}

The control of the first term is not difficult :

\begin{equation}
\ntsl{\Lambda^{-a}\underline{A}}{\infty}{2}\lesssim\ntsl{\underline{A}}{\infty}{4}\lesssim M_0 + \mathcal{R}\end{equation}

For the second term :
\begin{equation}
\nlsl{\Lambda^{-a} A \cdot \underline{A}}{2}^2 \lesssim \nlsl{\underline{A}}{2} \nlsl{\Lambda^{-a}( A \cdot \underline{A})}{4} \nlsl{A}{4}
\end{equation}

The difficult term is $\ntsl{\Lambda^{-a}\rho}{\infty}{2}$ which requires the use of the Bianchi equation in $\rho$ :
\begin{equation}
L(\rho) + \frac{3}{2} tr \chi \rho = div \bet -\frac{1}{2}\underline{\hat{\chi}} \cdot \alpha + \e \cdot \bet + 2 \hb \cdot \beta
\end{equation}

and the following calculous inequality (see \cite{krc}) :
\begin{equation}
\ntsl{\Lambda^{-a}\rho}{\infty}{2} \lesssim \ntsl{\Lambda^{-2a}\frac{d}{dt}\rho}{2}{2} + \ntsl{\rho}{2}{2} + \ntsl{[\frac{d}{dt}, \Lambda^{-a}]\rho}{1}{2}
\end{equation}

but similarly to the control proved in \cite{krg}, we have :
\begin{equation}
\ntsl{[\frac{d}{dt}, \Lambda^{-a}]\rho}{1}{2}\lesssim (1+K_a^{\frac{1}{1-a}}+K_a^\demi)^\e \ntsl{\rho}{2}{2}
\end{equation}
 for a small $\e$.
 
 It remains to control $\ntsl{\Lambda^{-2a}\frac{d}{dt}\rho}{2}{2}$.

 We have :
 \begin{equation}
 \ntsl{\Lambda^{-2a}\frac{d}{dt}\rho}{2}{2}\lesssim \ntsl{div \hspace{2mm}\bet}{2}{2} +  \ntsl{\Lambda^{-2a}A\cdot \underline{A}}{2}{2} 
 \end{equation}
 
 we have already proved : 
 
\begin{equation}
\ntsl{\Lambda^{-2a}A\cdot \underline{A}}{2}{2} \lesssim M_0^2 + M_0 \mathcal{R}
\end{equation}
 
 But we have :

 \begin{lem}
For a one form F and $a \rs \demi$,
\begin{equation}
\ntsl{\Lambda^{-2a}div \hspace{2mm}F}{2}{2}\lesssim  \ntsl{F}{2}{2}
 \end{equation}
 \end{lem}
 
 Finally, we get :
 
 \begin{equation}
 K_a \lesssim M_0^2+M_0 \mathcal{R}+ M_0(1+ K_a^{\frac{1}{1-a}}+ K_a )^\e
 \end{equation}

and then if we take $\e$ small enough we get the desired estimate.

\vspace{5 mm}

\begin{thm}
As a consequence of the bootstrap assumptions we have the following estimates :
\begin{eqnarray}
\sup_{t \in [0,1]} \|k\|_{H^\demi(S_t)}& \lesssim&  \mathcal{R}\\
\sup_{t \in [0,1]}\|n\|_{H^{\frac{3}{2}}(S_t)} & \lesssim&  \mathcal{R}
\end{eqnarray}
\end{thm}

In particular it implies the following estimates :

\begin{eqnarray}
\ntsl{k}{\infty}{4}& \lesssim&  \mathcal{R}\\
\ntsl{\nabla n}{\infty}{4} & \lesssim&  \mathcal{R}\\
\ntsl{\underline{\h}}{\infty}{4}& \lesssim&  \mathcal{R}\\
\ntsl{\deb}{\infty}{4}& \lesssim&  \mathcal{R}
\end{eqnarray}

that is for a generic element of type $C$ :
\begin{equation}
\norme{C}_{L_t^\infty(H^\demi)} \lesssim \mathcal{R}
\end{equation}

We also have according to trace theorems :

\begin{prop}
For $\kappa$, $0\ls \kappa < \demi$, we have :
\begin{equation}
\norme{C}_{\mathcal{B}^\kappa} \lesssim \mathcal{R}
\end{equation}
\end{prop}

it implies directly the following :
\begin{prop}
For $\kappa$, $0\ls \kappa < \demi$, we have :
\begin{eqnarray}
\norme{\underline{A}}_{\mathcal{B}^\kappa} &\lesssim& \mathcal{R} + M_0\\
\norme{A \cdot \underline{A}}_{\mathcal{P}^\kappa} &\lesssim& \mathcal{R}M_0 + M_0^2
\end{eqnarray}
\end{prop}

{\bf Proof :} It's sufficient to apply the localized version of the trace estimates of the end of section 1 and remark that the hypothesis on $tr\theta$ is a consequence of the a priori estimate for $k$ (consequence of the $3D$ Hodge estimates on the leaves) and the estimate of $tr\chi$ consequence of the bootstrap assumptions. 

{\it Remark :} We remark that the terms $\underline{A}$ are more regular that in \cite{krc}. This will make some error terms easier to treat later.

\subsection{Estimates for Hodge operators :}
We begin this part by recalling basic $L^2$ estimates for $2D$ Hodge systems :

\begin{prop}
Let $(S,\gamma)$ be a compact manifold with Gauss curvature K.
\begin{itemize}
\item i) For an arbitrary vectorfield $F$ on $S$, we have :
$$\int_S\big( \abs{\nabla F}^2 + K \abs{F}^2 \big) = \int_S\abs{\dope_1 F}^2$$

\item ii) For an arbitrary symmetric, traceless, 2 tensorfield $F$ on $S$, we have :
$$\int_S\big( \abs{\nabla F}^2 + 2K \abs{F}^2 \big) = 2\int_S\abs{\dope_2 F}^2$$

\item iii) For an arbitrary pair of functions $(\sigma, \rho)$ on $S$ we have :
$$\int_S\big( \abs{\nabla \sigma}^2 + \abs{\nabla \sigma}^2 ) = \int_S\abs{{}^{\star}\dope_1 F}^2$$

\item ii) For an arbitrary vectorfield $F$ on $S$, we have :
$$\int_S\big( \abs{\nabla F}^2  - K \abs{F}^2 \big) = 2\int_S\abs{{}^{\star}\dope_2 F}^2$$

\end{itemize}
\end{prop}

\begin{thm}{\bf 2 dimensional Hodge estimates}
The following estimates hold uniformly on an arbitrary 2-surface $S_t$ for $t_0 \ls t\ls T\ls t_0+ 1$, the definitions of the operators are given in section 2.10 :
\begin{itemize}
\item The operator $\dope_1$ is invertible on its range and its inverse $\dope_1^{-1}$ takes pair of functions $f=(\rho,\sigma)$ in the rande of $\dope_1$ (see above) into S-tangent 1-forms F with (div F = $\rho$, curl F = $\sigma$) with the following estimate :
$$\nd{\nabla \cdot \dope_1^{-1} F} + \nd{r^{-1}\dope_1^{-1} F} \lesssim \nd{F}$$

\item  The operator $\dope_2$ is invertible on its range and its inverse $\dope_2^{-1}$ takes S-tangent 1-forms F in the range of $\dope_2$ into S-tangent symmetric, traceless, 2-tensorfields Z with div Z = F with the following estimate : 
$$\nd{\nabla \cdot \dope_2^{-1} F} + \nd{r^{-1}\dope_2^{-1} F} \lesssim \nd{F}$$

\item The operator $\Delta$ is invertible on its range and its inverse  $\Delta^{-1}$ verifies the following estimate for an arbitrary scalar function f on $S_t$:
$$\nd{\nabla^2( -\Delta)^{-1} f} + \nd{ r^{-1}\nabla( -\Delta)^{-1} f} \lesssim \nd{f}$$

\item The operator ${}^{\star}\dope_1$ is invertible as an operator defined from the pairs of $H^1$ functions of vanishing mean (canonically identified to the quotient of $H^1$ by the pairs of constant functions). Moreover its inverse denoted by  ${}^{\star}\dope^{-1}_1$ takes S-tangent an 1-forms into a pair of function $(\rho,\sigma)$ with mean zero, such that $-\nabla \rho + (\nabla \sigma)^{\star} = F$ with the following estimate :
$$\nd{\nabla \cdot {}^{\star}\dope_1^{-1} F} \lesssim \nd{F}$$

\item The operator ${}^{\star}\dope_2$ is invertible as an operator defined from the quotient of $H^1$ vector fields by the kernel ${}^{\star}\dope_2$ (see above). Its inverse denoted by ${}^{\star}\dope^{-1}_2$ takes S-tangent $L^2$ 2-forms Z into S tangent 1-forms F such that ${}^{\star}\dope_2 F = Z$. Furthermore we have the following estimate :
$$\nd{\nabla \cdot {}^{\star}\dope_1^{-1} Z } \lesssim \nd{Z}$$
\end{itemize}
\end{thm}

\subsection{Estimates in Besov norms :}
\hs Due to the non-locality of operators like $\nabla \dope_2^{-1}$, $KR$ had to deal with Besov norms. Even if most of the following inequalities are trivial in a numerical space, they require in our setting use of the theory which was developed in \cite{krg}. In particular all the standard calculations involving Bony's paraproduct are to be redone but with weaker properties of the dyadic blocs (remark for instance the weak Bernstein inequalities, see appendix of section 1). A crucial question in this discussion is the  regularity of the gauss curvature $K$ involved in all these properties. A recurrent difficulty is that we do not have pointwise bound for $K$. It is a cause of non-sharpness in the trace theorems.

Let's begin with Sobolev embeddings :

\begin{thm}{\bf Embeddings theorems}
We have the following sharp Sobolev embedding for any scalar field $f$ on $S_t$ :
\begin{equation}
\nlsl{f}{\infty}\lesssim \|f\|_{B^1_{2,1}(S_t)}
\end{equation}
and also:
\begin{equation}
 \|f\|_{B^1_{2,1}(S_t)} \lesssim \nlsl{f}{2} +  \| \nabla f\|_{B^0_{2,1}(S_t)}
\end{equation}
thus :
\begin{equation}
\ntsl{f}{\infty}{\infty} \lesssim \|f\|_{\mathcal{B}^1}
\end{equation}
\end{thm}

The following version of classical product rules are also true in our weakly regular setting :
\begin{thm}
Let f be a scalar and U an arbitrary tensor on a fixed weakly spherical two dimensional compact surface $S$. We have the following estimates :

\begin{eqnarray}
\norme{fU}_{B^0_{2,1}(S)} &\lesssim& \big(\norme{\nabla f}_{L^2(S)} + \norme{f}_{L^\infty(S)} \big)\norme{U}_{B^0_{2,1}(S)}\\
\norme{fU}_{\mathcal{B}^0} &\lesssim& \big(\norme{\nabla f}_{L^2(S)} + \norme{f}_{L^\infty(S)} \big)\norme{U}_{\mathcal{B}^0}
\end{eqnarray}
\end{thm}

{\bf Proof :} See \cite{krg}.

The following embeddings are also great use 

\begin{thm}
Let F be a S-tangent tensorfield on $\Hyp$. We have for any $0\ls \kappa < \demi$ :
\begin{equation}
\norme{F}_{\mathcal{B}^\kappa}\lesssim \mathcal{N}_1(F)
\end{equation}
\end{thm} 

As a consequence of the bootstrap assumptions :
\begin{thm}
\begin{equation}
\norme{A}_{\mathcal{B}^\kappa}\lesssim M_0
\end{equation} 
\end{thm} 

We have now to state estimates for the transport equations. As usually in this section, each statement can be restated relative to either $L$, $L'$ or $\frac{d}{dt}$ and we will obtain the same estimates relative to the norms $L^p_x(L^q_t)$. Each time we come back to a transport equation relative to $t$ using the relations between the three parameters along null geodesics.

\begin{thm}
If we consider the following transport equation along $\Hyp$ for an arbitrary $S$-tangent tensor U :
\begin{equation}
\nabla_LU=F
\end{equation}
where F denotes a tensor of same type. If we assume that F is of one of the following forms :

i) $F=G \cdot \nabla_L P$. Then,
\begin{equation}
\norme{U}_{\mathcal{B}^0} \lesssim \big(\mathcal{N}_1(G) + \nl{G}{\infty}{2} \big) \mathcal{N}_1(P) + \norme{U(0)}_{B^0_{2,1}(S_0)}
\end{equation} 
ii) $F=G \cdot P$. Then,
\begin{equation}
\norme{U}_{\mathcal{B}^0} \lesssim \big(\mathcal{N}_1(G) + \nl{G}{\infty}{2} \big) \norme{P}_{\mathcal{P}^0} +\norme{U(0)}_{B^0_{2,1}(S_0)}
\end{equation}
iii) if $F=0$,
\begin{equation}
\norme{U}_{\mathcal{B}^0} \lesssim \norme{U(0)}_{B^0_{2,1}(S_0)}
\end{equation}
iv) we have also :
\begin{equation} 
\norme{G\cdot U}_{\mathcal{P}^0} \lesssim \big( \mathcal{N}_1(G) + \nl{G}{\infty}{2}\big)\big(\norme{U(0)}_{B^0_{2,1}(S_0)}+ \norme{F}_{\mathcal{P}^0}\big)
\end{equation}
\end{thm}

{\bf Proof :} see \cite{krs}. We have to say a few words about why the result of \cite{krs} remains true in our setting. In fact it should be redone mainly because of the new terms that appear in the commutator formulas. However, we will show in the next subsection how these additional terms can be estimated.
\vspace{5mm}
We make use of the following product estimates (see \cite{krg}) :
\begin{prop}
For any $S$-tangent tensors $F$, $G$, we have:
\begin{eqnarray}
 \norme{F\cdot G }_{\mathcal{P}^0} & \lesssim & \big( \ntsl{r^{-1}F}{2}{2} + \ntsl{\nabla F}{2}{2} \big)\cdot\norme{G}_{\mathcal{B}^0}\\
 \norme{F\cdot G }_{\mathcal{P}^0} & \lesssim &\mathcal{N}_2(r^\demi F) \cdot   \norme{F}_{\mathcal{P}^0}\\
 \norme{F\cdot G }_{\mathcal{P}^0} & \lesssim & \big( \ntsl{\nabla F}{2}{2} + \norm{\nabla F}_{L^\infty_\omega}{L^2_t} \big) \cdot \mathcal{N}_1(r^\demi G) 
\end{eqnarray}
We will in particular make use of (268) with $F$ of type $A$ and $G$ of type $C$, for which we will have :
\begin{equation}
 \norme{A\cdot C }_{\mathcal{P}^0} \lesssim  \big( r^{-1}\ntsl{A}{2}{2} + \ntsl{\nabla A}{2}{2} \big) \cdot\norme{C}_{\mathcal{B}^0}
\end{equation}
\end{prop}

As a consequence of the main Lemma :

\begin{prop}
Assume that the scalar U satisfies the transport equation along $\Hyp$ for some $k$ :

\begin{equation}
L(U) + k tr \chi U= F_1\cdot \nabla_L P +F_2 \cdot W.
\end{equation}
Then we have the following control :
\begin{eqnarray*}
\norme{U}_{\mathcal{B}^0}&\lesssim& \norme{U(0)}_{B^0_{2,1}(S_0)} + \big( \mathcal{N}_1(F_1) + \nl{F_1}{\infty}{2} \big) \cdot \mathcal{N}_1(P) +  \big( \mathcal{N}_1(F_2) \\
&&+ \nl{F_2}{\infty}{2} \big) \cdot \norme{W}_{\mathcal{P}^0}
\end{eqnarray*}
\end{prop}

\begin{prop}
Assume that the $S$-tangent tensor U satisfies the transport equation along $\Hyp$ for some $k$ :

\begin{equation}
\nabla_L(U) + k tr \chi U=E
\end{equation}
Then we have the following control for any other S-tangent tensorfield G:
\begin{equation} 
\norme{G\cdot U}_{\mathcal{P}^0} \lesssim \big( \mathcal{N}_1(G) + \nl{G}{\infty}{2}\big)\big(\norme{U(0)}_{B^0_{2,1}(S_0)}+ \norme{E}_{\mathcal{P}^0}\big)
\end{equation}
\end{prop}

\begin{prop}
Let F be a $S$-tangent tensor which admits a decomposition of the form $\nabla F = \nabla_LP+E$, with $P$,$E$ tensors. Then,
\begin{equation}
\nl{F}{\infty}{2}\lesssim \nun{F} +  \nun{P} + \norme{E}_{\mathcal{P}^0}
\end{equation}
\end{prop}

{\bf Proof :}

Let us consider $f(s)=\int_0^s\abs{F}^2$, we have :
\begin{equation}
\nabla _L f = \abs{F}^2, \ F(0)=0
\end{equation}
then we apply the commutation formula :

\begin{equation}
\nabla_L (\nabla f)+ \demi tr \chi (\nabla f) = 2 F \cdot \nabla F - \hat{\chi}\cdot (\nabla f) +(n^{-1}\nabla n) \nabla_L f 
\end{equation}

As usual we pay particular attention to the additional term. 

Using the transport lemma, we get schematically:

\begin{eqnarray}
\normeb{\nabla f} &\lesssim& \big( \nun{F} + \nl{F}{\infty}{2} \big) \cdot \nun{P}\\
&&+  \big( \nun{F} + \nl{F}{\infty}{2} \big) \cdot \normep{E}\\
&&+  \big( \nun{A} + \nl{A}{\infty}{2} \big) \cdot \normep{\nabla f}\\
&&+  \big( \nun{F} + \nl{F}{\infty}{2} \big) \cdot \nun{F} \cdot \normeb{C}
\end{eqnarray}

in view of the  control on $C$ and the bootstrap assumptions :
\begin{eqnarray*}
\normeb{\nabla f} &\lesssim& \big( \nun{F} + \nl{F}{\infty}{2} \big) \cdot \big( \nun{P} + \normep{E} + \mathcal{R}\nun{F} \big)\\
&& + M_0\normeb{\nabla f} 
 \end{eqnarray*}
 
 For $M_0$ small enough (not depending on $(t,u)$) : 
 
 \begin{eqnarray*}
\normeb{\nabla f} &\lesssim& \big( \nun{F} + \nl{F}{\infty}{2} \big) \cdot \big( \nun{P} + \normep{E} + \mathcal{R}\nun{F} \big)
\end{eqnarray*}

but :

\begin{eqnarray*}
\ntsl{f}{\infty}{\infty} & \lesssim & \|f\|_{\mathcal{B}^1} \\
& \lesssim &  \|\nabla f\|_{\mathcal{B}^0} + \ntsl{f}{\infty}{2}\\
& \lesssim & \ntsl{f}{\infty}{2} +  \big( \nun{F} + \nl{F}{\infty}{2} \big) \cdot \big( \nun{P} + \normep{E} + \mathcal{R}\nun{F} \big)\\
& \lesssim & \ntsl{F}{2}{4}^2 +  \big( \nun{F} + \nl{F}{\infty}{2} \big) \cdot \big( \nun{P} + \normep{E} + \mathcal{R}\nun{F} \big)
\end{eqnarray*}

then using $\nl{F}{2}{4}\lesssim \nun{F}$ :

\begin{equation*}
\nl{F}{\infty}{2}^2 \lesssim  \big( \nun{F} + \nl{F}{\infty}{2} \big) \cdot \big( \nun{P} + \normep{E} + \mathcal{R}\nun{F} \big) + \nun{F}^2
\end{equation*}

as a consequence :

\begin{eqnarray*}
\nl{F}{\infty}{2} &\lesssim&  \nun{F}(1 + \mathcal{R})+  \nun{P} + \normep{E} \\
&\lesssim&  \nun{F}+  \nun{P} + \normep{E}
\end{eqnarray*}

\begin{prop}
Let $\mathcal{D}^{-1}$ denote the inverse of one of the four previously defined operators. Then any $S$-tangent $F$ on $\Hyp$, and any $0\ls \kappa <1$ :
\begin{equation}
\|\nabla \mathcal{D}^{-1}F\|_{\mathcal{P}^\kappa}  \lesssim \|F\|_{\mathcal{P}^\kappa}
\end{equation}
\end{prop}

{\bf Proof :} see \cite{krg}

We will also need :

\begin{prop}
Let $\dope^{-1}$ be either $\dope_1^{-1}$, ${}^{\star}\dope_1^{-1}$, ${}^{\star}\dope_2^{-1}$, $\nabla \cdot\dope_2^{-1}\cdot \dope_1^{-1}$ or $\nabla \cdot \dope_1^{-1}\cdot {}^{\star}\dope_1^{-1}$ then for any $S$-tangent tensor belonging to the corresponding definition set and any $1<p\ls 2$, 

\begin{equation}
\|\dope^{-1} F\|_{B^0_{2,1}(S)} \lesssim \|r^{2-2/p} F\|_{L^p(S)}
\end{equation}

also along $\Hyp$ :

\begin{equation}
\normeb{\dope^{-1} F} \lesssim \ntsl{r^{2-2/p}  F}{\infty}{p}
\end{equation}

for  and $\frac{2}{2-\kappa}<p\ls 2$,
\begin{equation}
\norm{\dope^{-1} F}_{\mathcal{P}^\kappa} \lesssim \ntsl{r^{2-2/p-\kappa}F}{2}{p}
\end{equation}

\end{prop}

and the following lemma :

\begin{prop}
Let F be a S-tangent tensor on $\Hyp$ and $\dope^{-1}$ one of the operators as in last proposition. Then, we have the estimate :
\begin{equation}
\normeb{F} \lesssim \ntsl{F}{2}{2}+\ntsl{r\nabla F}{2}{2}
\end{equation}
%In particular,
%\begin{equation}
%\normeb{\dope^{-1}F} \lesssim \ntsl{F}{2}{2} + \ntsl{\nabla_L \cdot \dope^{-1} F}{2}{2}
%\end{equation}
\end{prop}

\subsection{Error terms :}

\hspace{5mm}In this subsection, we will study the various error terms appearing in the proof of the estimates. Some of them are easier to treat that it was in the proof in the case of geodesic foliation. It is typically the case of $A \cdot \underline{A}$. However, some new terms appear due to the non-cancellation in the commutation formula, typically in terms of the form $[\nabla_L,\dope]$.

We recall the following notations :

\begin{itemize}
\item $R'$ denotes either $(\rho, \sigma)$ or $\underline{\bet}$
\item $\dope^{-1}R'$ either  $\dope_1^{-1}(\rho, \sigma)$ or ${}^{\star}\dope_1^{-1}\underline{\bet}$
\item $\dope^{-2}R'$ either  $\dope_2^{-1} \dope_1^{-1}(\rho, \sigma)$ or $\dope_1^{-1}{}^{\star}\dope_1^{-1}\underline{\bet}$
\item $\dope^{-1}\nabla_LR'$ either  $\dope_1^{-1}\nabla_L(\rho, \sigma)$ or ${}^{\star}\dope_1^{-1}\nabla_L\underline{\bet}$
\item $\dope^{-2}\nabla_LR'$ either  $\dope_2^{-1} \dope_1^{-1}\nabla_L(\rho, \sigma)$ or $\dope_1^{-1}{}^{\star}\dope_1^{-1}\nabla_L\underline{\bet}$
\end{itemize} 

As before, we don't need to renormalize the coefficients.
Using the Bianchi equations, we get the following schematic form :

\begin{equation} 
\dope{-1}\cdot \nabla_L R' = R_0 + \dope^{-1}(\frac{1}{r} R_0
 + A \cdot (R_0+ \nabla A+\frac{1}{r}A + A\cdot \underline{A} + \underline{A} \cdot C) 
 \end{equation} 

Remark the presence of new terms $\underline{A} \cdot C$.

We now set the definitions for the commutators :

\begin{defi}
\begin{eqnarray*}
C_1(R')&=& \nabla \cdot \dope_2^{-1} \cdot [\nabla_L, \dope_1^{-1}](\rho, - \sigma) \text{ or } \nabla\cdot [\nabla_L, \dope_1^{-1}] \cdot {}^{\star}\dope_1^{-1}(\underline{\bet})\\
C_2(R')&=& \nabla \cdot [ \dope_2^{-1},\nabla_L] \cdot \dope_1^{-1}(\rho, - \sigma) \text{ or } \nabla\cdot [\nabla_L, \dope_1^{-1}] \cdot {}^{\star}\dope_1^{-1}(\underline{\bet})\\
C_3(R')&=& [\nabla,  \dope_2^{-1}]\cdot  \nabla_L\cdot \dope_1^{-1}(\rho, - \sigma) \text{ or }[\nabla, \nabla_L] \cdot \dope_1^{-1}] \cdot {}^{\star}\dope_1^{-1}(\underline{\bet})
\end{eqnarray*}
\end{defi}

We now make the distinction between strong and weak error types :
\begin{defi}
E is said to be of strong error type if for a fixed $\kappa >0$ :

\begin{equation}
\norme{E}_{\mathcal{P}^\kappa} \lesssim M_0^2 + \mathcal{R}M_0
\end{equation}
and of weak error if this inequality holds with $\kappa=0$.

\end{defi}

\begin{prop}
The following are strong error types :

$\dope^{-1}(A \cdot \nabla A)$, $\nabla \dope^{-2}(A \cdot A)$, $\dope^{-1}(A \cdot R)$, $\nabla \dope^{-2}(A \cdot R)$, $\dope^{-1}(A \cdot \nabla \cdot \dope{-1} R)$, $\dope^{-1}(A \cdot A \cdot A)$, $\dope^{-1}(A \cdot A \cdot \underline{A})$, $\nabla \dope^{-2}(A \cdot A \cdot A)$, $\nabla \dope^{-2}(A \cdot A \cdot \underline{A})$, $A \cdot A$, $\nabla \cdot \dope^{-1}A \cdot A$, $A \cdot \underline{A}$, $\nabla \cdot \dope^{-1}A \cdot \underline{A}$.

\end{prop}

{\bf Proof :} The proof is exactly the same that in \cite{krc} with the exception of the last to terms $A \cdot \underline{A}$, $\nabla \cdot \dope^{-1}A \cdot \underline{A}$ which are in our case of strong error type whereas they were in weak type in \cite{krc}.

\begin{prop}
The commutator $[\dope^{-1}, \nabla_L](R')$ is of strong error type
\end{prop}

{\bf Proof :}
We have schematically :
\begin{equation}
[\nabla_L, \dope^{-1}]= \dope^{-1}\cdot[\nabla_L, \dope] \cdot  \dope^{-1}
\end{equation}

and for an arbitrary $S$ tangent tensor :
\begin{equation}
[\nabla_L, \nabla] F= A \cdot \nabla F + A \cdot A   \cdot F +  \bet \cdot F + C \cdot \nabla_L F
\end{equation}

that leads to :
\begin{eqnarray*}
[\nabla_L, \dope^{-1}] R'&=& \dope^{-1} \big( A \cdot (\nabla \cdot \dope^{-1} R') + A \cdot A   \cdot (\dope^{-1} R')  +  \bet \cdot  (\dope^{-1} R')\\
&&+ C \cdot \nabla_L  (\dope^{-1} R') \big)\\
&=& I_1+I_2+I_3+I_4
\end{eqnarray*}
\begin{eqnarray*}
\norme{I_1}_{\mathcal{P}^\kappa} &\lesssim& \ntsl{A \cdot \nabla_L  (\dope^{-1} R')}{2}{4/3} \lesssim \ntsl{A}{\infty}{4}\cdot \ntsl{\nabla  (\dope^{-1} R')}{2}{2}\\
& \lesssim & \mathcal{R}M_0\\
\norme{I_2}_{\mathcal{P}^\kappa} &\lesssim& \ntsl{A \cdot A \cdot \dope^{-1} R'}{2}{4/3} \lesssim \ntsl{A}{\infty}{4}^2 \cdot \ntsl{\dope^{-1} R'}{2}{2}\\
& \lesssim & \mathcal{R}M_0^2\\
\norme{I_3}_{\mathcal{P}^\kappa} &\lesssim& \ntsl{\bet \cdot \dope^{-1} R'}{2}{4/3} \lesssim \ntsl{\dope^{-1}R}{\infty}{4} \cdot \ntsl{R}{2}{2}\\
%& \lesssim & \mathcal{R}M_0+M_0^2\\
\norme{I_4}_{\mathcal{P}^\kappa} &\lesssim& \ntsl{C \cdot \nabla_L  (\dope^{-1} R')}{2}{4/3} \lesssim \ntsl{C}{\infty}{4}\cdot \ntsl{\nabla_L  (\dope^{-1} R')}{2}{2}\\
%& \lesssim & \mathcal{R}M_0+M_0^2
\end{eqnarray*}

moreover :
\begin{eqnarray}
\ntsl{\dope^{-1}R}{\infty}{4} &\lesssim& \nun{\dope^{-1}R'}\\
\ntsl{\nabla_L  (\dope^{-1} R')}{2}{2} &\lesssim& \nun{\dope^{-1}R'}
\end{eqnarray}

Thus, we have :
\begin{equation}
\norme{[\nabla_L, \dope^{-1}]R'}_{\mathcal{P}^\kappa} \lesssim M_0^2 + \mathcal{R}M_0 + (M_0+ \mathcal{R})  \nun{\dope^{-1}R'}
\end{equation}

Where we have used the following estimates :

\begin{prop}
We also have the following estimates :
\begin{eqnarray}
\ntsl{\dope^{-1}\cdot \nabla_L R}{2}{2} &\lesssim&\mathcal{R} + M_0^2 + \mathcal{R}^2M_0\\
\ntsl{[\dope^{-1},\nabla_L] R}{2}{2}&\lesssim&\mathcal{R}+ M_0^2\\
\nun{\dope^{-1}R}&\lesssim&\mathcal{R}+ M_0^2
\end{eqnarray}
\end{prop}

{\bf Proof :}

We use :
\begin{equation*} 
\dope^{-1}\cdot \nabla_L R' = R_0 + \dope^{-1}\big(\frac{1}{r} R_0
 + A \cdot (R_0+ \nabla A+\frac{1}{r}A + A\cdot \underline{A} + \underline{A} \cdot C) \big)
 \end{equation*} 

but all the terms in $\dope^{-1}\big(\frac{1}{r} R_0 + A \cdot (R_0+ \nabla A+\frac{1}{r}A + A\cdot \underline{A} + \underline{A} \cdot C) \big)$ are of strong error type.

Finally :
\begin{equation}
\ntsl{\dope^{-1}\cdot \nabla_L R}{2}{2} \lesssim \mathcal{R} + M_0^2 + \mathcal{R}^2M_0
\end{equation}

Coming back to (2.149) :
\begin{equation}
\norme{[\nabla_L, \dope^{-1}]R'}_{\mathcal{P}^\kappa} \lesssim M_0^2 + \mathcal{R}M_0 + (M_0+ \mathcal{R})  \nun{\dope^{-1}R'}
\end{equation}

we use :
\begin{eqnarray*}
\nun{\dope^{-1}R'} &=& \ntsl{\dope^{-1}R'}{2}{2} +\ntsl{\nabla \cdot \dope^{-1}R'}{2}{2}+\ntsl{\nabla_L \cdot\dope^{-1}R'}{2}{2}\\
&\lesssim& \mathcal{R} + M_0^2 + \mathcal{R}^2M_0+\ntsl{[\nabla_L,\dope^{-1}]R'}{2}{2}
\end{eqnarray*}

Putting everything together :

\begin{equation*}
\norme{[\nabla_L, \dope^{-1}R']}_{\mathcal{P}^\kappa} \lesssim M_0^2 + \mathcal{R}M_0 + (M_0+ \mathcal{R})  \big(\mathcal{R} + M_0^2 + \mathcal{R}^2M_0+\ntsl{[\nabla_L,\dope^{-1}]R'}{2}{2}\big)
\end{equation*}

choosing $0<\mathcal{R} \ls M_0 \ls \demi$ gives :

\begin{equation*}
\norme{[\nabla_L, \dope^{-1}R']}_{\mathcal{P}^\kappa} \lesssim \mathcal{R}+ M_0^2
\end{equation*}

and :

\begin{equation*}
\norme{\dope^{-1}R'}_{\mathcal{P}^\kappa} \lesssim \mathcal{R}+ M_0^2 
\end{equation*}

\subsubsection{Commutator estimates :}

The commutator terms $C_i(R')$ can be written under the following form :
\begin{eqnarray*}
C_1(R')&=& \nabla \cdot \dope^{-2} \big(A \cdot \nabla \cdot ( \dope^{-1}(R'))+ A \cdot A \cdot ( \dope^{-1}(R')) + \bet \cdot ( \dope^{-1}(R')) + C \cdot \nabla_L ( \dope^{-1}(R'))\big)\\
C_2(R')&=& \nabla \cdot \dope^{-1} \big(A \cdot \nabla \cdot ( \dope^{-2}(R'))+ A \cdot A \cdot ( \dope^{-2}(R')) + \bet \cdot ( \dope^{-2}(R')) + C \cdot \nabla_L ( \dope^{-2}(R'))\big)\\
C_3(R') &=& A \cdot \nabla \cdot ( \dope^{-2}(R'))+ A \cdot A \cdot ( \dope^{-2}(R')) + \bet \cdot ( \dope^{-2}(R')) + C \cdot \nabla_L ( \dope^{-2}(R'))
\end{eqnarray*}

The additional terms are of the same level of regularity than the others. We have :

\begin{prop}
The commutator $C_1(R')$ is of strong type, moreover there exists a strong error terms $E_\kappa$ and $E'_\kappa$ such that :
\begin{eqnarray}
C_2(R')&=& \nabla \cdot \dope^{-1}\cdot ( \bet \cdot \dope^{-2}(R'))+ E \\
C_3(R')&=&\bet \cdot \dope^{-2}(R')+ E'
\end{eqnarray}
\end{prop}

{\bf Proof :}

$C_1(R')$ can be estimated exactly as $[\nabla_L, \dope^{-1}]R'$. 

For the two remaining commutators, let us begin by remarking that due to the boundness of $\nabla \cdot \dope^{-1}$ on $\mathcal{P}^\kappa$. The estimates on $C_3(R')$ imply the one on $C_2(R')$.

Now we have :
\begin{eqnarray*}
\norme{A \cdot (\nabla \cdot  \dope^{-2}(R'))}_{\mathcal{P}^\kappa}&\lesssim&\nun{A} \cdot \norme{( \nabla \cdot\dope^{-2}(R'))}_{\mathcal{B}^\kappa}\lesssim M_0 \norme{( \nabla \cdot \dope^{-2}(R'))}_{\mathcal{B}^\kappa}\\
&\lesssim& M_0 \hspace{1mm} \nun{ \nabla \cdot \dope^{-2}(R')}\\
 \norme{A \cdot A \cdot ( \dope^{-2}(R'))}_{\mathcal{P}^\kappa} &\lesssim&M_0^2 \mathcal{N}_2(\dope^{-2}R') \\
 \norme{C \cdot \nabla_L ( \dope^{-2}(R'))}_{\mathcal{P}^\kappa}&\lesssim& \mathcal{R} \hspace{1mm}  \mathcal{N}_2(\dope^{-2}R')
\end{eqnarray*}
It remains to control $ \nun{ \nabla \cdot \dope^{-2}(R')}$ and ${N}_2(\dope^{-2}(R'))$.

We have the following estimates :

\begin{prop}
We have the following estimates :
\begin{eqnarray}
\ntsl{\nabla \cdot \dope^{-2}\cdot \nabla_L R'}{2}{2}&\lesssim&\mathcal{R}+ M_0^2 \\
\ntsl{C_i(R')}{2}{2}&\lesssim&M_0^2\\
 \nun{ \nabla \cdot \dope^{-2}(R')} &\lesssim&\mathcal{R}+ M_0^2\\
{N}_2(\dope^{-2}(R')) &\lesssim&\mathcal{R}+ M_0^2
\end{eqnarray}
\end{prop}

{\bf Proof :} 
The proof of (306) is the same as in (2.13.2).
For the $L^2$ estimates, we have :

\begin{eqnarray*}
\ntsl{C_1(R')}{2}{2}&\lesssim& M_0^2\\
\ntsl{C_2(R')}{2}{2}&\lesssim& \ntsl{\nabla \cdot \dope^{-1} \cdot (\bet \cdot\dope^{-2}(R')}{2}{2} + M_0 \nun{\nabla \cdot \dope^{-2}R'}\\
&&+ M_0^2 \mathcal{N}_2( \dope^{-2}R')+ \ntsl{C\cdot \nabla_L \dope^{-2}(R')}{2}{2}\\
&\lesssim& M_0^2 + M_0 \big( \nun{\nabla \cdot \dope^{-2}R'}+  \mathcal{N}_2( \dope^{-2}R')\big)
\end{eqnarray*} 

and the same estimate for $C_3(R')$.

A this step we have to control  $ \nun{\nabla \cdot \dope^{-2}R'}$ +  $\mathcal{N}_2( \dope^{-2}R')$.

Using the estimates, we get :

\begin{eqnarray*}
\ntsl{\nabla \cdot \nabla_L \dope^{-2}(R')}{2}{2}&\lesssim& \mathcal{R}+ M_0^2 + \ntsl{C_i(R')}{2}{2}\\
\ntsl{\nabla_L \cdot \nabla \dope^{-2}(R')}{2}{2}&\lesssim& \mathcal{R}+ M_0^2 + \ntsl{C_i(R')}{2}{2}\\
\end{eqnarray*}

now using the elliptic estimates, we have :

\begin{equation}
\ntsl{\nabla^2\cdot \dope^{-2}(R')}{2}{2}\lesssim \mathcal{R}+ M_0 \ntsl{C_i(R')}{2}{2}
\end{equation}

it implies :
\begin{eqnarray}
\nun{\nabla \cdot \dope^{-2}R'}&\lesssim &\mathcal{R}+ M_0 \ntsl{C_i(R')}{2}{2}+ M_0^2\\
\mathcal{N}_2( \dope^{-2}R')&\lesssim &\mathcal{R}+ M_0 \ntsl{C_i(R')}{2}{2}+ M_0^2
\end{eqnarray}

Using finally that $M_0 \ls \demi$, we get the desired estimates.

\subsubsection{Decomposition of the $C_i(R')$ :}

\begin{prop}
Let F be an arbitrary tensor on $\Hyp$ verifying :
\begin{equation}
\mathcal{N}_2(F)<\infty
\end{equation}
Then we have that :
\newline
i) $\bet$ enjoys the following structure :
\begin{equation}
\bet= \nabla_L(\dope_1^{-1}(R'))+ E
\end{equation}

with $\norme{E}_{\mathcal{P}^\kappa} \ls M_0^2$

ii) $\bet \cdot F$  has the following structure :
\begin{equation}
\bet \cdot F= \nabla_LP+ E
\end{equation}

with $\nun{P} + \norme{E}_{\mathcal{P}^\kappa} \lesssim M_0 \mathcal{N}_2(F)$.

iii) $\nabla \cdot \dope^{-1}\cdot \bet \cdot F$ admits a structure similar to the structure of $\bet \cdot F$.

iv) $\dope^{-1}(\bet \cdot F)$  has the following structure :
\begin{equation}
\dope^{-1}(\bet \cdot F)= \nabla_LP+ E
\end{equation}

with $\mathcal{N}_2(P) + \norme{\nabla E}_{\mathcal{P}^\kappa} \lesssim M_0 \mathcal{N}_2(F)$.
\end{prop}

and we have as a consequence :
\begin{prop}
All the commutators $C_i(R')$ enjoy the following form :
\begin{equation}
C_i(R')= \nabla_L P + E 
\end{equation}
where $\nun{P} + \norme{E}_{\mathcal{P}^\kappa} \lesssim M_0^2$.
\end{prop}

{\bf Proof :} Once we have proved that the additional terms in the commutators due to the $(t,u)$-setting instead of the geodesic one, can be estimated at the same level of regularity that in \cite{krc}, the proof of the decomposition Lemma and its consequence is exactly the same. This proof is based on an elaborate iteration scheme. 

\subsection{Proof of the primary estimates :}

\subsubsection{Estimate on $\chi$ :}

{\bf estimate on $tr\chi-\frac{2}{s}$ :}
We will first record the following lemma.

Let us note :
\begin{eqnarray*}
V&=&b^{-1}tr\chi -\overline{b^{-1}tr\chi} \\
W&=&\overline{b^{-1}tr\chi}-\frac{2}{r}
\end{eqnarray*}

then we have :

\begin{lem}
\begin{eqnarray*}
\frac{d}{ds'}W+\demi b^{-1}tr\chi W&=& \demi \big(V \cdot W + \overline{V^2}) \big) - \overline{b^{-2}\abs{\hat{\chi}}^2}\\
\frac{d}{ds'}V+\demi b^{-1}tr\chi V&=& - \demi \overline{b^{-1}tr\chi} + \overline{V^2}-\big( b^{-2}\abs{\hat{\chi}}^2 -\overline{b^{-2}\abs{\hat{\chi}}^2}\big)\\
\end{eqnarray*}
\end{lem}

With the corresponding reformulations in term of $t$ and $s$.

{\it Remark :}
It is clearly not a good idea to try to use the $(t,u)$ structure for this equation. The natural inner structure is easily seen considering the geodesic foliation. In this kind of case, we re-give the good structure by considering the quantities corresponding to the affine geodesic parameter $s'$.

We first recall that we have :

\begin{eqnarray}
L'(b)&=&-\deb\\
L'(b^{-1}tr\chi)&=&-\demi b^{-2}\abs{tr \chi}^2 - b^{-2}\abs{\hat{\chi}}^2\\
L'(r)&=&\frac{2}{r}\overline{b^{-1}tr\chi}
\end{eqnarray}  

In view of these relations, we get :

\begin{eqnarray}
L'(\overline{b^{-1}tr\chi})&=&-\overline{b^{-1}tr\chi}^2 +\overline{ b^{-2}tr\chi} +\overline{ L'(tr\chi)}\\
&=& -\overline{b^{-1}tr\chi}^2+\demi \overline{b^{-2}tr\chi^2} - \overline{b^{-2}\abs{\hat{\chi}}^2}\\
&& - \demi \overline{b^{-1}tr\chi}^2+\demi \overline{(b^{-1}tr \chi -\overline{b^{-1}tr\chi})^2 }- \overline{b^{-2}\abs{\hat{\chi}}^2}
\end{eqnarray}

Thus :
\begin{equation}
L'(\overline{b^{-1}tr \chi}-\frac{2}{r})= - \demi \overline{b^{-1}tr\chi}(\overline{b^{-1}tr\chi}-\frac{2}{r})+\demi \overline{(b^{-1}tr \chi -\overline{b^{-1}tr\chi})^2 }- \overline{b^{-2}\abs{\hat{\chi}}^2}
\end{equation}

From another hand, we have the following relation :

\begin{eqnarray}
L'(b^{-1}tr\chi- \overline{b^{-1}tr\chi}) &=& -\demi \big( b^{-1}tr\chi - \overline{b^{-1}tr\chi} \big) \big( b^{-1}tr\chi + \overline{b^{-1}tr\chi} \big) \\
&=&+ \demi \overline{(b^{-1}tr \chi -\overline{b^{-1}tr\chi})^2 } -\big( b^{-2}\abs{\hat{\chi}}^2 -\overline{b^{-2}\abs{\hat{\chi}}^2}\big)
\end{eqnarray}

\vspace{5mm}
Thus applying the transport lemma an using both {\bf (BA1)}, {\bf (BA2)} and the comparison theorem, we get :

\begin{equation}
\ntsl{b^{-1}tr\chi-\frac{2}{r}}{\infty}{\infty}\lesssim \mathcal{I}+ M_0^2
\end{equation}
which implies :
\begin{equation}
\ntsl{tr\chi-\frac{2}{s}}{\infty}{\infty}\lesssim \mathcal{I}+ \mathcal{R}+M_0^2
\end{equation}

{\bf Estimate for $L(tr\chi-\frac{2}{s})$ :}
The estimation of  $\nabla_L(tr \chi -\frac{2}{s})$ comes when writing :

\begin{equation*}
L(tr\chi-\frac{2}{s})+(\deb +\frac{2}{s})tr\chi=-\abs{\hat{\chi}}^2-\demi (tr\chi- \frac{2}{s})^2
\end{equation*}

which gives :
\begin{equation}
\nl{tr\chi-\frac{2}{s}}{\infty}{1}\lesssim \mathcal{R}+ M_0^2
\end{equation}

{\bf Estimate for $\nabla tr\chi$ :}

We recall  :
\begin{eqnarray*}
\nabla_L(\nabla tr \chi) &=& -\frac{3}{2}tr \chi \nabla tr\chi-\hat{\chi} \cdot \nabla tr \chi - 2 \hat{\chi}\cdot \nabla\hat{\chi}-\deb \nabla tr \chi - \deb tr \chi \\
&&- n^{-1}\nabla n \big(\abs{\hat{\chi}}^2+ \demi (tr\chi)^2 + \deb tr \chi)
\end{eqnarray*}

We have moreover the following Codazzi equation :
\begin{equation}
\nabla \cdot \hat{\chi} + \hat{\chi} \cdot \e  =-\bet+\frac{1}{2}\nabla tr \chi + \frac{1}{2} tr \chi \e
\end{equation}

and the following Bianchi equations :
\begin{eqnarray*}
L(\rho) + \frac{3}{2} tr \chi \rho &=& div \hspace{2mm} \bet -\frac{1}{2}\underline{\hat{\chi}} \cdot \alpha + \e \cdot \bet + 2 \hb \cdot \beta\\
L(\sigma) +  \frac{3}{2} tr \chi \sigma&=& -  div \hspace{2mm} \bet + \frac{1}{2} \hat{\underline{\chi}} \cdot \alpha - \e \wedge \beta-2\hb \wedge \bet
\end{eqnarray*}

Thus :

\begin{equation}
\bet= \dope_1^{-1}L(\rho,-\sigma)+  \dope_1^{-1}(R_0+ A' \cdot R + C \cdot R)
\end{equation}

then :
\begin{eqnarray}
 \hat{\chi}&=& \dope_2^{-1}\dope_1^{-1}L(\rho,-\sigma)+   \dope_2^{-1}\dope_1^{-1}(R_0+ A' \cdot R + C \cdot R)\\
& & + \dope_1^{-1}(A \cdot C + \nabla tr \chi)
\end{eqnarray}

and as a consequence :
\begin{eqnarray*}
\nabla \hat{\chi}&=& \nabla \dope_2^{-1}\dope_1^{-1}L(\rho,-\sigma)+   \nabla \dope_2^{-1}\dope_1^{-1}(R_0+ A' \cdot R + C \cdot R)\\
& & + \nabla \dope_1^{-1}(A \cdot C + \nabla tr \chi)
\end{eqnarray*}

Schematically we thus have :
\begin{eqnarray*}
\nabla \hat{\chi}&=& \nabla \dope_2^{-1}\dope_1^{-1}L(R_0)+ F + \nabla \dope_1^{-1}(\nabla tr \chi)
\end{eqnarray*}
where 

\begin{equation}
\|F \|_{\mathcal{P}_\e} \lesssim \mathcal{R} + M_0^2
\end{equation}

In view of the structure estimate, we can write :
\begin{equation}
\nabla \dope_2^{-1}\dope_1^{-1}L(R_0)=\nabla_L( \nabla \dope_2^{-1}\dope_1^{-1}R_0) + C(R_0)
\end{equation}

and according to proposition (2.13.7) :
\begin{eqnarray}
C(R_0)&=& \nabla_L P+E\\
\mathcal{N}_1(P) + \|E \|_{\mathcal{P}_\e} &\lesssim& M_0^2
\end{eqnarray}

Therefore :
\begin{eqnarray*}
\nabla_L(\nabla tr \chi)+\frac{3}{2}tr \chi \nabla tr\chi &=&A \big( \nabla_LP+ \nabla tr \chi + \nabla \cdot \dope^{-1}M +E) \\
& & + C\big(A^2+ C \cdot A\big)
\end{eqnarray*}

We have here to be careful with the additional term $ C\big(A^2+ C \cdot A\big)$.

To apply the transport lemma in Besov norm, it's sufficient to control :

\begin{equation}
\big( \mathcal{N}_1(A) + \ntsl{A}{2}{2})\|A.C\|_{\mathcal{P}_0}
\end{equation}

but :
\begin{eqnarray}
\big( \mathcal{N}_1(A) + \ntsl{A}{2}{2})&\lesssim& M_0\\
\normep{A.C}\lesssim\big( \ntsl{A}{2}{2} +  \ntsl{\nabla A}{2}{2} \big) \|C\|_{\mathcal{B}_0} &\lesssim& \mathcal{R}M_0
\end{eqnarray}

Then applying the transport lemma, we get :
\begin{equation}
\|\nabla tr \chi \|_{\mathcal{B}_0} \lesssim \mathcal{I}+M_0^2
\end{equation}

For the estimate of $\nl{\nabla tr \chi}{2}{\infty}$, we come back to :

\begin{eqnarray*}
\nabla_L(\nabla tr \chi) &=& -\frac{3}{2}tr \chi \nabla tr\chi-\hat{\chi} \cdot \nabla tr \chi - 2 \hat{\chi}\cdot \nabla\hat{\chi}-\deb \nabla tr \chi - \deb tr \chi \\
&&- n^{-1}\nabla n \big(\abs{\hat{\chi}}^2+ \demi (tr\chi)^2 + \deb tr \chi)
\end{eqnarray*}

and applying the standard transport lemma, we get  :
\begin{eqnarray*}
\nl{\nabla tr \chi}{2}{\infty} &\lesssim&  \nli{\nabla tr \chi(0)}{2}+ \nl{A\cdot M+A\cdot A  }{2}{1}\\
&& +\nl{C \cdot C \cdot tr \chi + C \cdot A \cdot A}{2}{1}\\
&\lesssim& \mathcal{I}+\nl{M}{2}{\infty} \cdot \nl{A}{\infty}{1} + \nl{A}{\infty}{2}^2\\
&&+ \nl{C}{4}{2}^2\nl{tr \chi}{\infty}{\infty} +  \nl{C}{\infty}{1}\nl{A}{4}{\infty}^2
\end{eqnarray*}

it implies :

\begin{equation}
\nl{\nabla tr \chi}{2}{\infty} \lesssim \mathcal{I}+ M_0^2 + \mathcal{R}M_0
\end{equation}

{\bf Estimates for $\hat{\chi}$ :}

The estimate of $\mathcal{N}_1(\hat{\chi})$ is a direct consequence of Codazzi equation :

\begin{equation}
\nabla \cdot \hat{\chi} + \hat{\chi} \cdot \e  =-\bet+\frac{1}{2}\nabla tr \chi + \frac{1}{2} tr \chi \e
\end{equation}

In view of the elliptic estimate :
\begin{eqnarray}
\ntsl{\hat{\chi}}{2}{2} + \ntsl{\nabla \hat{\chi}}{2}{2} & \lesssim & \ntsl{\bet}{2}{2}\\
&& + \ntsl{\nabla tr \chi}{2}{2} + \ntsl{A \cdot C}{2}{2}\\
& \lesssim &\mathcal{I}+\mathcal{R}M_0 + M_0^2
\end{eqnarray}

The last estimate is the one concerning $\nl{\hat{\chi}}{\infty}{2}$,

With the notations that we took in the previous part we have :

\begin{eqnarray}
\nl{\hat{\chi}}{\infty}{2} &\lesssim& \mathcal{N}_1 + \mathcal{N}_1(P) \|\nabla \dope^{-1}M\|_{\mathcal{P}_0}+  \| E \|_{\mathcal{P}_0}+ \| A \cdot C \|_{\mathcal{P}_0} \\
&\lesssim& \mathcal{N}_1(\hat{\chi}) + \mathcal{N}_1(P) + \|M\|_{\mathcal{B}_0}+  \| E \|_{\mathcal{P}_0}\\
&\lesssim& \mathcal{R} + \mathcal{I} +\mathcal{R}M_0+ M_0^2
\end{eqnarray}

\subsubsection{Secondary estimates :}
The estimates of $\chiu$ are consequences of the one proved for $\chi$ and the following equality :
\begin{equation*}
{}^{(2)}k=-\demi (\chi + \chiu)
\end{equation*}

Moreover, we have due to the maximality of the $t$-foliation :

\begin{equation*}
tr_{(3)}k=tr_{(2)}k + \delta=0
\end{equation*}

The a priori estimates together with what we have just proved leads to :
\begin{eqnarray}
\left\| tr \underline{\chi} + \frac{2}{s}\right\| _{L^{\infty}_{[t_0,t_0+1]}L^\infty_x +L^{\infty}_{[0,1]}H^\demi_x}  & \lesssim &  \mathcal{R} + \mathcal{I}+ M_0^2\\
\|\hat{\chiu}\|_{L^{\infty}_{[t_0,t_0+1]}H^\demi_x} & \lesssim &  \mathcal{R} + \mathcal{I}+M_0^2
\end{eqnarray}

%
%Estimating the angular derivatives of  $tr \chiu$ in the form (2.55) is thus reduced to proving the following estimate :
%\begin{equation*}
%\ntsl{\nabla \delta}{2}{2}\lesssim  \mathcal{R} + \mathcal{I}
%\end{equation*}
%We shall remark that it is clearly the good level of differentiability as $\nabla k$ is of the level of $\bo{R}$. However we lose some control on $\bo{R}$ when estimating the flux on $\Hyp$ in comparison with the estimates on the leaves (we lose the control on $\|\underline{\al}\|_{L^2}$) thus we can expect not to control every first derivatives of $k$ and/or every second derivatives of $n$.  

%But we have (see \cite{ck} p.312 equation (11.1.2d)) the following relation :

%\begin{equation*}
%\div{\hat{k_{(2)}}}=\demi(\beta - \underline{\beta}) -\demi \nabla \delta+ \hat{\theta}\cdot \e -\demi tr \theta \e
%\end{equation*}

%but using, we have for all $t$ :
%\begin{equation*}
%\int_{S_t}\big(  \abs{\nabla \hat{k_{2}}}^2 + K \abs{\hat{k_{2}}}^2 \big)= 2 \int_{S_t} \abs{ div \hat{k_{2}}}^2
%\end{equation*}
%Integrating in time we find  :

%\begin{equation*}
%\int_{\Hyp}\abs{div\hspace{2mm} \hat{{}^{(2)}k}}^2 \lesssim \mathcal{R} + \mathcal{I} 
%\end{equation*}

%The estimate of $\theta$ are implied by the one on $\chi$ and . In particular :

%\begin{equation*}
%\int_{\mathcal{\Hyp}}\abs{\hat{\theta}\cdot \e}^2 \lesssim  \ntsl{\hat{\theta}}{2}{4} \ntsl{\e}{\infty}{4} \lesssim
% \mathcal{R} + \mathcal{I}
% \end{equation*}
%we can also the same estimates to $ tr \theta \e$. 
%Using the flux estimate for $\bet$ and $\underline{\bet}$ gives the result.

\subsubsection{Estimates for $\tilde{\mu}$ :}
As explained before the mass $\tilde{\mu}$ will play the role of $\nabla tr \chi$ in the following estimates. We will work on the 2 dimensional Hodge system for $\h$ :
\begin{eqnarray*}
curl \hspace{.5 mm} \h&=& \demi \hat{\chi} \wedge \hat{\chiu}- \sigma \\
div  \hspace{.5 mm} \h&=& \tilde{\mu} + \demi \hat{\chi} \cdot\hat{\chiu}  - \rho + \demi \delta tr \chi - \abs{\h}^2 
\end{eqnarray*}

together with the transport equation for the reduced mass :
\begin{eqnarray*}
L(\mub) + tr\chi \mub&=&  2 \hat{\chi} \cdot \nabla \h + (\h-\hb)(b\nabla(b^{-1} tr\chi) + tr \chi \h)\\
&& -\demi tr \chi (\hat{\chi}\cdot \hat{\chiu}-2 \rho + 2 \hb \cdot \h) + 2 \h \cdot \hat{\chi}\cdot \ \h -  \delta\abs{\hat{\chi}}^2 -\demi \deb (tr \chi)^2 
\end{eqnarray*}
\hs The main problem comes from the term $\hat{\chi} \cdot \nabla \h$ which requires the application of a trace theorem on $\Hyp$ which is the subject of \cite{krs}. We used this mass to avoid the need for additional control on $\hb$ other than the ones provided by the trace theorems. 

Schematically : 
\begin{eqnarray*}
L(\mub) + tr\chi \mub& \lesssim &   \hat{\chi} \cdot \nabla\h + tr \chi \cdot \rho+ A \cdot A  \cdot A +A \cdot A  \cdot \underline{A} \\
&&+ (A+C) \cdot \nabla tr \chi + C \cdot A \cdot A.
\end{eqnarray*}

which leads to with $M=(-\tilde{\mu},0)$ to :

\begin{equation}
\h= \dope_1^{-1}(\rho,-\sigma) -  \dope_1^{-1}M +  \dope_1^{-1}(A\cdot \underline{A}) +\dope_1^{-1}(A\cdot C)
\end{equation}

from another hand :

\begin{equation}
\nabla_L \underline{\bet}= {}^{\star}\dope_1(\rho, \sigma) + \frac{1}{r}R_0 + A \cdot (R_0+ \nabla A+ A \cdot \underline{A}) + C \cdot (A + R_0)
\end{equation}

denoting by $J(\rho, \sigma)=(-\rho, \sigma)$, and schematically :
\begin{eqnarray*}
\h&=& - \dope_1^{-1}\cdot J \cdot {}^{\star}\dope_1^{-1}(\nabla_L \underline{\bet}+ \frac{1}{r}R_0)  -  \dope_1^{-1}M +  \dope_1^{-1}(A\cdot \underline{A}) +\dope_1^{-1}(A\cdot C)\\
&& + \dope_1^{-1}\cdot J \cdot {}^{\star}\dope_1^{-1}(A \cdot (R_0+ \nabla A+ A \cdot \underline{A}) + C \cdot (A + R_0)
)
\end{eqnarray*}

If we denote by $\dope^{-2}:= \dope_1^{-1}\cdot J \cdot {}^{\star}\dope_1^{-1}$ and recalling :

\begin{equation}
\norme{A \cdot \underline{A}}_{\mathcal{P}^\kappa} \lesssim M_0^2
\end{equation}

using the boundness of $\nabla \cdot \dope_1^{-1}$ on $\mathcal{P}^\kappa$, we get :

\begin{equation}
\nabla \h = - \nabla \cdot \dope^{-2}(\nabla_L \underline{\bet}) +- \nabla \cdot \dope_1^{-1}M + E
\end{equation}

where $\norme{E}_{\mathcal{P}^0} \lesssim \mathcal{R}+ M_0^2$.

but :
\begin{equation}
\nabla_L \cdot ( \nabla \cdot \dope^{-2}R') = \nabla \cdot \nabla_L R'
+ C(R')
\end{equation} 
where $C(R')$ admits the following decomposition :
\begin{equation}
 C(R')= \nabla_L P' + E'
\end{equation} 
with $\nun{P'}+ \norme{E}_{\mathcal{P}^\kappa} \lesssim M_0^2$.

Setting $P=P'+ \nabla \cdot \dope^{-2}(R')$ and recalling $\nun{\nabla \cdot \dope^{-1}(R')}$, we have :
\begin{equation}
\nabla \h = \nabla_L(P)+\nabla \cdot \dope^{-2}( \frac{1}{r}R_0) + \nabla \cdot \dope_1^{-1}M + E
\end{equation}

with $\nun{P}+ \norme{E}_{\mathcal{P}^0} \lesssim \mathcal{R}+M_0^2$.

We have now to deal with $tr\chi \rho$, by using a proof similar to the one we have just done, we get that $\rho$ admits a decomposition of the form :

\begin{equation}
\rho= \nabla_L p+ e
\end{equation}
with $\nun{p}+ \norme{e}_{\mathcal{P}^0} \lesssim \mathcal{R}+M_0^2$.

\begin{eqnarray*}
L(\mub) + tr\chi \mub& \lesssim &  (A+\frac{1}{r})\nabla_L(P)+ \frac{1}{r}E+ E+ A \cdot A  \cdot A +A \cdot A  \cdot \underline{A} \\
&&+ (A+C) \cdot \nabla tr \chi + C \cdot A \cdot A + A \cdot \dope_1^{-1}M
\end{eqnarray*}
with $\nun{P}+ \norme{E}_{\mathcal{P}^0} \lesssim \mathcal{R}+M_0^2$.

The terms $A\cdot A \cdot A$ are better then $A\cdot A \cdot \underline{A}$ which are of form $A\cdot A \cdot C$. So we write schematically :

\begin{eqnarray*}
L(\mub) + tr\chi \mub& \lesssim &  (A+\frac{1}{r})\nabla_L(P)+ \frac{1}{r}E+ E + (A+C) \cdot \nabla tr \chi\\
&& + C \cdot A \cdot A + A \cdot \dope_1^{-1}M
\end{eqnarray*}

that we write :
\begin{equation}
\nabla_LM + tr \chi M= F_1\cdot \nabla_L P + F_2 \cdot E + A\cdot W+ C \nabla tr \chi 
\end{equation}

then using the Besov version of the transport lemma :
\begin{eqnarray*}
\normeb{M}&\lesssim& \norme{M(0)}_{B^0_{2,1}(S_0)} + \big(\nun{F_1}+ \nl{F_1}{\infty}{2} \big)\nun{P}\\
&&+ \big(\nun{F_2}+ \nl{F_2}{\infty}{2} \big)  \norme{E}_{\mathcal{P}^0}+ \big(\nun{A}+ \nl{A}{\infty}{2} \big)  \norme{W}_{\mathcal{P}^0} \\
&&+\ntsl{C}{2}{\infty}   \normeb{\nabla tr \chi}\\
&\lesssim& \mathcal{I}+ \mathcal{R}+ M_0^2+ M_0 \norme{W}_{\mathcal{P}^0}
\end{eqnarray*}

where :
\begin{eqnarray*}
\norme{W}_{\mathcal{P}^0} &\lesssim& \norme{\nabla \cdot \dope^{-1}M + \nabla tr\chi+E}_{\mathcal{P}^0} \\
& \lesssim & \mathcal{I}+ \mathcal{R}+ M_0^2+ \normeb{M}
\end{eqnarray*}

putting everything together :

\begin{equation}
\normeb{M}\lesssim \mathcal{I}+ \mathcal{R}+ M_0^2
\end{equation}

We can also derive directly a $L^2_xL^\infty_t$ of $M$ :

\begin{eqnarray*}
\nl{M}{2}{\infty}&\lesssim& \norme{M(0)}_{L^2(S_0)}+ \mathcal{R}+ \nl{\hat{\chi}\cdot \nabla \h}{2}{1}\\
&&+ \nl{(A+C)\nabla tr \chi}{2}{1}+ \nl{A \cdot A \cdot C}{2}{1}\\
&\lesssim& \mathcal{I}+\mathcal{R}+M_0^2
\end{eqnarray*}

\subsubsection{Estimates for $\h$ :}

Applying the elliptic estimates :

\begin{eqnarray*}
\ntsl{\h}{2}{2}+\ntsl{\nabla\h}{2}{2} &\lesssim& \ntsl{R}{2}{2}+\ntsl{M}{2}{2}+ \ntsl{A \cdot \underline{A}}{2}{2}\\
&\lesssim& \mathcal{I}+ \mathcal{R}+ M_0^2
\end{eqnarray*}

From another hand :

\begin{equation}
\nabla_L \h+ tr\chi \h= R + A\cdot A + A\cdot C
\end{equation}

it implies direclty :
\begin{equation}
\ntsl{\nabla_L \h}{2}{2}\lesssim \mathcal{I}+ \mathcal{R}+ M_0^2
\end{equation}

As usually the trace estimates are more difficult. We have proved :

\begin{equation}
\nabla \h = \nabla_L(P)+\nabla \cdot \dope^{-2}( \frac{1}{r}R_0) + \nabla \cdot \dope_1^{-1}M + E
\end{equation}

with $\nun{P}+ \norme{E}_{\mathcal{P}^0} \lesssim \mathcal{R}+M_0^2$.

But then we can apply the same control as previously and finally :
\begin{eqnarray*}
\nl{\h}{\infty}{2}&\lesssim& \nun{\h} + \nun{P} + \normep{\nabla \cdot \dope^{-2}( \frac{1}{r}R_0) + \nabla \cdot \dope_1^{-1}M + E}\\
&\lesssim&  \mathcal{I}+ \mathcal{R}+ M_0^2
\end{eqnarray*}

These results together give :
\begin{prop}
Assuming the bootstrap assumptions, we get :
\begin{eqnarray}
\nl{\tilde{\mu}}{2}{\infty} &\lesssim&\mathcal{I}+ \mathcal{R}+ M_0^2\\ 
\nun{\h}&\lesssim&\mathcal{I}+ \mathcal{R}+ M_0^2\\
\nl{\h}{\infty}{2}&\lesssim&\mathcal{I}+ \mathcal{R}+ M_0^2\\
\normeb{\tilde{\mu}}&\lesssim&\mathcal{I}+ \mathcal{R}+ M_0^2
\end{eqnarray}
\end{prop}

\vspace{5mm}
{\bf Conclusion :}
Taking $\mathcal{I}$ and $\mathcal{R}$ small enough and depending only on our fundamental constants enable us to close the bootstrap argument.

\newpage
\bibliographystyle{plain}
\bibliography{integral_breakdown_v2}
\end{document}